\documentclass[11pt]{article}
\usepackage{latexsym,amssymb,amstext,amsmath}
\usepackage{hyperref}
\renewcommand{\baselinestretch}{1.2}
\allowdisplaybreaks
\def\slash#1{\rlap{\hbox{$\mskip 1 mu /$}}#1}      
\def\Slash#1{\rlap{\hbox{$\mskip 3 mu /$}}#1}      
\def\oneone{\rlap 1\mkern4mu{\rm l}} 
\newcommand{\ft}[2]{{\textstyle\frac{#1}{#2}}}

\begin{document}
%
\begin{titlepage}
\begin{flushright} \small
Nikhef-2017-028
\end{flushright}
\bigskip

\begin{center}
 {\LARGE\bfseries  Euclidean Supergravity}
\\[10mm]
\textbf{Bernard de Wit$^{a,b}$ and Valentin Reys$^{c,d}$ }\\[5mm] 
\vskip 4mm
$^a${\em Nikhef Theory Group, Science Park 105, 1098 XG Amsterdam, The
  Netherlands}\\
$^b${\em Institute for Theoretical Physics, Utrecht  University,} \\
  {\em Leuvenlaan 4, 3584 CE Utrecht, The Netherlands}\\
$^c${\em Dipartimento di Fisica, Universit\'a di Milano-Bicocca,}\\
  {\em Piazza della Scienza 3, I-20126 Milano, Italy} \\
$^d${\em INFN, Sezione di Milano-Bicocca,}\\
  {\em Piazza della Scienza 3, I-20126 Milano, Italy}\\ [3mm]
{\tt B.deWit@uu.nl}\,,\; {\tt valentin.reys@unimib.it }  
\end{center}
\vspace{3ex}

\begin{center} {\bfseries Abstract}
\end{center}
\begin{quotation} \noindent Supergravity with eight supercharges in a
  four-dimensional Euclidean space is constructed at the full
  non-linear level by performing an off-shell time-like reduction of
  five-dimensional supergravity. The resulting four-dimensional theory
  is realized off-shell with the Weyl, vector and tensor
  supermultiplets and a corresponding multiplet
  calculus. Hypermultiplets are included as well, but they are
  themselves only realized with on-shell supersymmetry.  The off-shell
  reduction leads to a full understanding of the Euclidean theory.  A
  complete multiplet calculus is presented along the lines of the
  Minkowskian theory. Unlike in Minkowski space, chiral and
  anti-chiral multiplets are real and supersymmetric actions are
  generally unbounded from below. Precisely as in the Minkowski case,
  where one has different formulations of Poincar\'e
  supergravity by introducing different compensating supermultiplets, one can
  also obtain different versions of Euclidean supergravity.
\end{quotation}

\vfill

\flushleft{\today}
\end{titlepage}

\section{Introduction}
\label{sec:introduction}
\setcounter{equation}{0}
Euclidean versions of supersymmetric field theories are of interest in
the study of a variety of physical problems. An early application
concerned the many-instanton problem, which was first discussed by
Zumino on the basis of a supersymmetric version of Euclidean
Yang-Mills theory \cite{Zumino:1977yh}. This theory is not equivalent
to the Wick-rotated version of super-Yang-Mills theory. As is well
known, Majorana spinors cannot exist in a Euclidean four-space, but
there exists a version with eight supercharges that constitute a
symplectic Majorana spinor, whose R-symmetry group equals
$\mathrm{SU}(2)\times\mathrm{SO}(1,1)$. The presence of the
non-compact factor, which acts on spinors by chiral transformations,
was already noted by Dirac when discussing Euclidean
quantum-electrodynamics \cite{Schwinger:1959zz}. Subsequent studies of
the multi-instanton solutions in the context of supersymmetry were,
for instance, reported in \cite{D'Adda:1977ur,Novikov:1983ek}.
 
Also superconformal transformations played a role in the analysis of
\cite{Zumino:1977yh}. The conformal group associated with $4D$
Euclidean space equals $\mathrm{SO}(5,1)$, whose Lie algebra can be
extended to a superalgebra known as $\mathrm{SU}^\ast(4|2)$,
according to the classification presented in
\cite{Frappat:1996pb}.\footnote{
  See in particular table 16 of this work. We thank G. Inverso for
  pointing this out to us.} 
Neither conformal nor non-conformal supergravities have been fully
constructed in the context of Euclidean space-times. It is the purpose
of this paper to fill this gap, motivated by many other applications
where Euclidean supergravity is relevant. This will be done by
carrying out an {\it off-shell} time-like reduction on
five-dimensional (off-shell) Minkowski supergravity. In principle this
yields a complete off-shell dictionary of the $5D$ Minkowski fields in
terms of the $4D$ Euclidean fields. Because the reduction is
off-shell, it can also be used in the context of higher-derivative
couplings, as was, for instance, done in
\cite{Banerjee:2011ts,Banerjee:2015uee}. The same strategy for
studying aspects of the $4D$ Euclidean supergravities by reduction
from the $5D$ Minkowski theory was already used in
\cite{Cortes:2003zd,Cortes:2009cs}, but there the reduction was
performed {\it on-shell} and the emphasis was more directed toward
target-space aspects.

As we already mentioned, the above $4D$ Euclidean field theories are
not equivalent to the theories that one obtains by analytic
continuation to imaginary times of a $4D$ Minkowski theory. The latter
play an important role in the more rigorous treatment of quantum field
theories in the context of the Euclidean postulate, as well as in
field theories at finite temperature and lattice gauge theories. The
earliest discussion of such an analytically continued supersymmetric
theory was presented in \cite{Nicolai:1978vc}.  In a $4D$ Minkowski
theory one has at least four supercharges which constitute a Majorana
spinor, and under analytic continuation the number of supercharges
will remain the same.  In contradistinction, the $4D$ Euclidean theories
discussed in this paper have at least twice as many supercharges,
which constitute a symplectic Majorana spinor. 

Euclidean supersymmetric field theories have already played a role in
many applications and we briefly review some of them. First of all, a
consistent definition of a Euclidean supersymmetric gauge theory was
introduced in \cite{Witten:1988ze,Witten:1994ev}.  Upon twisting the
$\mathrm{SU}(2)$ R-symmetry group with one of the $\mathrm{SU}(2)$
factors belonging to the $\mathrm{SO}(4)$ tangent space group, the
Lagrangian remains invariant under a singlet supersymmetry for
non-trivial Euclidean spaces. The singlet supersymmetry closes into a
uniform gauge transformation and this property leads to a topological
theory with a $Q$-exact energy-momentum tensor, so that correlation
functions and the partition function become independent of the
metric. Subsequent work on the relation between the supersymmetric
field theories and their twisted version can be found in
e.g. \cite{Karlhede:1988ax,Galperin:1990qf}.

More recently, Euclidean supersymmetric theories became of interest
because of the possibility of applying supersymmetric localization,
which potentially leads to a way of obtaining exact results in
quantum field theories. In the seminal work of \cite{Pestun:2007rz},
$(9+1)$-dimensional super-Yang-Mills theory was dimensionally reduced
to 4-dimensional gauge theory in flat Euclidean space with 16
supersymmetries. This approach is similar to the one followed in
\cite{Cortes:2003zd} and in the present paper. Under a subsequent
conformal mapping $\mathbb{R}^4$ is changed to $S^4$. Because the path
integral plays a central role in localization, the convergence of the
latter was then ensured by deforming the integration contour of the
scalar fields originating from the time component of the $10D$ gauge
field to be over imaginary values.

A rather general analysis of $N=2$ conformal supergravity theories on
four-manifolds was conducted in \cite{Klare:2013dka} with a view
towards localization. In most cases one is interested in coupling the
matter multiplets to a given supergravity background that will be
frozen so that some rigid supersymmetry will remain. Only the matter
fields are involved in the localization and the residual rigid
supersymmetry is essential for setting up actual calculations. The
strategy for defining the Euclidean theory was to perform a Wick
rotation of the Minkowski theory to imaginary time, followed by a
doubling of the field components and the supersymmetry spinor
parameters. Subsequently one restricts the spinor parameters to
symplectic Majorana spinors and adopts a set of appropriate reality
conditions on the fields to ensure $N=2$ Euclidean
supersymmetry. Reality and boundedness of the resulting actions for
vector multiplets and hypermultiplets were not discussed. The
structure of the Euclidean supergravity only plays an ancillary role,
as one is primarily interested in a suitable fixed supergravity
background. This approach for deriving the Euclidean supergravity from
the Minkowskian one has become quite common in this context (in four
dimensions see, for instance, \cite{Hama:2012bg}; for applications in
other than four-dimensional Euclidean spaces, the reader may consult
the review \cite{Pestun:2016zxk}).  However, the approach based on
time-like reduction, used in \cite{Cortes:2003zd,Pestun:2007rz} and in
this paper, is perhaps more systematic, especially in the case where
the non-linear aspects of the theory will be important. In most cases
the various methods seem to lead to equivalent results.

The approach of \cite{Festuccia:2011ws} also makes use of a Wick
rotation to imaginary times, followed by an adjustment of the various
reality conditions on the fields of the Minkowski theory. A more explicit
construction was presented in the follow-up paper
\cite{Dumitrescu:2012ha}. This approach differs in that it restricts
itself to {\it four} supersymmetries, which seemingly excludes to make
reference to a consistent version of Euclidean supersymmetry, as that
would require at least {\it eight} supersymmetries.  It is quite
conceivable that the resulting theory can nevertheless be related to
the full Euclidean supergravity in the context of suitable truncations
or by considering supergravity solutions with partially broken
supersymmetry. Another topic addressed in \cite{Festuccia:2011ws}
concerns the extra terms that one needs in the action when coupling to
curved spaces (which were determined by making use of supercurrent
techniques). Having the full supergravity available, these terms will
alternatively follow from standard supergravity Lagrangians. Here we
emphasize that the standard introduction of compensating fields to
conformal supergravity proceeds in a way that is completely equivalent
to what has been done in the Minkowskian theories.

In most cases localization has been applied to compact spaces, but
there are also applications to anti-de Sitter spaces
\cite{Dabholkar:2010uh,Dabholkar:2014wpa}. In the first reference
localization was applied to the evaluation of the quantum entropy
function \cite{Sen:2008vm} to obtain the exact entropy of certain
supersymmetric black holes, which can be compared to exact results
that follow from string theory, such as given in
\cite{Dijkgraaf:1996it, Maldacena:1999bp}.  The second reference used
localization to compute the partition function of anti-de Sitter space
in four dimensions and related it to the partition function of the
dual three-dimensional boundary conformal field theory
\cite{Aharony:2008ug}. The reader may also consult
\cite{Kapustin:2009kz,Fuji:2011km} for additional contributions. As
demonstrated by these applications, localization does not necessarily
involve only matter fields in a fixed supergravity background. This
fact provides a specific motivation for the construction of the full
non-linear Euclidean $4D$ supergravity that we intend to present in
this paper.

The construction of $4D$ Euclidean supergravity will be based on an
{\it off-shell} time-like reduction of $5D$ $N=1$ Minkowski
supergravity to four Euclidean dimensions. This reduction is based on a
corresponding reduction of the off-shell supersymmetry algebra and can
therefore be performed systematically on separate supermultiplets. To
accomplish this one maps a supermultiplet in a higher dimension on a
corresponding, not necessarily irreducible, supermultiplet in lower
dimension. When considering the supersymmetry algebra in the context
of a lower-dimensional space-time, the dimension of the automorphism
group of the algebra (the R-symmetry group) usually increases, and
this has to be taken into account when casting the resulting
supermultiplet in its standard form. The fact that an irreducible
multiplet in higher dimension can possibly become reducible in lower
dimensions tends to complicate the reduction procedure. The same technique
has been followed in \cite{Banerjee:2011ts} to establish the precise
relation between the $5D$ and $4D$ Minkowskian supergravities with
higher-order derivatives.

The construction is facilitated by the fact that the spinor
dimension is the same in five and in four dimensions. In fact, both
the $5D$ Minkowski and $4D$ Euclidean supergravities are based on
symplectic Majorana spinors, which share a common $\mathrm{SU}(2)$
factor in the R-symmetry group. We will exhibit in detail how the
additional $\mathrm{SO}(1,1)$ factor will emerge in four
dimensions. Here we recall that in conformal supergravity, R-symmetry
is realized as a local symmetry.

The whole reduction scheme is subtle, especially in view of the fact
that the $5D$ Weyl multiplet decomposes into a $4D$ Weyl multiplet and
an additional vector multiplet. In spite of this, both in five and in
four dimensions, the matter multiplets are defined in a generic
superconformal background expressed in the Weyl
multiplet fields. To fully establish this fact requires
to also consider the transformation rules beyond the linearized
approximation. Once we obtain the $4D$ Euclidean off-shell
transformation rules, we write them in close analogy with the Minkowskian
fields in similar notation and normalization factors.  After this it
is in principle straightforward to obtain the effective actions and
other results, either by substituting the Euclidean fields or by
comparison to the Minkowski formulae. However, there remain some
subtle differences that require special care, such as the fact that
chiral and anti-chiral multiplets are real in the Euclidean setting.

This paper is organized as follows. Section
\ref{sec:euclid-superg-from-dim-red} presents a brief summary of $5D$
supergravity with Minkowski signature, followed by a brief explanation
of some key features of the time-like reduction.  The actual details
of the reduction have been relegated to appendices, so that
section \ref{sec:euclidean-sg-4D} will start with the definition of
the $4D$ Euclidean fields and their transformation rules. Apart from
various definitions this section presents a full discussion of the
Euclidean Weyl supermultiplet.  The subsequent sections also make
heavy use of the Euclidean transformation rules that follow from the
reduction.  Chiral and anti-chiral multiplets are discussed in section
\ref{sec:chiral-multipletl} and section \ref{sec:matter-multiplets}
gives a summary of the vector and tensor supermultiplets and the
hypermultiplet. Vector multiplet systems are discussed in section
\ref{sec:vector-multiplet-actions}, including the Euclidean
electric-magnetic duality group. Finally our conclusions and summary
are presented in section \ref{sec:summary-conclusions}. 

The details of the off-shell reduction are relegated to two
appendices. Appendix \ref{sec:off-shell-dim-red-Weyl} deals with the
Weyl multiplet and the disentanglement of a $4D$ Kaluza-Klein vector
multiplet. Appendix \ref{sec:shell-dimens-reduct-matter} deals with
the various matter multiplets. For completeness we also include the
transformations of the so-called `non-linear multiplet' in five
dimensions, which was originally discovered in four dimensions, to
demonstrate that the various formulations of $4D$ Minkowskian
supergravity each have their corresponding counterparts in $4D$
Euclidean supergravity.

\section{Euclidean supergravity from off-shell time-like
  dimensional reduction}
\label{sec:euclid-superg-from-dim-red}
\setcounter{equation}{0}
A convenient and systematic way to obtain the full Euclidean
supergravity in four dimensions is to start from five-dimensional
Minkowskian supergravity with eight supersymmetries and perform a
time-like reduction. To obtain a complete off-shell formulation of
Euclidean supergravity, the dimensional reduction has to be applied
without referring to specific Lagrangians or to field equations.  The
primary purpose is to fully uncover the structure of the
four-dimensional Euclidean theory whose off-shell features have not
been pursued systematically so far.

Let us therefore start by briefly summarizing the salient features of
the $5D$ Weyl multiplet. Subsequently we will indicate the starting
points for the time-like dimensional reduction scheme. The more
detailed derivation of the off-shell reduction is relegated to
appendix \ref{sec:off-shell-dim-red-Weyl}. The resulting formulation
of $4D$ off-shell Euclidean supergravity is presented in the next
section \ref{sec:euclidean-sg-4D}.

The independent fields of the $5D$ Weyl multiplet consist of the
f\"unfbein $e_M{}^A$, the gravitino fields $\psi_M{}^i$, the
dilatational gauge field $b_M$, the R-symmetry gauge fields $V_{M
  i}{}^j$ (which is an anti-hermitian, traceless matrix in the
$\mathrm{SU}(2)$ indices $i,j$),  a tensor $T_{AB}$, a scalar
$D$ and a spinor $\chi^i$. All spinor fields are
symplectic Majorana spinors. Our conventions are as in
\cite{Banerjee:2011ts}. The three gauge fields $\omega_M{}^{AB}$,
$f_M{}^A$ and $\phi_M{}^i$, associated with local Lorentz
transformations, conformal boosts and S-supersymmetry, respectively,
are composite fields as will be discussed later. The infinitesimal Q, S
and K transformations of the independent fields, parametrized by
spinors $\epsilon^i$ and $\eta^i$ and a vector
$\Lambda_\mathrm{K}{}^A$, respectively, are as
follows,\footnote{
  In five dimensions we consistently use world indices $M,N,\ldots$
  and tangent space indices $A,B,\ldots$. For fields that do not carry
  such indices the distinction between $5D$ and $4D$ fields may not
  always be manifest, but it will be indicated in the text whenever
  necessary.} 
\begin{align}
  \label{eq:Weyl-susy-var}
  \delta e_M{}^A =&\,  \bar\epsilon_i \gamma^A \psi_M{}^i\,,
  \nonumber\\ 
  \delta \psi_{M}{}^i  =&\, 2\,
  {\cal  D}_M \epsilon^i + \tfrac1{2}\mathrm{i}\,
  T_{AB}( 3\,\gamma^{AB}\gamma_M-\gamma_M\gamma^{AB}) \epsilon^i
  -\mathrm{i}  \gamma_M\eta^i \,, \nonumber\\ 
  \delta V_{M i}{}^j =&\, 6 \mathrm{i}\,
  \bar\epsilon_{i} \phi_{M}{}^{j}
  -16\, \bar\epsilon_{i}\gamma_M\chi^{j} -3 \mathrm{i}\,
  \bar\eta_{i}\psi_M{}^{j} + 
  \delta^i{}_j\,[-{3}\mathrm{i}\,\bar\epsilon_{k}\phi_{M}{}^{k}
  +8\,\bar\epsilon_{k}\gamma_M\chi^{k}+\ft{3}{2}\mathrm{i}\,
  \bar\eta_{k}\psi_M{}^{k}] \,, \nonumber \\
  \delta b_M =&\,
  \mathrm{i} \bar\epsilon_i\phi_M{}^i -4 \bar\epsilon_i\gamma_M
  \chi^i + \ft12\mathrm{i} \bar\eta_i\psi_M{}^i +2\,\Lambda _\mathrm{K}{\!}^A\,
  e_{MA} \,, \nonumber\\
  \delta T_{AB} =&\,  \ft43 \mathrm{i}\, \bar\epsilon_i \gamma_{AB}
  \chi^i -\ft{1}{4} \mathrm{i}\, \bar\epsilon_i R_{AB}{}^i(Q)\,,
  \nonumber\\  
  \delta \chi^i =&\,  
  \ft 12 \epsilon^i D +\ft{1}{64} 
  R_{MN j}{}^{i}(V) \,\gamma^{MN} \epsilon^j 
  + \tfrac3{64}\mathrm{i}(3\, \gamma^{AB} \Slash{D}
  +\Slash{D}\gamma^{AB})T_{AB} \, \epsilon^i \nonumber\\
  &\,
  -\tfrac 3{16} T_{AB}T_{CD}\gamma^{ABCD}\epsilon^i 
  +\tfrac3{16} T_{AB}\gamma^{AB} \eta^i  \,, \nonumber\\
  \delta D =&\,
  2\, \bar\epsilon_i \Slash{D} \chi^i - 2\mathrm{i}\,
  \bar\epsilon_i  T_{AB}\,\gamma^{AB} \chi^i - \mathrm{i}
  \bar\eta_i\chi^i \,. 
\end{align}
Under local scale transformations the fields and transformation
parameters transform as indicated in table
\ref{tab:weyl-multiplet}. The derivatives $\mathcal{D}_M$ are
covariant with respect to all the bosonic gauge symmetries with the
exception of the conformal boosts. In particular we note
\begin{equation}
  \label{eq:D-epsilon}
\mathcal{D}_{M} \epsilon^i = \big( \partial_M - \tfrac{1}{4}
\omega_M{}^{CD} \, \gamma_{CD} + \tfrac1{2} \, b_M\big)
\epsilon^i + \tfrac1{2} \,{V}_{M j}{}^i \, \epsilon^j  \,, 
\end{equation}
where the gauge fields transform under their respective gauge
transformations according to
$\delta\omega_M{}^{AB}=\mathcal{D}_M\epsilon^{AB}$, $\delta b_M=
\mathcal{D}_M\Lambda_D$ and $\delta V_{M i}{}^j= -2 \,\mathcal{D}_M
\Lambda_i{}^j$, with $(\Lambda_i{}^j)^\ast\equiv \Lambda^i{}_j=
- \Lambda_j{}^i$. The derivatives $D_M$ are covariant with
respect to all the superconformal symmetries. 
  
%
\begin{table}[t]
\centering
\begin{tabular*}{14.9cm}{@{\extracolsep{\fill}}
|c||ccccccc|ccc||ccc|}
\hline 
 & &\multicolumn{8}{c}{Weyl multiplet} & &
 \multicolumn{2}{c}{parameters} & \\  \hline 
 field & $e_M{}^{A}$ & $\psi_M{}^i$ & $b_M$ &
 ${V}_{M\,i}{}^j$ & $T_{AB} $ & 
 $ \chi^i $ & $D$ & $\omega_{M}{}^{AB}$ & $f_M{}^A$ &$\phi_M{}^i$&
 $\epsilon^i$ & $\eta^i$  
 & \\ \hline
$w$  & $-1$ & $-\tfrac12 $ & 0 &  0 & 1 & $\tfrac{3}{2}$ & 2 & 0 &
1 & $\tfrac12 $ & $-\tfrac12$  & $\tfrac12$ & \\ \hline 
\end{tabular*}
\vskip 2mm
\renewcommand{\baselinestretch}{1}
\parbox[c]{14.8cm}{\caption{\footnotesize
    Weyl weights $w$ of the $5D$ 
    Weyl multiplet component fields and the supersymmetry
    transformation parameters. \label{tab:weyl-multiplet}}}    
\end{table}

The above supersymmetry variations and also the conventional
constraints involve a number of supercovariant curvature tensors,
denoted by $R(P)_{MN}{}^A$, $R(M)_{MN}{}^{AB}$, $R(D)_{MN}$,
$R(K)_{AB}{}^A$ $R(V)_{MN i}{}^j$, $R(Q)_{MN}{}^i$ and $R(S)_{MN}{}^i$
whose explicit form can be found in \cite{Banerjee:2011ts}. The
so-called conventional constraints,
\begin{align}
  \label{eq:conv-constraints-5}
  R(P)_{MN}{\!}^A = 0\,,\qquad 
  \gamma^M R(Q)_{MN}{\!}^i = 0\,, \qquad
  e_A{\!}^M\, R(M)_{MN}{\!}^{AB} = 0 \,, 
\end{align}
determine the gauge fields $\omega_M{}^{AB}$, $f_M{}^A$ and
$\phi_M{}^i$. These constraints lead to additional conditions on the
curvatures when combined with the Bianchi identities. In this way one
can derive $R(M)_{[ABC]D} =0= R(D)_{AB}$ and the pair-exchange
property $R(M)_{ABCD}=R(M)_{CDAB}$ from the first and the third
constraint.  The second constraint, which implies also that
$\gamma_{[MN} R(Q)_{PQ]}{}^i =0$, determines the curvature
$R(S)_{MN}{}^i$.

The reduction to four space-time dimensions is effected by first
carrying out the standard Kaluza-Klein decompositions on the various
fields that will ensure that the resulting $4D$ fields will transform
consistently under four-dimensional diffeomorphisms. The $5D$
space-time coordinates $x^M$ are decomposed into four coordinates
$x^\mu$ and a fifth coordinate $x^{\hat 5}$. The dependence on this
fifth coordinate will be suppressed in the reduction. Likewise the
tangent-space indices $A$ decompose into the four indices $a=1,2,3,4$
and a fifth index $A=5$. In Pauli-K\"all\'en notation one of the
coordinates is imaginary so that the $5D$ space-time signature will be
a permutation of $(-++++)$. In \cite{Banerjee:2011ts} the fifth
coordinate $x^{\hat 5}$ was real, so that the reduced theory was based
on a four-dimensional Minkowskian space-time. In this paper we
consider the time-like reduction where the fifth coordinate is purely
imaginary. Upon the reduction, where the dependence on the fifth
coordinate is suppressed, the resulting theory will then be based on a
four-dimensional Euclidean space. An important observation is that the
results of \cite{Banerjee:2011ts} were obtained with Pauli-K\"all\'en
convention, which enables a direct conversion into the Euclidean theory
by an appropriate change of the reality conditions on the fields. One
simply has to include factors $\pm \mathrm{i}$ whenever dealing with
the fifth world or tangent-space component. For instance, the fifth
coordinate of $x^M$ takes the form $x^{\hat5}= \mathrm{i} x^0$, so
that the fifth component of a contravariant vector field $V^{\hat5}$
will be imaginary and can be written as $\mathrm{i} V^0$, where $V^0$
is real. For a covariant vector the fifth component $W_{\hat 5}$ will
instead be equal to $-\mathrm{i} W_0$, where $W_0$ is real. A
corresponding rule applies to tangent-space vectors.

After this general introduction we will exhibit the consequences of
the above strategy. As is standard, the vielbein field and the
dilatational gauge field are first written in special form, by means
of an appropriate local Lorentz transformation and a conformal boost
in the time direction, respectively. In obvious notation,
\begin{equation}
  \label{eq:kk-ansatz}
  e_M{}^A= \begin{pmatrix} e_\mu{}^a & \mathrm{i} B_\mu\phi^{-1} \\[4mm]
    0 & \phi^{-1}
    \end{pmatrix} \;,\qquad
    e_A{}^M= \begin{pmatrix} e_a{}^\mu & -\mathrm{i} e_a{}^\nu B_\nu \\[4mm]
    0 & \phi
    \end{pmatrix}\;,\qquad
    b_M = \begin{pmatrix} b_\mu \\[4mm]  0
  \end{pmatrix} \,.
\end{equation}
Note that the vielbein field is not real because we will keep using
the tangent-space indices $A=1,\ldots,5$.  As compared to the
space-like reduction the field $\phi$ has remained unchanged while the
Kaluza-Klein gauge field $B_\mu$ requires a factor $\mathrm{i}$ so
that it remains real.  All the fields on the right-hand side of
\eqref{eq:kk-ansatz} are now real and possible sign factors depend on
whether we have suppressed an upper coordinate $A=5$ and/or a lower
coordinate $M=\hat5$. The fields now carry only four-dimensional world
and tangent-space indices, $\mu,\nu,\ldots$ and $a,b,\ldots$, taking
four values while the components referring to the fifth direction will be
suppressed. Observe that the scaling weights for $e_M{}^A$ and
$e_\mu{}^a$ are equal to $w=-1$, while for $\phi$ we have $w=1$. The
fields $b_M$, $b_\mu$ and $B_\mu$ carry weight $w=0$. 

For the fermions there is yet no need to introduce new notation,
because the spinors have an equal number of components in five and
four space-time dimensions. We will thus employ symplectic Majorana
spinors $\psi^i$ with $i=1,2$ subject to the reality
constraint,\footnote{ 
  The charge conjugation matrix $C$ has the properties $C\gamma_A
  C^{-1} = \gamma_A{}^{\rm T}$, with $C^{\rm T} = -C$ and $ C^\dagger
  = C^{-1}$. The $5D$ hermitian gamma matrices in Pauli-K\"all\'en notation
  satisfy $\gamma_{ABCDE} = {\bf 1}\,\varepsilon_{ABCDE}$.
} 
\begin{equation}
  \label{eq:Majorana}
  C^{-1} \,\bar\chi_i {}^{\rm T}= \varepsilon_{ij}\,\chi^j\,.
\end{equation}
The Dirac conjugate is defined by $\bar\psi= \psi^\dagger\gamma^5$,
where $\gamma^5=\mathrm{i} \gamma^0$.  Observe that we adhere to the
convention according to which raising or lowering of $\mathrm{SU}(2)$
indices is effected by complex conjugation. For fermionic bilinears,
with spinor fields $\psi^i$ and $\varphi^i$ and a spinor matrix
$\Gamma$ constructed from products of gamma matrices, we note the
following result,
\begin{equation}
  \label{eq:bilinear}
  (\bar\varphi_j\,
  \gamma_5\,\Gamma^{\dagger}\,\gamma_5 \,\psi^i)^\dagger = \bar
  \psi_i\Gamma \varphi^j= - \delta_i{}^j \,
  \bar\varphi_k\,  
  C^{-1}\, \Gamma^{\rm T}\,C\, \psi^k   + \bar\varphi_i\,  
  C^{-1}\, \Gamma^{\rm T}\,C\, \psi^j\,. 
\end{equation}
Hence the bilinears $O_i{}^j$ equal to $\mathrm{i}\,\bar \psi_i\,
\varphi^j$, $\bar \psi_i\gamma_A \varphi^j$ and $\mathrm{i}\,\bar
\psi_i\gamma_{AB} \varphi^j$ are pseudo-hermitean: $O^i{}_j=
\varepsilon^{ik}\varepsilon_{jl} \,O_k{}^l$, provided $A,B,\ldots=
1,\ldots,4$. In the context of the spinors special care is required in
converting the Minkoswki ones into the Euclidean signature, because (Fierz)
reordering of the spinors depends sensitively on whether the spinor is
a Majorana or an anti-Majorana field. Observe that the gravitino field
$\psi_{\hat 5}$, with its world index in the fifth direction, will be
an anti-Majorana field. This will be properly accounted for in the
Kaluza-Klein ans\"atze, which will include the proper factors of the
imaginary unit, as can be seen in appendix
\ref{sec:off-shell-dim-red-Weyl}.

\section{Supergravity with eight supercharges in four Euclidean 
  dimensions}
\label{sec:euclidean-sg-4D}
\setcounter{equation}{0}
Here we present the superconformal transformation rules in $4D$
Euclidean supergravity that have been derived in appendix
\ref{sec:off-shell-dim-red-Weyl} by means of an off-shell reduction of
the $5D$ theory. The $5D$ R-symmetry is then extended to chiral
$\mathrm{SU}(2)\times\mathrm{SO}(1,1)$. Unlike in $4D$ Minkowski
supergravity the spinors of the Euclidean theory will be symplectic
Majorana spinors. For Euclidean spinors it is natural to define the
Dirac conjugate by hermitian conjugation, so that $\bar \chi_i \equiv
\chi_i^\dagger$. With this definition the symplectic Majorana spinors
satisfy the condition
\begin{equation}
  \label{eq:Majorana-4D}
  C^{-1} \,\bar\chi_i {}^{\rm T}= \varepsilon_{ij}\,\chi^j\,,
\end{equation}
The charge conjugation matrix $C$ thus differs from the $5D$
one. While it is still anti-symmetric and unitary, just
as in five dimensions, the (hermitian) gamma matrices $\gamma_a$
now satisfy  the relation 
\begin{equation}
  \label{eq:gamma-C}
  C\,\gamma_a\,C^{-1} = -\gamma_a{\!}^\mathrm{T}\,, \qquad
  (a=1,2,3,4). 
\end{equation}
Note here that $\gamma_5\equiv \gamma_1
\,\gamma_2\,\gamma_3\,\gamma_4$ satisfies 
\begin{equation}
  \label{eq:gamma-5}
  C\,\gamma_5\,C^{-1} = \gamma_5{\!}^\mathrm{T}\,. 
\end{equation}
In appendix \ref{sec:mod-conventional-and chiral} it is explained how
these definitions emerge in the reduction from five dimensions (see,
in particular, \eqref{eq:new-C}).  As always, raising or lowering of
$\mathrm{SU}(2)$ indices is effected by complex conjugation. For
fermionic bilinears $\bar\psi_i \Gamma \varphi^j$ we note the
following result,
\begin{equation}
  \label{eq:euclidean-bilinear}
  (\bar\varphi_j\,
  \Gamma^{\dagger}\,\,\psi^i)^\dagger = \bar
  \psi_i\Gamma \varphi^j=-\varepsilon_{ik}\, \varepsilon^{jl} \, 
  \bar\varphi_l\,  
  C^{-1}\, \Gamma^{\rm T}\,C\, \psi^k\,, 
\end{equation}
which is rather similar but not identical to the $5D$ result
\eqref{eq:bilinear}. We note that \eqref{eq:euclidean-bilinear}
applies also to chiral spinors. For convenience we also specify the
Fierz rearrangement formulae for chiral spinors,
\begin{align}
  \label{eq:Fierz}
  \varphi^j{\!}_\pm \; \bar\psi_{i\pm} =&\, -\tfrac14(1\pm\gamma^5) \,
  \big(\bar\psi_{i\pm} \,\varphi^j{\!}_\pm\big) + \tfrac1{8} \gamma^{ab}
  \, \big(\bar\psi_{i\pm} \gamma_{ab}\varphi^j{\!}_\pm\big) \,,\nonumber\\[1mm]
    \varphi^j{\!}_\mp \; \bar\psi_{i\pm} =&\, -\tfrac14 \gamma^a (1\pm\gamma^5) \,
  \big(\bar\psi_{i\pm} \gamma_a \varphi^j{\!}_\mp \big)\,. 
\end{align}

Let us now consider the transformations of the fields of the Euclidean
$4D$ Weyl multiplet. The superconformal algebra comprises the
generators of the spatial diffeomorphisms, local tangent space rotations,
dilatation, special conformal, chiral $\mathrm{SU}(2)$ and
$\mathrm{SO}(1,1)$, supersymmetry (Q) and special supersymmetry (S)
transformations.  The gauge fields associated with general-coordinate
transformations ($e_\mu{}^a$), dilatations ($b_\mu$), chiral symmetry
($\mathcal{V}_\mu{}^i{}_j$ and $A_\mu$) and Q-supersymmetry
($\psi_\mu{}^i$) are independent fields.  The remaining gauge fields
associated with the Lorentz ($\omega_\mu{}^{ab}$), special conformal
($f_\mu{}^a$) and S-supersymmetry transformations ($\phi_\mu{}^i$) are
composite fields. Obviously, world and tangent-space indices $\mu$ and
$a$, respectively, take the values $1,2,3,4$.  The corresponding
supercovariant curvatures and covariant fields are contained in a
tensor chiral multiplet, which comprises $24 \oplus 24$ off-shell degrees of
freedom. This relation with the chiral supermultiplet is decribed in
section \ref{sec:chiral-multipletl}. In addition to the independent
superconformal gauge fields, it contains three other fields: a
symplectic Majorana spinor doublet $\chi^i$, a scalar $D$, and a
Lorentz antisymmetric tensor $T_{ab}$, which decomposes into a
self-dual and an anti-self-dual field. The Weyl and chiral weights
have been collected in table \ref{table:weyl}.

As derived in detail in appendix \ref{sec:off-shell-dim-red-Weyl} the
Euclidean fields, which comprise $24\oplus 24$ bosonic and fermionic degrees
of freedom, transform as follows under the Q-, S- and K-
transformations of the superconformal algebra,
\begin{align}
    \delta e_\mu{}^a =&\,  \bar\epsilon_i\,\gamma^5\gamma^a\psi_\mu{}^i
  \,, \nonumber\\[.1ex]
    \delta\psi_\mu{}^i =&\,2\,\mathcal{D}_\mu \epsilon^i  
    + \tfrac1{16} \mathrm{i}  \,(T_{ab}{\!}^+ +T_{ab}{\!}^- ) \gamma^{ab} \gamma_\mu
    \epsilon^i    -\mathrm{i} \gamma_\mu \eta^i \,, \nonumber\\[1mm]
   \delta b_\mu =&\, \tfrac12 \mathrm{i}\, \bar\epsilon_i\,\gamma^5
   \phi_\mu{}^i -\tfrac34 \bar\epsilon_i\,\gamma^5\gamma_\mu \chi^i
   +\tfrac12 \mathrm{i}\,  \bar\eta_i \,\gamma^5\psi_\mu{\!}^i  +
   \Lambda_\mathrm{K}{\!}^a \,e_{\mu a} \,,
   \nonumber\\[1mm]
   \delta{A}_\mu =&\,
   -\tfrac12\mathrm{i}\,\bar{\epsilon}_i \,\phi_\mu{}^i -
   \tfrac34\,\bar{\epsilon}_i\,\gamma_\mu \chi^i 
  -\tfrac12 \mathrm{i}\,\bar\eta_i\,\psi_\mu{\!}^i \, ,
  \nonumber\\[1mm]
        \delta\mathcal{V}_{\mu}{}^i{}_j =&\, 
      2\mathrm{i}\,\bar{\epsilon}_j\,\gamma^5\phi_\mu{}^i
      - 3\, \bar{\epsilon}_j\,\gamma^5\gamma_\mu{\chi}^i - 2\mathrm{i}\,\bar\eta_j
      \,\gamma^5\psi_\mu{}^i\nonumber\\
      &\,- \tfrac12\delta^i{}_j\bigl(2\mathrm{i} \,
      \bar{\epsilon}_k\,\gamma^5{\phi}_\mu{}^k - 3\,
      \bar{\epsilon}_k\,\gamma^5\gamma_\mu{\chi}^k
      - 2\mathrm{i}\,\bar\eta_k\,\gamma^5  \psi_\mu{}^k \bigr) \, ,  \nonumber\\[1mm]
      \delta{T}_{ab}{\!}^\pm =&\, -8\mathrm{i}\,
    \bar\epsilon_i\,\gamma^5 R(Q)_{ab}{\!}^{i\,\pm} \,,  \nonumber \\[1mm]
    \delta\chi^i=&\,  \tfrac1{24}
  \mathrm{i} \,\gamma^{ab} \, \Slash{D}
  (T_{ab}^+ + T_{ab}^-)  \epsilon^i+
  \tfrac1{6}R(\mathcal{V})_{ab}{\!}^i{}_j \,\gamma^{ab} \epsilon^j
  -\tfrac13 R(A)_{ab} \,\gamma^{ab} \gamma^5 \epsilon^i \nonumber\\ 
  &\,  + D\,
  \epsilon^i +\tfrac1{24}(T_{ab}^+ + T_{ab}^-) \gamma^{ab} \eta^i \,,
  \nonumber\\[1mm] 
    \delta{D} =&\, \bar{\epsilon}_i \,\gamma^5 \Slash{D}\chi^i \,,
\end{align}
Note that we have refrained from decomposing the spinors into chiral
components, to keep the equations as compact as possible. Only when
referring specifically to (anti)chiral spinors will we add the
subscript $\pm$. Note that $\epsilon^i$ denotes the symplectic
Majorana parameter of Q-supersymmetry, $\eta^i$ the symplectic
Majorana parameter of S-supersymmetry, and $\Lambda_\mathrm{K}{\!}^a$ is
the transformation parameter for special conformal boosts. The full
superconformal covariant derivative is denoted by $D_\mu$, while
$\mathcal{D}_\mu$ denotes a derivative covariant with respect to
Lorentz, dilatation, chiral $\mathrm{SO}(1,1)$, and $\mathrm{SU}(2)$
transformations. In particular,
\begin{align}
  \label{eq:D-epsilon-eta}
  \mathcal{D}_{\mu} \epsilon^i =&\,  \big(\partial_\mu -
  \tfrac{1}{4}\omega_\mu{}^{ab} \, \gamma_{ab} + \tfrac1{2} \, b_\mu +
  \tfrac{1}{2}\, A_\mu\gamma^5\big) \epsilon^i + \tfrac1{2} \,
  \mathcal{V}_{\mu}{}^i{\!}_j \, \epsilon^j \,, \nonumber\\
  \mathcal{D}_{\mu} \eta^i =&\, \big(\partial_\mu -
  \tfrac{1}{4}\omega_\mu{}^{ab} \, \gamma_{ab} - \tfrac1{2} \, b_\mu -
  \tfrac{1}{2}\, A_\mu\gamma^5\big) \eta^i + \tfrac1{2} \,
  \mathcal{V}_{\mu}{}^i{\!}_j \, \eta^j \, . 
\end{align}

\begin{table}[t]
\centering
\begin{tabular*}{14.8cm}{@{\extracolsep{\fill}}
    |c||cccccccc|ccc||ccc| }
\hline
 & &\multicolumn{9}{c}{Weyl multiplet} & &
 \multicolumn{2}{c}{parameters} & \\[1mm]  \hline \hline
 field & $e_\mu{}^{a}$ & $\psi_\mu{\!}^i$ & $b_\mu$ & $A_\mu$ &
 $\mathcal{V}_\mu{}^i{}_j$ & $T_{ab}^\pm $ &
 $ \chi^i $ & $D$ & $\omega_\mu^{\,ab}$ & $f_\mu{}^a$ & $\phi_\mu{\!}^i$ &
 $\epsilon^i$ & $\eta^i$
 & \\[.5mm] \hline
$w$  & $-1$ & $-\tfrac12 $ & 0 &  0 & 0 & 1 & $\tfrac{3}{2}$ & 2 & 0 &
1 & $\tfrac12 $ & $ -\tfrac12 $  & $ \tfrac12  $ & \\[.5mm] \hline
$c$  & $0$ & $\mp\tfrac12 $ & 0 &  0 & 0 & $\pm1$ & $\mp\tfrac{1}{2}$ & 0 &
0 & 0 & $\pm\tfrac12 $ & $ \mp\tfrac12 $  & $ \pm\tfrac12  $ & \\[.5mm] \hline
 $\tilde\gamma_5$   &  & $\pm$ &   &  $\pm$  &   &  $\pm$ & $\pm$ &  &  $\pm$&
 & $\pm$ & $\pm$  & $\pm $ & \\ \hline
\end{tabular*}
\vskip 2mm
\renewcommand{\baselinestretch}{1}
\parbox[c]{14.8cm}{\caption{\label{table:weyl}{\footnotesize
      Weyl weights $w$ and chiral $\mathrm{SO}(1,1)$ weights $c$, 
      chirality/duality $\tilde\gamma_5$ of the spinors and the
      antisymmetric tensor field 
      for the $4D$ Euclidean Weyl multiplet.}}}
\end{table}

There are three additional gauge fields, namely $\omega_\mu{\!}^{ab}$,
$\phi_\mu{\!}^i$ and $f_\mu{\!}^a$, associated with $\mathrm{SO}(4)$
tangent space rotations, S-supersymmetry and conformal boosts,
respectively, with corresponding parameters $\epsilon^{ab}$, $\eta^i$
and $\Lambda_\mathrm{K}{\!}^a$. These fields are composite and depend
on the other fields. This is expressed in terms of so-called
conventional constraints which contain various superconformal
curvature tensors that will be presented momentarily. We first present
the explicit solutions for $\omega_{\mu}{\!}^{ab}$, $\phi_{\mu}{\!}^i$
and $f_{\mu}{\!}^a$,
\begin{align}
  \label{eq:dependent-gf}
  \omega_\mu{\!}^{ab} =&\, -2\,e^{\nu[a} \,(\partial_{[\mu} + b_{[\mu}) e_{\nu]}{\!}^{b]}
     -e^{\nu[a}e^{b]\rho}\,e_{\mu c}\,(\partial_\rho +b_\rho) e_\nu{\!}^c
      -\ft{1}{4}\bigl(2\,\bar{\psi}_{\mu\,i}\,\gamma^5\gamma^{[a}\psi^{b]i}
     + \bar{\psi}^a{}_i\,\gamma^5\gamma_\mu\psi^{b\,i}\bigr) \, , \nonumber\\ 
     \phi_{\mu}{\!}^i =& \, -\tfrac12\mathrm{i}\left( \gamma^{\rho
         \sigma} \gamma_\mu - \tfrac{1}{3} \gamma_\mu \gamma^{\rho
         \sigma} \right) \left(\mathcal{D}_\rho \psi_{\sigma}{}^i +
       \tfrac{1}{32}\mathrm{i}\,(T_{ab}^+ + T_{ab}^-) \gamma^{ab}
       \gamma_\rho\psi_{\sigma}{}^i + \tfrac{1}{4} \gamma_{\rho
         \sigma} \chi^i \right) \, , \nonumber\\ 
   f_\mu{\!}^{a} =& \, \tfrac12\,R(\omega,e)_\mu{}^a -
  \tfrac14\,\bigl(D+\tfrac13 R(\omega,e)\bigr) e_\mu{}^a -
  \tfrac12\,\widetilde{R}(A)_\mu{}^a - \tfrac1{32}\,T_{\mu
    b}^-\,T^{+\,ba} + \cdots\,,
\end{align}
where the ellipses in the last equation denote fermionic
contributions. Here and elsewhere we use the notation where $\widetilde
R_{ab}$ equals the dual tensor $\tfrac12
\varepsilon_{abcd}\,R^{cd}$. Furthermore, $R(\omega,e)_\mu{}^a=
R(\omega)_{\mu\nu}{}^{ab} e_b{}^\nu$ is the non-symmetric Ricci
tensor, and $R(\omega,e)$ the corresponding Ricci scalar. The
uncontracted curvature $R(\omega)_{\mu\nu}{\!}^{ab}$ is defined by
\begin{equation}
  \label{eq:R-omega}
   R(\omega)_{\mu\nu}{\!}^{ab} = \partial_\mu\omega_\nu{\!}^{ab} 
   -\partial_\nu\omega_\mu{\!}^{ab} -\omega_\mu{\!}^{ac} \,
   \omega_{\nu{c}}{\!}^{b } 
   +\omega_\nu{\!}^{ac} \,   \omega_{\mu{c}}{\!}^{b }\,. 
\end{equation}
The Q- and S-supersymmetry variations of the composite gauge fields,
as well as their transformations under conformal boosts, take the
following form, 
\begin{align}
    \label{eq:delta-omega-phi-f}
  \delta\omega_\mu{}^{ab} =&\,
  -\tfrac12\mathrm{i}\,\bar{\epsilon}_i\gamma^5\gamma^{ab}\phi_\mu{}^i
  + \tfrac14\mathrm{i}\,\bar{\epsilon}_i\gamma^5\psi_\mu{}^i (T^{+ab} + T^{-ab}) +
  \tfrac34\,\bar{\epsilon}_i\gamma^5\gamma_\mu\gamma^{ab}\chi^i
  \nonumber\\  
  &\, +\bar{\epsilon}_i\gamma^5\gamma_\mu R(Q)^{ab\,i} +
  \tfrac12\mathrm{i}\,\bar{\eta}_i\gamma^5\gamma^{ab}\psi_\mu{}^i +
  2\,\Lambda_K{}^{[a}e_\mu{}^{b]} \, ,   \nonumber\\[1mm] 
  \delta\phi_\mu{}^i =&\, 2\,\mathcal{D}_\mu \eta^i + 2\mathrm{i}
  \,f_\mu{}^a\gamma_a\epsilon^i 
  +\tfrac1{16}\Slash{D}  (T_{ab}{\!}^++T_{ab}{\!}^-) \gamma^{ab}\gamma_\mu 
 \,\epsilon^i \nonumber\\
  &\, -\tfrac14\mathrm{i}\gamma^{ab}\gamma_\mu
  R(\mathcal{V})_{ab}{\!}^i{}_j \,\epsilon^j
  - \tfrac12\mathrm{i}\gamma^{ab}\gamma_\mu \gamma^5 
  R(A)_{ab}\,\epsilon^i \nonumber\\
  &\,
  -\tfrac32 \mathrm{i} [ ( \bar{\chi}_j \gamma^5 \gamma^a \epsilon^j )
  \gamma_a \psi_\mu{}^i - ( \bar{\chi}_j \gamma^5 \gamma^a
  \psi_\mu{}^j ) \gamma_a \epsilon^i ] 
  -\mathrm{i}
  \Lambda_\mathrm{K}{\!}^a\,\gamma_a   \psi_\mu{\!}^i  \, ,  \nonumber\\[1mm]
  \delta f_\mu{\!}^a=&\,
  \tfrac14\mathrm{i}\,\bar\epsilon_i\,\gamma^5\psi_\mu{\!}^i D_b (T^{+ba} + T^{-ba})
  -   \tfrac34 e_\mu{\!}^a\, \bar\epsilon_i \gamma^5\Slash{D}\chi^i -\tfrac34
  \bar\epsilon_i \,\gamma^5\gamma^a \psi_\mu{\!}^i \,D \nonumber\\
   &\, +\bar\epsilon_i\,\gamma^5 \gamma_\mu D_b R(Q)^{ba\,i} 
   + \tfrac12\,\bar\eta_i \,\gamma^5\gamma^a\phi_\mu{\!}^i 
   + \mathcal{D}_\mu\Lambda_\mathrm{K}{\!}^a \, . 
\end{align}
A systematic way to derive these variations is by noting that
the composite gauge fields follow from imposing three conventional
constraints (whose solutions are given by \eqref{eq:dependent-gf}), 
\begin{align}
  \label{eq:conv-constraints}
  &R(P)_{\mu \nu}{}^a =  0 \, , \nonumber \\[1mm]
  &\gamma^\mu R(Q)_{\mu \nu}{}^i + \tfrac32 \gamma_{\nu}
  \chi^i = 0 \, , \nonumber \\[1mm]
  &
  e^{\nu}{\!}_b \,R(M)_{\mu \nu}{\!}^{ab} - \widetilde{R}(A)_{\mu}{\!}^a 
  + \tfrac1{16} T_{\mu{b}}{\!}^-  \,T^{ab+}  
  -\tfrac{3}{2}\, D \, e_\mu{\!}^a = 0 \, .
\end{align}
In principle these constraints are not invariant under the original
symmetries associated with the $\mathrm{SU}^\ast(4|2)$
superalgebra. However, they are invariant under all symmetries except
Q-supersymmetry. Therefore they will only induce changes in the
Q-supersymmetry transformations and in the supersymmetry commutators
that involve the Q-supersymmetry generators. This phenomenon is well
known from the $4D$ Minkowskian conformal supergravities.

For completeness we present the definitions of all the supercovariant
curvature tensors,
\begin{align}
  \label{eq:curvatures}
  R(P)_{\mu \nu}{}^a  =&\; 2 \, \mathcal{D}_{[\mu} \, e_{\nu]}{}^a
  - \tfrac1{2}\,\bar{\psi}_{i[\mu}\gamma^5\gamma^a\psi_{\nu]}{}^i \, ,
  \nonumber\\[.2ex] 
  R(Q)_{\mu \nu}{}^i = & \; 2 \, \mathcal{D}_{[\mu} \psi_{\nu]}{}^i -
  \mathrm{i}\,\gamma_{[\mu} \phi_{\nu]}{}^i +
  \tfrac{1}{16}\mathrm{i}\,(T_{ab}^+ + T_{ab}^-) \,
  \gamma^{ab}\gamma_{[\mu} \psi_{\nu]}{}^i \, , \nonumber\\[.2ex] 
  R(D)_{\mu \nu} = & \;2\,\partial_{[\mu} b_{\nu]} - 2\,f_{[\mu}{}^a
  e_{\nu]a} -
  \tfrac{1}{2}\mathrm{i}\,\bar{\psi}_{i[\mu}\gamma^5\phi_{\nu]}{}^i +
  \tfrac{3}{4}\,\bar{\psi}_{i[\mu}\gamma^5\gamma_{\nu]} \chi^i \, ,
  \nonumber\\[.2ex] 
  R(A)_{\mu \nu} = &\; 2 \, \partial_{[\mu} A_{\nu]} +
  \tfrac12\mathrm{i}\,\bar{\psi}_{i[\mu}{}\phi_{\nu]}{}^i +
  \tfrac34\,\bar{\psi}_{i[\mu}\gamma_{\nu]}\chi^i \, ,
  \nonumber\\[.2ex] 
  R(\mathcal{V})_{\mu \nu}{}^i{}_j =& \;
  2\, \partial_{[\mu}\mathcal{V}_{\nu]}{}^i{}_j +
  \mathcal{V}_{[\mu}{}^i{}_k \, \mathcal{V}_{\nu]}{}^k{}_j  \nonumber \\ 
  &\,-  2\mathrm{i}\,\bar{\psi}_{j[\mu}\gamma^5\phi_{\nu]}{}^i +
  3\,\bar{\psi}_{j[\mu}\gamma^5\gamma_{\nu]} \chi^i +
  \tfrac12\delta^i{}_j\bigl(2\mathrm{i}\,\bar{\psi}_{k[\mu}\gamma^5\phi_{\nu]}{}^k
  -3\,\bar{\psi}_{k[\mu}\gamma^5\gamma_{\nu]} \chi^k \bigr) \, ,
  \nonumber \\[.2ex] 
  R(M)_{\mu \nu}{\!}^{ab} =& \; 2 \,\partial_{[\mu} \omega_{\nu]}{}^{ab}
  - 2\, \omega_{[\mu}{}^{ac}\omega_{\nu]c}{}^b - 4 f_{[\mu}{}^{[a}
  e_{\nu]}{}^{b]} +
  \tfrac12\mathrm{i}\,\bar{\psi}_{i[\mu}\gamma^5\phi_{\nu]}{}^i
  \nonumber\\
  &\,
  -\tfrac18\mathrm{i}\,\bar{\psi}_{\mu\,i}\gamma^5\psi_\nu{}^i\,(T^{ab+}
  + T^{ab-}) -
  \tfrac34\bar{\psi}_{i[\mu}\gamma^5\gamma_{\nu]}\gamma^{ab}\chi^i -
  \bar{\psi}_{i[\mu}\gamma^5\gamma_{\nu]}R(Q)^{ab\,i} \, ,
  \nonumber\\[.2ex] 
  R(S)_{\mu\nu}{\!}^i  =& \; 2 \,\mathcal{D}_{[\mu}\phi_{\nu]}{}^i +
  2\mathrm{i}\,f_{[\mu}{}^a\gamma_a \psi_{\nu]}{}^i +
  \tfrac1{16}\mathrm{i}\,\Slash{D}(T^+_{ab} + T^-_{ab}) \gamma^{ab}
  \gamma_{[\mu}\psi_{\nu]}{}^i \nonumber\\
  &\; +\tfrac32
  \mathrm{i}\,\gamma_a\psi_{[\mu}{}^i\bar{\psi}_{\nu]j} 
  \gamma^5\gamma^a\chi^j -
  \tfrac14\mathrm{i}\,R(\mathcal{V})_{ab}{}^i{}_j\gamma^{ab}
  \gamma_{[\mu}\psi_{\nu]}{}^j   -
  \tfrac12\mathrm{i}\,R(A)_{ab}\gamma^5\gamma^{ab}
  \gamma_{[\mu}\psi_{\nu]}{}^i
  \, , \nonumber\\[.2ex] 
  R(K)_{\mu\nu}{\!}^a =&\; 2\,\mathcal{D}_{[\mu}f_{\nu]}{}^a 
  - \tfrac14\,\bar{\phi}_{i[\mu}\gamma^5\gamma^a\phi_{\nu]}{}^i
  \nonumber\\
  &\:   -\tfrac18\Bigl[\mathrm{i}\,
  \bar{\psi}_{i[\mu}\gamma^5D_b(T^{+ba}+T^{-ba})\psi_{\nu]}{}^i  
  + 6\,e_{[\mu}{}^a\bar{\psi}_{\nu]i}\gamma^5\Slash{D}\chi^i
  \nonumber\\
  &\;\qquad - 3\,D\,\bar{\psi}_{i[\mu}\gamma^5\gamma^a\psi_{\nu]}{}^i 
  +  8 \,\bar{\psi}_{i[\mu}\gamma^5\gamma_{\nu]}D_b R(Q)^{ba\,i} \Bigr] \,. 
\end{align}
It is also convenient to introduce two modified curvatures by including 
suitable covariant terms,
\begin{align}
  \label{eq:modified-R}
  \mathcal{R}(M)_{ab}{}^{cd} =&\; R(M)_{ab}{}^{cd} 
  + \tfrac1{32}\bigl(T_{ab}^-\,T^{+cd} + T_{ab}^+\,T^{-cd}\bigr) \, ,
  \nonumber\\ 
  \mathcal{R}(S)_{ab}{}^i{}_\pm =&\; R(S)_{ab}{}^i{}_\pm 
  - \tfrac38\,T_{ab}^\pm\,\chi^i_\pm \, ,
\end{align}
where we note that $\gamma^{ab}\bigl(\mathcal{R}(S) - R(S)\bigr)_{ab}{}^i = 0$. 
By making use of the conventional constraints and the Bianchi identities 
of the superconformal algebra, one can show that the modified curvature 
$\mathcal{R}(M)_{ab}{}^{cd}$ satisfies the following relations:
\begin{align}
  \label{eq:curvature-rel}
  \mathcal{R}(M)_{\mu\nu}{}^{ab}\,e^\nu{}_b =&\;
  \widetilde{R}(A)_\mu{}^a + \tfrac32\,D\,e_\mu{}^a \, , \nonumber \\ 
\tfrac14\,\varepsilon_{ab}{}^{ef}\,\varepsilon^{cd}{}_{gh}\,\mathcal{R}(M)_{ef}{}^{gh} =&\; \mathcal{R}(M)_{ab}{}^{cd} \, , \nonumber \\
\varepsilon_{cdea}\,\mathcal{R}(M)^{cd\,e}{}_b =&\; \varepsilon_{becd}\,\mathcal{R}(M)_a{}^{e\,cd} = 2\,\widetilde{R}(D)_{ab} = 2\,R(A)_{ab} \, .
\end{align}
Note that this modified curvature does not satisfy the pair exchange property,
\begin{equation}
  \label{eq:lack-of-pair=exchange}
\mathcal{R}(M)_{ab}{}^{cd} - \mathcal{R}(M)^{cd}{}_{ab} = 
4\,\delta_{[a}^{[c}\,\widetilde{R}(A)_{b]}{}^{d]} \, .
\end{equation}
The fermionic conventional constraint implies that the chiral and anti-chiral 
projections of the $R(Q)_{ab}{}^i$ curvature are anti-self-dual and 
self-dual, respectively:
\begin{equation}
  \label{chiral-dual-RQ}
  \gamma^5 R(Q)_{ab}{}^i = - \widetilde{R}(Q)_{ab}{}^i \, .
\end{equation}
In addition, combining the constraint with the Bianchi identities yields
\begin{equation}
  \label{eq:R-S-curv-rel}
  \gamma^a\,\widetilde{\mathcal{R}}(S)_{ab}{}^i{}_\pm 
  = 2\mathrm{i}\,D^a R(Q)_{ab}{}^i{}_\mp \, .
\end{equation}

Finally we present the commutation relations of the superconformal
algebra. The first one is the commutator between two Q-supersymmetry
variations,
\begin{align}
  \label{eq:QQ}
  \big[\delta_\mathrm{Q}(\epsilon_1), \,
  \delta_\mathrm{Q}(\epsilon_2)\big] =&\; \xi^\mu D_\mu
  +\delta_\mathrm{M} (\varepsilon_{ab}) +
  \delta_\mathrm{S} ({\eta}^i)
  +\delta_\mathrm{K}({\Lambda}_\mathrm{K}{\!}^a) + \delta_\mathrm{gauge}\,,   
\end{align}
where the parameters of the various transformations appearing on the
r.h.s. are given by
\begin{align}
  \label{eq:parameters-QQ-commutators}
  \xi^\mu =&\, 2\,\bar\epsilon_{2i}\,\gamma^5\gamma^\mu
  \epsilon_1{}^i \, , \nonumber \\
  \varepsilon_{ab} =&\,
  \tfrac12\mathrm{i}\,\bar{\epsilon}_{2i-}\,\epsilon_1{\!}^i{\!}_- \, T_{ab}^- -
  \tfrac12\mathrm{i}\,\bar{\epsilon}_{2i+}\,\epsilon_1{\!}^i{\!}_+ \, T^+_{ab}
  \,, \nonumber \\
  {\eta}^i =&\,
  -6\mathrm{i}\,\bar{\epsilon}_{[1j-} \,\epsilon_{2] }{}^{i}{\!}_-\,
  \chi^j{\!}_+  
 +6\mathrm{i}\,\bar{\epsilon}_{[1j+} \,\epsilon_{2] }{}^{i}{\!}_+\,
  \chi^j{\!}_- \, ,
  \nonumber \\
  {\Lambda}_\mathrm{K}{\!}^a =&\,
  -\tfrac12\mathrm{i}\,\bar{\epsilon}_{2i-}\, \epsilon_1{\!}^i{\!}_- \,D_b T^{-ba} +
  \tfrac12\mathrm{i}\,\bar{\epsilon}_{2i+}\,\epsilon_1{\!}^i{\!}_+\, D_b T^{+ba} +
  \tfrac32\,\bar{\epsilon}_{2i}\gamma^a\gamma^5\epsilon_1{}^i\, D \, .
\end{align}

We also present the commutator of a Q-supersymmetry variation with
either an S-supersymmetry variation or a conformal boost variation,
\begin{align}
  \label{eq:QS+QK}
   \big[\delta_\mathrm{Q}(\epsilon), \, \delta_\mathrm{S}(\eta)\big] =&\;
  \delta_\mathrm{M}(\mathrm{i}\,\bar\epsilon_i\,\gamma^5\gamma^{ab} \eta^i ) 
  +\delta_\mathrm{D}(-\mathrm{i}\,\bar\epsilon_i\,\gamma^5\eta^i)
  +\delta_{\mathrm{SO}(1,1)} (\mathrm{i}\,\bar\epsilon_i
  \,\eta^i)    \nonumber\\
  &\,+  \delta_{\mathrm{SU}(2)}(2\mathrm{i}\,\bar\epsilon_j \,\gamma^5\eta^i
  -\delta_j{\!}^i \,\mathrm{i}\,\bar\epsilon_k
  \,\gamma^5\eta^k)\,, \nonumber\\[2mm] 
  \big[\delta_\mathrm{Q}(\epsilon), \, \delta_\mathrm{K}(\Lambda_K)\big] =&\;
  \delta_\mathrm{S}(-\mathrm{i} \Lambda_\mathrm{K}{\!}^a \gamma_a
  \epsilon^i) \,. 
\end{align}
In the presence of matter multiplets that contain gauge fields, such
as the vector and the tensor multiplets, the gauge
transformations of the latter will also appear in the commutator of two
Q-supersymmetry transformations. We will indicate these contributions
when discussing the vector and tensor multiplets in section \ref{sec:matter-multiplets}.

\section{Chiral and anti-chiral multiplets}
\label{sec:chiral-multipletl}
\setcounter{equation}{0}
Chiral multiplets are defined by the condition that the scalar field
of lowest Weyl weight $w$ transforms into a chiral spinor field. The
scalar also carries a chiral weight $c$ with respect to
$\mathrm{SO}(1,1)$ transformations, which must be equal to
$c=-w$. Likewise one may also define an anti-chiral multiplet where
the scalar field transforms into an anti-chiral spinor. In that case
one must have $c=w$. To explain the reason for this restriction on the
chiral weights one parametrizes the possible Q- and
S-supersymmetry transformations on the first two fields, denoted by
$A_\pm$ and $\Psi^i{\!}_\pm$, where the plus (minus) index refers to the
chiral  (anti-chiral) multiplet (and thus determines the chirality of the
spinor),
\begin{align}
  \label{eq:first-two-terms-achiral}
    \delta A_\pm =&\, \pm\mathrm{i} \bar\epsilon_{i\pm} \Psi^i{\!}_\pm\,, \nonumber\\
  \delta \Psi^i{\!}_\pm  =&\,- 2\mathrm{i}\,\Slash{D} A_\pm
  \epsilon^{i}{\!}_\mp  + \varepsilon_{jk}\, B^{ij}{\!}_\pm\, \epsilon^k{\!}_\pm -
  \tfrac12  F_{ab}{\!}^\mp \, \gamma^{ab}  \epsilon^{i}{\!}_\pm + 2w\,
  A_\pm \,\eta^{i}{\!}_\pm \,,
\end{align}
where we note the supercovariant derivative of $A_\pm$, 
\begin{equation}
  \label{eq:DA-pm}
  D_\mu A_\pm = (\partial_\mu -w\, b_\mu  - c_\pm \,A_\mu ) A_\pm  \mp
  \tfrac12\mathrm{i} \,\bar\psi_{\mu\,i\pm}\, \Psi^{i}{\!}_\pm\,. 
\end{equation}
Imposing the commutation relation \eqref{eq:QQ} on $A_\pm$ shows that
$\delta\Psi^{i}{\!}_\pm$ is correct, provided that $B^{ij}{\!}_\pm$ is
symmetric in $(ij)$. Furthermore the second commutation relation in 
\eqref{eq:QS+QK} shows that the S-supersymmetry variation of
$\Psi^i{\!}_\pm$ is correct, while the first one shows that the algebra
can only close provided that the chiral charges $c_\pm$ are equal to
$c_\pm= \mp w$, thus confirming the claim above.

%
\begin{table}[t]
\begin{center}
\begin{tabular*}{12.8cm}{@{\extracolsep{\fill}}|c||cccccc| }
\hline
 & & \multicolumn{4}{c}{Scalar (anti-)chiral multiplet} & \\  \hline \hline
 field & $A_\pm$ & $\Psi^i{\!}_\pm$ & $B^{ij}{\!}_\pm$ & $F_{ab}{\!}^\mp$&
 $\Lambda^{i}{\!}_\pm$ & $C_\pm$ \\[.5mm] \hline
$w$  & $w$ & $w+\tfrac12$ & $w+1$ & $w+1$ & $w+\tfrac32$ &$w+2$
\\[.5mm] \hline
$c$  & $\mp w$ & $\mp(w-\tfrac12)$ & $\mp(w-1)$ & $\mp(w-1)$ &
$\mp(w-\tfrac32) $ &$\mp(w-2)$
\\[.5mm] \hline
$\tilde\gamma_5$   & & $\pm$ & &  $\mp$ & $\pm$ & \\ \hline
\end{tabular*}
\vskip 2mm
\renewcommand{\baselinestretch}{1}
\parbox[c]{12.8cm}{\caption{\label{table:chiral}{\footnotesize Weyl
      and chiral weights ($w$ and $c$) and fermion chirality/two-form
      duality $(\tilde\gamma_5)$ of scalar (anti-)chiral multiplet
      component fields. Note that the chiral (anti-chiral) multiplet
      contains fermion components of positive (negative) chirality and
      Lorentz two-forms that are anti-self-dual (self-dual). All
      bosonic fields are (pseudo-)real.  }}}
\end{center}
\end{table}

It is clear that the (anti-)chiral multiplet contains $8\oplus 8$ bosonic
and fermionc degrees of freedom. Their Weyl and chiral weights are
straightforward to determine and are shown in  table
\ref{table:chiral}. Moreover the (anti-)chiral multiplets are real,
unlike in Minkowski space where they are complex and where the complex
conjugate of a chiral multiplet will constitute an anti-chiral
multiplet. Hence in the context of Euclidean supersymmetry the chiral
and anti-chiral fields should be treated as independent. 

The continuation of the above analysis straightforwardly leads to all the Q- and
S-supersymmetry transformations of the fields belonging to a chiral or
anti-chiral supermultiplet, 
\begin{align}
  \label{eq:conformal-chiral}
    \delta A_\pm =&\; \pm\mathrm{i} \bar\epsilon_{i\pm}
    \Psi^i{\!}_\pm\,, \nonumber\\[1mm]
  \delta \Psi^{i}{\!}_\pm  =&\;- 2\mathrm{i}\,\Slash{D} A_\pm\,
  \epsilon^{i}{\!}_\mp  + \varepsilon_{jk}\, B^{ij}{\!}_\pm\, \epsilon^{k}{\!}_\pm -
  \tfrac12  F_{ab}{\!}^\mp \, \gamma^{ab}  \epsilon^{i}{\!}_\pm + 2w\,
  A_\pm \,\eta^i{\!}_\pm \,, \nonumber\\[1mm]
   \delta B^{ij}{\!}_\pm =&\;\mp 2\,\bar\epsilon_{k\mp}\, \Slash{D} \Psi^{(i}{\!}_\pm
   \, \varepsilon^{j)k} \mp 2\mathrm{i}\,
  \bar\epsilon_{k\pm}  \Lambda^{(i}{\!}_\pm  \,\varepsilon^{j)k} \mp
  2\mathrm{i}(1-w)\,\bar\eta_k{\!}_\pm
  \Psi^{(i}{\!}_\pm\,\varepsilon^{j)k} \,, \nonumber\\[1mm]
  \delta F_{ab}{\!}^\mp =&\;\mp\tfrac12
  \bar\epsilon_{i\mp}\,\Slash{D}\gamma_{ab} \Psi^i{\!}_\pm \pm
  \tfrac12\mathrm{i}\, \bar\epsilon_{i\pm}\gamma_{ab}\Lambda^i{\!}_\pm 
  \pm \tfrac12\mathrm{i}(1+w)\, \bar\eta_{i\pm}\gamma_{ab}\, \Psi^i{\!}_\pm \,,
  \nonumber\\[1mm]
  \delta \Lambda^i{\!}_\pm =&\;\tfrac12\mathrm{i} 
  \gamma^{ab}\Slash{D}F_{ab}^\mp\,
   \epsilon^{i}{\!}_\mp  +\mathrm{i} \,\Slash{D}B^{ij}{\!}_\pm\,\varepsilon_{jk}\,
   \epsilon^k{\!}_\mp  + 
  C_\pm \,\epsilon^{i}{\!}_\pm \nonumber\\
  &\;   +\tfrac18\mathrm{i}\,\big(\Slash{D}A_\pm\,T_{ab}^\pm 
  +w\,A_\pm\,\Slash{D} T_{ab}^\pm  
  \big) \gamma^{ab} \,\epsilon^{i}{\!}_\mp 
  \pm\tfrac32\mathrm{i}\, \gamma_a\, \epsilon^{i}{\!}_\mp\; 
  \bar\chi_{j\mp} \,\gamma^a\,\Psi^j{\!}_\pm   \nonumber\\
  &\;  -(1+w)\,B^{ij}{\!}_\pm\,\varepsilon_{jk}\,\eta^k{\!}_\pm + 
  \tfrac12 (1-w)\,\gamma^{ab}\,
  F_{ab}^\mp \, \eta^{i}{\!}_\pm \,, \nonumber\\[1mm]
    \delta C_\pm =&\;\mp 2\,
    \bar\epsilon_{i\mp} \,\Slash{D}\Lambda^i{\!}_\pm 
    \pm 6\mathrm{i}\,\bar\epsilon_{i\mp} \,\chi^k{\!}_\mp \;
    \varepsilon_{kj} B^{ij}{\!}_\pm   \nonumber\\
    &\; \pm  \tfrac18 \big( (w-1)\,\bar{\epsilon}_{i\mp}\, \gamma^{ab}\,
    \Slash{D}T_{ab}^\pm \,\Psi^i_\pm + \bar{\epsilon}_{i\mp} \,\gamma^{ab}
    T_{ab}^\pm \,\Slash{D}\Psi^i{\!}_\pm \big) 
       \pm 2 \mathrm{i} w \,\bar{\eta}_{i \pm}\, \Lambda^i{\!}_\pm  \,.
\end{align}

At this point we should point out that there are three special values
for $w$. One is $w=1$ where the fields $B^{ij}{\!}_\pm$ and
$F_{ab}{\!}^\mp$ are both neutral under the $\mathrm{SO}(1,1)$
R-symmetry. In this situation it is possible to define a reduced
multiplet by first combining a chiral and an anti-chiral multiplet with
$w=1$, which together comprise $16\oplus 16$ bosonic and fermionic
fields. Subsequently one imposes a constraint which again reduces the
number of components to $8\oplus 8$. This constraint is only consistent with
supersymmetry provided $w=1$ and implies, for instance, that
$F_{ab}{\!}^\mp$ are subject to a Bianchi identity, which reads
$D_a(F^{ab+}-F^{ab-})=0$ modulo terms that depend non-linearly on the
fields . We will return to this in the next section when we discuss
the vector multiplet.

The second one is $w=2$. In that case the field $C_\pm$ has Weyl
weight equal to 4 and zero $\mathrm{SO}(1,1)$ weight. Both the
chiral and the anti-chiral multiplet with the proper weight will
define a superconformally invariant action. For this one makes use of
the following density formula,
\begin{align}
  \label{eq:chiral-density}
     \mathcal{L}_\pm =&\; e\Big[C_\pm \mp \bar\psi_{\mu i\mp}\, \gamma^\mu
   \Lambda^i{\!}_\pm \pm \tfrac1{16} \bar\psi_{\mu i\mp} \,
   T_{ab}{\!}^\pm \, \gamma^{ab} 
   \gamma^\mu  \Psi^i{\!}_{\pm}  + \tfrac1{16} A_\pm\, (T_{ab}{\!}^\pm)^2 
   \nonumber\\[1mm]
   &\; \quad \pm\tfrac12\mathrm{i} \, \bar\psi_{\mu i\mp} \, \gamma^{\mu\nu}\,
   \psi_{\nu}{\!}^j{\!}_\mp \,\varepsilon_{jk}   \, B^{ki}{\!}_\pm 
    \mp\mathrm{i}\,\bar\psi_{\mu i\mp} \,\psi_\nu{\!}^i{\!}_\mp\,
    (F^{\mu\nu\pm}  -\tfrac12 A_\pm \,T^{\mu\nu\,\pm}\big) \Big]
   \nonumber\\[1mm] 
   &\; \mp\tfrac12 \varepsilon^{\mu\nu\rho\sigma}\, \bar\psi_{\mu
     i\mp}\,\psi_\nu{\!}^i{\!}_\mp \,
   \big(\mathrm{i}\,\bar \psi_{\rho j\mp}\, \gamma_\sigma\Psi^j{\!}_\pm +
   \bar\psi_{\rho j\mp} \,\psi_\sigma{\!}^j{\!}_\mp \,  A_\pm\big) \,. 
\end{align}

Another special case is an (anti-)chiral multiplet with $w=0$. In that
case the highest component $C_\pm$ is invariant under
S-supersymmetry. Therefore $C_+$ itself can be regarded as the lowest
component of a new anti-chiral multiplet; likewise $C_-$ can be
regarded as the lowest component of a new chiral multiplet. Both these
new multiplets carry Weyl weight $w=2$. This multiplet is known as
the {\it kinetic chiral multiplet}. When applying the density formula
to this multiplet, the answer will be equal to a total derivative. On
the other hand, applying the density formula to the product of a $w=0$
(anti-)chiral multiplet times a kinetic (anti-)chiral multiplet, will
lead to a higher derivative action. We refer to
\cite{deWit:2010za,Butter:2013lta} for a more detailed discussion of
the kinetic multiplet and variants thereof in the context of the $4D$
Minkowski theory. We stress that the generic features of these
multiplets are the same as in the Euclidean theory. In the remainder
of this section we consider a number of additional noteworthy features
of chiral and anti-chiral supermultiplets.

\subsection{Product rule for Euclidean chiral multiplets}
\label{sec:chiral}

The product of two (anti-)chiral multiplets of weights $w_1$ and
$w_2$, with components denoted by
$(A_\pm,\,\Psi^i{\!}_\pm,\,B^{ij}{\!}_\pm,\,F_{ab}{\!}^\mp,\,\Lambda^i{\!}_\pm,\,C_\pm)$
and
$(a_\pm,\,\psi^i{\!}_\pm,\,b^{ij}{\!}_\pm,\,f_{ab}{\!}^\mp,\,\lambda^i{\!}_\pm,\,c_\pm)$,
yields an (anti-)chiral multiplet of weight $w=w_1 + w_2$ according to
the following product rule:
\begin{align}
\label{eq:chiral-prod-rule}
  (A_\pm,\,&\Psi^i{\!}_\pm,\,B^{ij}{\!}_\pm,\;F_{ab}{\!}^\mp,\,\Lambda^i{\!}_\pm,\,C_\pm) 
  \otimes (a_\pm,\,\psi^i{\!}_\pm,
  \,b^{ij}{\!}_\pm,\,f_{ab}{\!}^\mp,\,\lambda^i{\!}_\pm,\,c_\pm) =
  \nonumber \\ 
  &\,\Big(A_\pm \,a_\pm,\;A_\pm\,\psi^i{\!}_\pm + a_\pm\,\Psi^i{\!}_\pm,\;
  A_\pm\, b^{ij}{\!}_\pm + a_\pm \,B^{ij}{\!}_\pm 
  \mp\mathrm{i}\,\bar{\psi}_{k\pm}\,\varepsilon^{k(i}\Psi^{j)}{\!}_\pm, 
  \nonumber\\ 
  &\quad A_\pm \,f_{ab}{\!}^\mp + a_\pm\, F_{ab}{\!}^\mp \pm
  \tfrac14\mathrm{i}\,\bar{\psi}_{i\pm}\,\gamma_{ab}\Psi^i{\!}_\pm,\,
  \nonumber \\ 
  &\quad A_\pm\,\lambda^i{\!}_\pm + a_\pm\,\Lambda^i{\!}_\pm -
  \tfrac12\,\varepsilon_{jk}(B^{ij}{\!}_\pm\,\psi^k{\!}_\pm +
  b^{ij}{\!}_\pm\,\Psi^k{\!}_\pm) -
  \tfrac14(F_{ab}{\!}^\mp\gamma^{ab}\psi^i{\!}_\pm +
  f_{ab}{\!}^\mp\gamma^{ab}\Psi^i{\!}_\pm), \nonumber \\ 
  &\quad A_\pm \,c_\pm + a_\pm \,C_\pm -
  \tfrac12\,\varepsilon_{ik}\varepsilon_{jl}\,B^{ij}{\!}_\pm\,
  b^{kl}{\!}_\pm - f_{ab}{\!}^\mp\, F^{ab\mp}
  \pm\mathrm{i}\,\bar{\psi}_{i\pm}\Lambda^i{\!}_\pm
  \pm\mathrm{i}\,\bar{\Psi}_{i\pm}\lambda^i{\!}_\pm \Big) \, .
\end{align}
Note that the product of a chiral multiplet with an anti-chiral
multiplet does not yield a chiral or anti-chiral multiplet, because
$A_\pm a_\mp$ does not transform into an (anti-)chiral fermion under
supersymmetry, but into $a_\mp\Psi^i{\!}_\pm -
A_\pm\psi^i{\!}_\mp$. Therefore the result is a general scalar
supermultiplet which comprises $128\oplus 128$ bosonic and fermionic
components. 

A homogeneous function $\mathcal{G}^\pm(\Phi_\pm)$ of several (anti-)chiral
superfields $\Phi^\Lambda{\!}_\pm$ defines an \mbox{(anti-)chiral} superfield, whose
components take the following form,
\begin{align}
  \label{eq:chiral-function-comp}
  A_\pm\vert_{\mathcal{G}^\pm} =&\; \mathcal{G}^\pm(A_\pm) \, ,
  \nonumber \\[1mm]
  \Psi^i{\!}_\pm\vert_{\mathcal{G}^\pm} =&\,
  \mathcal{G}^\pm{\!}_\Lambda\,\Psi^{i}{\!}_\pm{\!}^\Lambda \, ,
  \nonumber \\[1mm] 
  B^{ij}{\!}_\pm\vert_{\mathcal{G}^\pm} =&\;
  \mathcal{G}^\pm{\!}_\Lambda\,B^{ij}{\!}_\pm{\!}^\Lambda \mp
  \tfrac12\mathrm{i}\,\mathcal{G}^\pm{\!}_{\Lambda\Sigma}\,
  \bar{\Psi}_{k\pm}{\!}^\Lambda\,\varepsilon^{k(i}\,\Psi^{j)}{\!}_\pm{\!}^\Sigma
  \, , \nonumber \\[1mm] 
  F_{ab}{\!}^\mp\vert_{\mathcal{G}^\pm} =&\; \mathcal{G}^\pm{\!}_\Lambda\,
  F_{ab}{\!}^{\mp\,\Lambda} \pm
  \tfrac18\mathrm{i}\,\mathcal{G}^\pm{\!}_{\Lambda\Sigma}\, 
  \bar{\Psi}_{i\pm}{\!}^\Lambda\gamma_{ab}\Psi^{i}{\!}_\pm{\!}^\Sigma
  \, , \nonumber \\[1mm] 
  \Lambda^i{\!}_\pm\vert_{\mathcal{G}^\pm} =&\; 
  \mathcal{G}^\pm{\!}_\Lambda \, \Lambda^{i}{\!}_\pm{\!}^\Lambda
  -  \tfrac12\,\mathcal{G}^\pm{\!}_{\Lambda\Sigma}\,
  [\varepsilon_{jk} \,B^{ij}{\!}_\pm{\!}^\Lambda \,\Psi^{k}{\!}_\pm{\!}^\Sigma +
  \tfrac12\,F_{ab}{\!}^{\mp\,\Lambda}\,\gamma^{ab}\Psi^i{\!}_\pm{\!}^\Sigma]
   \nonumber\\  
  &\, \mp
  \tfrac1{48}\mathrm{i}\,\mathcal{G}^\pm{\!}_{\Lambda\Sigma\Gamma}\,
  \gamma^{ab}\Psi^{i}{\!}_\pm{\!}^\Lambda\;
  \bar{\Psi}_{j\,\pm}{\!}^\Sigma\gamma_{ab}\Psi^{j}{\!}_\pm{\!}^\Gamma
  \, , \nonumber \\[1mm]  
  C_\pm\vert_{\mathcal{G}^\pm} =&\; \mathcal{G}^\pm{\!}_\Lambda\,
  C_\pm{\!}^\Lambda -
  \tfrac14\,\mathcal{G}^\pm{\!}_{\Lambda\Sigma}\,\big[\varepsilon_{ik}\varepsilon_{jl}
  \,B^{ij}{\!}_\pm{\!}^\Lambda \, B^{kl}{\!}_\pm{\!}^\Sigma +
  2\,F_{ab}{\!}^{\mp\,\Lambda} F^{ab \mp\,\Sigma} \pm
  4\mathrm{i}\,\bar{\Lambda}_i{\!}^\pm{\!}^\Lambda\,\Psi^i{\!} _\pm{\!}^\Sigma\big]
  \nonumber \\ 
  &\mp\tfrac14\mathrm{i}\,\mathcal{G}^\pm{\!}_{\Lambda\Sigma\Gamma}\,
  \big[\varepsilon_{jk}\,B^{ij}{\!}_\pm{\!}^\Lambda
  \,\bar{\Psi}_{i\pm}{\!}^\Sigma\, \Psi^k{\!}_\pm{\!}^\Gamma 
  + \tfrac12\,\bar{\Psi}_{i\,\pm}^\Lambda
  F_{ab}{\!}^{\mp\,\Sigma}\,\gamma^{ab}\,\Psi^i{\!}_\pm{\!}^\Gamma \big] \nonumber \\
  &\, \pm
  \tfrac1{192}\mathcal{G}^\pm{\!}_{\Lambda\Sigma\Gamma\Xi}\, 
  \bar{\Psi}_{i\pm}{\!}^\Lambda\gamma_{ab}\Psi^i{\!}_\pm{\!}^\Sigma \;
  \bar{\Psi}_{j\pm}{\!}^\Gamma\,\gamma^{ab}\,\Psi^j{\!}_\pm{\!}^\Xi  \, . 
\end{align}
Here $\mathcal{G}^\pm{\!}_{\Lambda\Sigma \cdots}$ denote multiple
derivatives of the function $\mathcal{G}^\pm$ with respect to the
$\Phi^\Lambda{\!}_\pm, \Phi^\Sigma{\!}_\pm,\ldots$

\subsection{The Weyl multiplet as a reduced chiral multiplet}
\label{sec:weyl-multiplet-as-chiral}
The Weyl multiplet is related to a sum of a chiral anti-self-dual
tensor multiplet and an anti-chiral self-dual multiplet. Because a
self-dual and an anti-self-dual tensor carry three components, both
these multiplets contain $24\oplus24$ degrees of freedom. The Weyl
multiplet is then obtained by imposing a constraint on this reducible
field representation that leads to a different supermultiplet of
$24\oplus 24$ degrees of freedom. From the transformation rules of the
Weyl multiplet it is easy to identify this structure. Namely, the
lowest components of the chiral and anti-chiral multiplets are the
fields $T_{ab}{\!}^\pm$ and the fermionic field strengths
$R(Q)_{ab}{\!}^{i\pm}$. As shown in \eqref{chiral-dual-RQ}, these
fermionic field strengths are (anti-)chiral and their chirality and
duality phases are opposite. The fact that the chiral and anti-chiral
components get more and more entangled for the higher components of
these multiplets, is an indication of the constraint that holds
between the chiral and the anti-chiral multiplet.

Because the Weyl multiplet is the only tensor (anti-)chiral multiplet that we
encounter, we will not exhibit the nature of its chiral multiplet
constraint in any detail. But the relation with the chiral multiplet
implies that one can construct a {\it scalar} chiral and
anti-chiral multiplet of weight $w=2$ by considering the square of the
Weyl multiplet. These multiplets can then easily be coupled to other
(anti-)chiral scalar multiplets. Therefore we present this square of
the Weyl multiplet, and list its explicit (anti-)chiral multiplet
components, 
\begin{align}
  \label{eq:weyl-squared-components}
  A_\pm\vert_{\mathrm{W}^2} =&\, \big(T_{ab}{\!}^\mp\big)^2 \, , \nonumber \\[1mm]
  \Psi^i{\!}_\pm\vert_{\mathrm{W}^2} =&\, -16\,T^{\mp\,ab}\,R(Q)_{ab}{\!}^i{\!}_\pm
  \, , \nonumber \\[1mm] 
  B^{ij}{\!}_\pm\vert_{\mathrm{W}^2} =&\, 16\,T^{\mp\,ab} \varepsilon^{k(i} \,
  R(\mathcal{V})_{ab}{}^{j)}{}_k \mp
  64\mathrm{i}\,\varepsilon^{k(i}\, \bar{R}(Q)_{ab\,k\,\pm}\,R(Q)^{ab\,j)}{\!}_\pm
  \, , \nonumber \\[1mm]
  F_{ab}{\!}^\mp\vert_{\mathrm{W}^2} =&\,
  -16\,\mathcal{R}(M)^{cd}{\!}_{ab}\,T _{cd}{\!}^{\mp} 
  \pm
  16\mathrm{i}\,\bar{R}(Q)_{cd\,i\,\pm}\,\gamma_{ab}\,R(Q)^{cd\,i}{\!}_\pm
  \, , \nonumber\\[1mm]
  \Lambda^i{\!}_\pm\vert_{\mathrm{W}^2} =&\, -32\,\gamma^{ab} R(Q)_{cd}{}^i{}_\pm
 \mathcal{R}(M)^{cd}{}_{ab} + 16\,\big(\mathcal{R}(S)_{ab}{}^i{}_\pm - 3\mathrm{i}\,\gamma_{[a}D_{b]}\,\chi^i_\mp
 \big)T^{\mp ab}  \nonumber\\
 &\, - 64\,R(\mathcal{V})_{ab}{}^i{}_j\,R(Q)^{ab\,j}{\!}_\pm \, , \nonumber\\[1mm]
  C_\pm\vert_{\mathrm{W}^2} =&\,
  -64\,\mathcal{R}(M)^{cd\mp}{\!}_{ab}\,\mathcal{R}(M)_{cd}{\!}^{\mp \,ab}
  - 32\,R(\mathcal{V})^{\mp\,ab\,i}{}_j
  \,R(\mathcal{V})_{ab}{\!}^{\mp j}{}_i 
  \nonumber\\
   &\, +16\, T^{\mp\,ab}D_a D^c T_{cb}{\!}^\pm 
    \mp
    128\mathrm{i}\,\bar{\mathcal{R}}(S)^{ab}{}_{i\pm}\,R(Q)_{ab}{}^i{}_\pm
    \mp 384\,\bar{R}(Q)^{ab}{}_{i\pm}\gamma_a D_b\chi^i{\!}_\mp \, .  
\end{align}
In view of its Weyl weight one can construct an invariant action that
is linearly proportional to the various components indicated above, by
making use of the density formula \eqref{eq:chiral-density}. This is
the action of conformal $N=2$ supergravity. It is worth noting that
the corresponding Lagrangians $\mathcal{L}_+$ and $\mathcal{L}_-$ will
differ by a total derivative.

\section{The vector and tensor multiplets and the hypermultiplet}
\label{sec:matter-multiplets}
\setcounter{equation}{0}
In this section we briefly discuss the three relevant matter
multiplets in $N=2$ supersymmetry. The vector and the tensor multiplets are
defined off-shell and each comprise $8\oplus 8$ off-shell
degrees of freedom. The hypermultiplet is only defined in an on-shell
version.  The field content of these multiplets is summarized in
table~\ref{table:w-weights-matter-4D} together with their Weyl and
chiral weights. The derivation follows again by dimensional reduction
of the $5D$ Minkowski theory, which is presented in appendix
\ref{sec:shell-dimens-reduct-matter}. Below we directly proceed to the
results for the $4D$ Euclidean supermultiplets.

\subsection{The vector supermultiplet}
\label{sec:vector-mult}
The $4D$ Euclidean vector supermultiplet involves two real scalar
fields $X_+$ and $X_-$, one symplectic Majorana spinor $\Omega^i$, the
gauge field $W_\mu$ and an auxiliary field $Y^{ij}$ obeying the
pseudo-reality condition $(Y^{ij})^* \equiv Y_{ij} =
\varepsilon_{ik}\varepsilon_{jl}\,Y^{kl}$. Here we present the
non-abelian version, where $g$ will denote the gauge coupling
constant. Therefore we write all the vector multiplet components as matrices
that take their values in the Lie algebra of the gauge
group. The Q- and S-supersymmetry transformations of the $4D$
Euclidean supermultiplet components are as follows,
\begin{align}
  \label{eq:vector-4D}
  \delta X_\pm=&\, \pm\mathrm{i}\,\bar\epsilon_{i\pm}\,\Omega^i{\!}_\pm
  \, , \nonumber\\[1mm] 
  \delta W_\mu=&\, \bar\epsilon_{i+} \big(\gamma_\mu \Omega^i{\!}_- - 
  2\mathrm{i}\,X_-\psi_\mu{\!}^i{\!}_+ \big) - \bar\epsilon_{i-}
  \big(\gamma_\mu \Omega^i{\!}_+ - 2\mathrm{i}\,X_+\psi_\mu{\!}^i
  {\!}_-\big)
  \nonumber\\[1mm]
  \delta \Omega^i{\!}_\pm =&\, -2\mathrm{i}\,\Slash{D} X_\pm\,
  \epsilon^i{\!}_\mp  -\tfrac12[\hat F(W)_{ab}{\!}^\mp -\tfrac14 X_\mp\,
  T_{ab}{\!}^\mp]\gamma^{ab} \epsilon^i{\!}_\pm  - \varepsilon_{kj}\, 
  Y^{ik}\epsilon^j{\!}_\pm \nonumber\\
  &\, + 2g[X_\pm,\,X_\mp]\epsilon^i{\!}_\pm + 2\,X_\pm\,\eta^i{\!}_\pm
  \, ,  \nonumber   \\[1mm] 
  \delta Y^{ij} =&\,
  2\,\varepsilon^{k(i}\,
  \bar{\epsilon}_{k+}\Slash{D}\Omega^{j)}{\!}_-  -
  2\,\varepsilon^{k(i}\,\bar{\epsilon}_{k-}
  \Slash{D}\Omega^{j)}{\!}_+\nonumber \\ 
  &\, -
  4\mathrm{i}g\,\varepsilon^{k(i}\,\bar{\epsilon}_{k+}
  [X_-,\,\Omega^{j)}_+]   +
  4\mathrm{i}g\,\varepsilon^{k(i}\,\bar{\epsilon}_{k-}
  [X_+,\,\Omega^{j)}_-]      \, . 
\end{align}
The supercovariant field strength $\hat F(W)_{\mu\nu}$ equals the
non-abelian version of the expression \eqref{eq:W-field-strength}, 
\begin{align}
  \hat{F}(W)_{\mu\nu} =&\,  2\,\partial_{[\mu} W_{\nu]} - g[W_\mu,\,W_\nu] +
  \bar{\psi}_{i[\mu}\gamma_{\nu]}\Omega^i{\!}_+ -
  \bar{\psi}_{i[\mu}\gamma_{\nu]}\Omega^i{\!}_- \nonumber\\
  &\, + \mathrm{i}\,X_-
  \bar{\psi}_{\mu\,i}\,\psi_\nu{\!}^i{\!}_+ -
  \mathrm{i}\,X_+\bar{\psi}_{\mu\,i}\,\psi_\nu{\!}^i{\!}_- \, . 
\end{align}
For convenience, we also note the following result for the superconformally invariant field strength,
\begin{equation}
  \label{eq:variation-vector-fieldstrength}
  \delta\bigl[\hat{F}(W)_{\mu\nu} - \tfrac14X_+\,T_{\mu\nu}{\!}^+ - 
  \tfrac14 X_-\,T_{\mu\nu}{\!}^-\bigr] =  
  -2\,\bar{\epsilon}_i\gamma^5\gamma_{[\mu}D_{\nu]}\Omega^i +
  \mathrm{i}\,\bar{\eta}_i\gamma^5\gamma_{\mu\nu}\Omega^i \,, 
\end{equation}
where on the right-hand side we refrained from writing the spinors in
terms of chiral components to make the expression shorter.

\begin{table}[t]
\begin{center}
\begin{tabular*}{12.5cm}{@{\extracolsep{\fill}}|c|cccc|cccc|cc| }
\hline
 & \multicolumn{4}{c|}{vector multiplet} & \multicolumn{4}{c|}{tensor multiplet} & 
 \multicolumn{2}{c|}{hypermultiplet} \\
 \hline \hline
 field & $X_\pm$ & $W_\mu$  & $\Omega^i{\!}_\pm$ & $Y^{ij}$& $L^{ij}$&$E_a$
 &$\varphi^i$ & $G_\pm$ & 
 $A_i{}^\alpha$ & $\zeta^\alpha{\!}_\pm$ \\[.5mm] \hline
$w$  & $1$ & $0$ & $\tfrac32$ & $2$ &$2$&$3$&$\tfrac52$ &$3$ &
 $1$ &$\tfrac32$
\\[.5mm] \hline
$c$  & $\mp1$ & $0$ & $\mp\tfrac12$ & $0$ &$0$&$0$& $\pm\tfrac12$
&$\pm1$ & $0$ &$\pm\tfrac12$ 
\\[.5mm] \hline
$\gamma_5$   & && $\pm$  &&&& $\pm$&   &  & $\pm$ \\ \hline
\end{tabular*}
\vskip 2mm
\renewcommand{\baselinestretch}{1}
\parbox[c]{12.5cm}{\caption{\footnotesize 
      Weyl weights $w$, chiral $\mathrm{SO}(1,1)$ weights $c$, 
      chirality $\gamma_5$ of the spinors for the $4D$ Euclidean
      vector multiplet, the tensor multiplet and the
      hypermultiplet. \label{table:w-weights-matter-4D}  }} 
\end{center}
\end{table}

The transformation rules \eqref{eq:vector-4D} close according to the
superconformal algebra indicated in the equations \eqref{eq:QQ},
\eqref{eq:parameters-QQ-commutators} and \eqref{eq:QS+QK}, where the
gauge transformation $\delta_\mathrm{gauge}$ in \eqref{eq:QQ} is
associated with the vector multiplet, whose field-dependent parameter
$\theta$ equals
\begin{equation}
\label{eq:gauge-vector-QQ}
  \theta  = 4\mathrm{i}\,\bar{\epsilon}_{2i-}\epsilon_1{}^i{\!}_- X_+ - 
  4\mathrm{i}\,\bar{\epsilon}_{2i+}\epsilon_1{}^i{\!}_+ X_- \, ,
\end{equation}
where $X_\pm$ are the two real scalar fields of the vector
multiplet. For every independent vector multiplet there will be a
corresponding gauge transformation.

\subsection{The tensor supermultiplet}
\label{sec:tensor-mult}
The $4D$ Euclidean tensor supermultiplet comprises a triplet of scalar
fields $L_{ij} = \varepsilon_{ik} \varepsilon_{jl}\, L^{kl}$, a
symplectic Majorana spinor $\varphi^i$, an anti-symmetric tensor gauge
field $E_{\mu\nu}$ and two (auxiliary) scalars $G_\pm$. Their Q- and
S-supersymmetry transformations are
\begin{align}
  \label{eq:tensor-4D}
  \delta L^{ij} =&\, 2\mathrm{i} \, \bar\epsilon_{k+}\,
  \varphi^{(i}{\!}_+ \,\varepsilon^{j)k} - 2\mathrm{i} \, \bar\epsilon_{k-}\,
  \varphi^{(i}{\!}_- \,\varepsilon^{j)k} \,,\nonumber\\
  \delta E_{\mu\nu}  =&\, \mathrm{i}\,\bar\epsilon_{i+}\,\gamma_{\mu\nu}
  \,\varphi_+^i + \mathrm{i}\,\bar\epsilon_{i-}\,\gamma_{\mu\nu}
  \,\varphi_-^i + 2\,\varepsilon_{jk}\,L^{ij}\big(\bar{\epsilon}_{i+}\,
  \gamma_{[\mu}\,\psi_{\nu]-}{\!}^k + \bar{\epsilon}_{i-}\,
  \gamma_{[\mu}\,\psi_{\nu]+}{\!}^k\big)\,, \nonumber\\
  \delta\varphi^i{\!}_\pm =&\, -\mathrm{i}\,\varepsilon_{jk} \Slash{D}L^{ij}
  \,\epsilon^k{\!}_\mp - \mathrm{i}\,\hat{\Slash{E}}\,\epsilon^i{\!}_\mp + 
  G_\pm \,\epsilon^i{\!}_\pm + 2\, L^{ij} \varepsilon_{jk} \,
  \eta^k{\!}_\pm \,, \nonumber\\
  \delta G_\pm =&\, \mp2\,\bar{\epsilon}_{i\mp}\Slash{D}\varphi^i{\!}_\pm
  \pm 6\mathrm{i}\,\varepsilon_{jk}L^{ij}\,\bar{\epsilon}_{i\mp}\chi^k{\!}_\mp
  \mp\tfrac18\mathrm{i}\,T_{ab}^\pm\,\bar{\epsilon}_{i\mp}\gamma^{ab}\varphi^i{\!}_\mp
  \pm2\mathrm{i}\,\bar{\eta}_{i\pm}\varphi^i{\!}_\pm \, .
\end{align}
where $\hat{E}^\mu$ denotes the supercovariant field strength
associated with $E_{\mu\nu}$,
\begin{equation}
  \label{eq:tensor-field-strength}
  \hat{E}^\mu = \tfrac12\,e^{-1}\varepsilon^{\mu\nu\rho\sigma}\big[\partial_\nu 
  E_{\rho\sigma} - \tfrac12\mathrm{i}\,\bar{\psi}_{\nu i}\gamma_{\rho\sigma}\varphi^i
  -\tfrac12\,\varepsilon_{jk}L^{ij}\,\bar{\psi}_{\nu i}\gamma_\rho\psi_\sigma{}^k\big] \, .
\end{equation}
This vector defines the fully supersymmetric field strength of the tensor
field, which is subject to a Bianchi identity that involves $D_\mu \hat E^\mu$. 

The above transformation rules close according to the superconformal
algebra, and just as in the case of the vector multiplet, there is a new
contribution to the gauge transformation $\delta_\mathrm{gauge}$ in
\eqref{eq:QQ} related to a tensor gauge transformation $\delta
E_{\mu\nu} = 2\,\partial_{[\mu} \theta_{\nu]}$  with parameter, 
\begin{equation}
\label{eq:gauge-tensor-QQ}
  \theta_\mu = 2\,\bar{\epsilon}_{2i} \gamma_\mu\epsilon_1{}^k
  \,\varepsilon_{kj} \, L^{ij} \,,
\end{equation}
where $L^{ij}$ is the scalar triplet field of the tensor multiplet. For every
independent tensor multiplet there will be such a corresponding tensor
gauge transformation.

There is an intriguing relationship between the abelian vector
multiplet and the tensor multiplet. Namely, the vector multiplet field
$Y^{ij}$ transforms identically as the scalar $L^{ij}$ of the tensor
field: they are both $\mathrm{SU}(2)$ triplets, their Weyl and chiral
weights are identical and they are invariant under
S-supersymmetry. Furthermore, under Q-supersymmetry they transform into
a spinor that transforms as an $\mathrm{SU}(2)$ doublet
$\varphi^i{\!}_\pm$ and $\mathrm{i}\,\Slash{D} \Omega^i{\!}_\mp$,
respectively. This shows that $Y^{ij}$ transforms as the lowest
component of a tensor multiplet. This is analogous to the situation
that we encountered for a chiral supermultiplet of weight $w=0$, whose
highest component transforms as a $w=2$ anti-chiral supermultiplet.

A related feature is that there exists a supersymmetric density for
the product of a vector with a tensor multiplet, 
\begin{align}
  \label{eq:vector-tensor-coupling}
  e^{-1} \mathcal{L}_\mathrm{VT} =&\,  X_+\,G_+ + X_-\,G_-  + \tfrac12
  Y^{ij} L_{ij} +\tfrac14 e^{-1} \varepsilon^{\mu\nu\rho\sigma} \,
  F_{\mu\nu}\,E_{\rho\sigma}  \nonumber\\
  &\, 
  +\big(\mathrm{i}\,\bar\Omega_{i+}  + X_+\,\bar\psi_{\mu i -}\gamma^\mu\big)\,  
  \varphi^i{\!}_+  
  -(\mathrm{i}\,\bar\Omega_{i-}  + X_-\,\bar\psi_{\mu i +} \gamma^\mu)\,
  \varphi^i{\!}_-   \nonumber\\
  &\,
  +\tfrac12 \big(\bar\Omega_{i+} \gamma^\mu \psi_\mu{\!}^k{\!}_-
  - \bar\Omega_{i-} \gamma^\mu \psi_\mu{\!}^k{\!}_+ \nonumber\\
  &\,\qquad 
  +\mathrm{i}\,X_+\,\bar\psi_{\mu i -} \gamma^{\mu\nu} \psi_\nu{\!}^k{\!}_-  
  - \mathrm{i}\,X_-\,\bar\psi_{\mu i +} \gamma^{\mu\nu} \psi_\nu{\!}^k{\!}_+\big)
  \,\varepsilon^{ij}  \, L_{jk}   \,. 
\end{align}
Observe that the term proportional to the vector field strength times
the tensor gauge field is invariant under tensor gauge transformations
up to a total derivative. This expression takes a similar form as in the 
Minkowski theory \cite{deWit:1980tn}.

We should point out that it is also possible to define a vector
multiplet in terms of the components of a tensor multiplet. However,
in that case one has to multiply by non-trivial functions of the
tensor multiplet scalars in order to bridge the gap in the conformal
dimensions. The resulting Lagrangians will then
contain higher-derivative couplings \cite{deWit:2006gn}. 

\subsection{Hypermultiplets}
\label{sec:hypers}
The supersymmetry transformations for the $4D$ Euclidean
hypermultiplets read as follows,
\begin{align}
  \label{eq:hyper-4D}
  \delta A_i{}^\alpha = &\, 2\mathrm{i}\,\bar\epsilon_{i+}
  \,\zeta^\alpha{\!}_+
  -2\mathrm{i}\,\bar{\epsilon}_{i-}\,\zeta^\alpha{\!}_- \,,\nonumber\\
  \delta \zeta^\alpha_\pm =&\, -\mathrm{i}\,\Slash{D} A_i{\!}^\alpha\,
  \epsilon^i{\!}_\mp - 2g\,X_\mp{\!}^\alpha{\!}_\beta\, A_i{\!}^\beta
  \epsilon^i{\!}_\pm + A_i{}^\alpha \eta^i{\!}_\pm \,.
\end{align}
Here, the $A_i{}^\alpha$ are the local sections of an
$\mathrm{Sp}(r)\times\mathrm{Sp}(1)$ bundle.  We note the existence of
a covariantly constant anti-symmetric tensor $\Omega_{\alpha\beta}$
(and its complex conjugate $\Omega^{\alpha\beta}$ satisfying the
reality condition $\Omega_{\alpha\gamma}\,\Omega^{\beta\gamma} =
\delta_\alpha{}^\beta$), which in principle depends on the
scalars. The symplectic Majorana condition for the spinors reads as
$C^{-1}\bar\zeta_\alpha{}^\mathrm{T} = \Omega_{\alpha\beta}
\,\zeta^\beta$. Covariant derivatives contain the $\mathrm{Sp}(r)$
connection $\Gamma_A{}^\alpha{\!}_\beta$, associated with
field-dependent $\mathrm{Sp}(r)$ rotations of the fermions, while the
$\mathrm{Sp}(1)$ connections are provided by the $\mathrm{SU}(2)$
R-symmetry gauge fields $\mathcal{V}_\mu{\!}^i{}_j$. The
sections $A_i{}^\alpha$ are pseudo-real, i.e. they are subject to the
constraint, $A_i{}^\alpha \varepsilon^{ij} \Omega_{\alpha\beta} =
A^j{}_\beta\equiv (A_j{}^\beta)^\ast$.  The target-space is a
hyper-K\"ahler cone whose metric is encoded in the so-called
hyper-K\"ahler potential. In \eqref{eq:hyper-4D} we wrote the
transformation rules for the case that the hyper-K\"ahler cone is
flat. The extension to non-trivial hyper-K\"ahler cone geometries is
straightforward and has been described in \cite{deWit:1999fp,deWit:2001brd}.
We have also included an optional coupling to a number of abelian or
non-abelian vector multiplets with a generic gauge coupling $g$. The
corresponding vector multiplets are then written as matrices which
take their values in the Lie algebra associated with the gauge
group. For instance, we have the vector multiplet scalars
$X_\pm{\!}^\alpha{\!}_\beta \equiv
X_\pm{\!}^I\,(t_I)^\alpha{\!}_\beta$, where the $t_I$ denote the gauge
group generators. Of course, this requires that the tensor
$\Omega_{\alpha\beta}$, as well as related quantities, will be
invariant under the gauge group. In addition the covariant derivatives
must be covariantized also with respect to the gauge group. Note that
the latter must be a subgroup of $\mathrm{Sp}(r)$.

It is possible to construct a tensor multiplet from the product of two
hypermultiplets \cite{deWit:1980tn}. This product is defined in terms
of a numerical  tensor $\eta_{\alpha\beta}$ satisfying
\begin{equation}
  \label{eq:tensor-hyper-square}
  \eta_{\alpha\beta} =
  \Omega_{\gamma\alpha}\,\eta^{\gamma\delta}\,\Omega_{\delta\beta} \,, 
\end{equation}
with $\eta^{\alpha\beta} \equiv (\eta_{\alpha\beta})^*$. The tensor
multiplet components are then given by the following combinations of
hypermultiplet components $(A_i{}^\alpha,\,\zeta^\alpha)$ and
$(A'_i{}^\alpha,\,{\zeta'}^\alpha)$:
\begin{align}
  \label{eq:tensormult-from-hypers}
  L^{ij} =&\,
  -\varepsilon^{k(i}\varepsilon^{j)l}\,A_k{}^\alpha\,A'_l{}^\beta\,\eta_{\alpha\beta}
  \, ,  \nonumber \\
  \varphi^i =&\, \varepsilon^{ij}\bigl(A_j{}^\alpha\,\zeta'^\beta 
  + A'_j{}^\beta\,\zeta^\alpha\bigr)\eta_{\alpha\beta} \, , \nonumber \\
  \hat{E}_\mu =&\, \tfrac12\,\varepsilon^{ij}\bigl(A_i{}^\alpha\,D_\mu
  A'_j{}^\beta - D_\mu
  A_i{}^\alpha\,A'_j{}^\beta\bigr)\eta_{\alpha\beta} +
  \Omega^{\alpha\gamma}\eta_{\alpha\beta}\,\bar{\zeta}_{\gamma
    -}\gamma_\mu\,{\zeta'}_+^\beta -
  \Omega^{\alpha\gamma}\eta_{\alpha\beta}\,\bar{\zeta}_{\gamma
    +}\gamma_\mu\,{\zeta'}_-^\beta \, , \nonumber \\
  G_\pm =&\,
  \mp2\mathrm{i}\,\Omega^{\alpha\gamma}\eta_{\alpha\beta}\,\bar{\zeta}_{\gamma
    \pm}{\zeta'}_\pm^\beta \, 
\end{align}
The hypermultiplet is only defined on shell, and this will reflect
itself as a constraint on the field $\hat E_\mu$, which can actually
be identified with the Bianchi identity for the tensor field that we
mentioned earlier. Note that in the above equations we have suppressed
the optional gauging for the underlying hypermultiplets.

Finally we present the supersymmetric 
Lagrangian density for hypermultiplets that transform under a
certain local gauge group,
\begin{align}
  \label{eq:hyper-lagrangian}
  e^{-1}\mathcal{L}_\mathrm{H} =&\;
  \tfrac12\,\varepsilon^{ij}\,\Omega_{\alpha\beta}\,A_i{}^\alpha(D^a
  D_a + \tfrac32\,D)
  A_j{}^\beta \nonumber \\
  &\; - (\bar{\zeta}_\alpha + \tfrac12\mathrm{i}\,\bar{\psi}_{\mu
    i}\gamma^\mu A^i{}_\alpha)\gamma^5 [\Slash{D}\zeta^\alpha -
  \mathrm{i}(\tfrac32\,\chi^k A_k{}^\alpha -
  \tfrac14\,T_{ab}\gamma^{ab}\zeta^\alpha)
  +\mathrm{i}\,g\,\Omega^{k\,\alpha}{}_\beta A_k{}^\beta] \nonumber \\
  &\;+2\mathrm{i}\,g\,\bar{\zeta}_{\alpha
    -}\,X_-{}^\alpha{}_\beta\,\zeta^\beta_- - 2\mathrm{i}\,g\,
  \bar{\zeta}_{\alpha+}\,X_+{}^\alpha{}_\beta\,\zeta^\beta_+ \nonumber \\
  &\;+\Omega_{\alpha\beta}(2\,g^2\,\varepsilon^{ij}A_i{}^\alpha
  X_-{}^\beta{}_\gamma X_+{}^\gamma{}_\delta\,A_j{}^\delta
  -\tfrac12\,g\,A_i{}^\alpha\,Y^{ij\,\beta}{}_\gamma\,A_j{}^\gamma) \nonumber \\
  &\;+\tfrac12\mathrm{i}\,g\,A^i{}_\alpha\,
  \bar{\Omega}_i{}^\alpha{}_\beta\,\gamma^5\,\zeta^\beta  \, ,
\end{align}
where the covariant derivatives contain the gauge fields associated
with the optional gauge invariance.

\subsection{The vector multiplet as a reduced chiral  multiplet}
\label{sec:vector-as-chiral}
The vector multiplet is related to the reducible combination of a chiral
and an anti-chiral scalar multiplet, just as the Weyl multiplet was
related to a chiral and an anti-chiral tensor multiplet. On this
reducible multiplet one imposes a supersymmetric constraint. This can
only be done provided the (anti-)chiral multiplets carry weight $w=1$. 
Since a scalar (anti-)chiral multiplet carries $8\oplus 8$ degrees of
freedom, one ends up again with $8\oplus 8$ degrees of freedom. 

Denoting the components of the (anti-)chiral multiplets by
$(A_\pm,\,\Psi^i{\!}_\pm,\,B^{ij}{\!}_\pm,\,F_{ab}^\mp,
\,\Lambda^i{\!}_\pm,C_\pm)\vert_V$, the supersymmetric constraint
identifies the fields $B^{ij}{\!}_+$ and $B^{ij}{\!}_-$, and it leads
to a Bianchi identity that involves $F_{ab}{\!}^+$ and $F_{ab}{\!}^-$,
modified by extra terms induced by supergravity; the higher
components $\Lambda^i{\!}_\pm$ and $C_\pm\vert_V$ are expressed in
terms of the lower components.

Conversely, this construction enables one to define a chiral and
an anti-chiral scalar multiplet in terms of the vector multiplet
components. This leads to the following result, 
\begin{align}
  \label{eq:vect-embedd-in chiral}
  A_\pm\vert_\mathrm{V} =&\, X_\pm \, , \nonumber \\
  \Psi^i{\!}_\pm\vert_\mathrm{V} =&\, \Omega^i{\!}_\pm \, , \nonumber \\
  B^{ij}{\!}_\pm\vert_\mathrm{V} =&\, -Y^{ij} \, , \nonumber\\
  F_{ab}{\!}^\mp\vert_\mathrm{V} =&\, \hat{F}_{ab}{\!}^\mp -
  \tfrac14\,X_\mp \,T_{ab}{\!}^\mp \, , \nonumber \\ 
  \Lambda^i{\!}_\pm\vert_\mathrm{V} =&\, \mathrm{i}\,\Slash{D}\Omega^i
  {\!}_\mp \, , \nonumber \\ 
  C_\pm\vert_\mathrm{V} =&\, 2\,\Box_c X_\mp + \tfrac14\,(\hat{F}_{ab}{\!}^\pm -
  \tfrac14\,X_\pm T_{ab}{\!}^\pm)\,T^{ab\,\pm} 
  \mp 3\mathrm{i}\,\bar{\chi}_{i\,\mp}\,\Omega^i{\!}_\mp \, , 
\end{align}
where $\Box_c \equiv D^a D_a$ is the superconformal d'Alembertian obtained
by the product of two fully superconformal derivatives. With these
identifications, the transformation rules \eqref{eq:vector-4D} induce
the proper transformations for the (anti-)chiral multiplets with
$w=1$. We will make use of this result in the next section.

\section{Locally supersymmetric actions of vector multiplets}
\label{sec:vector-multiplet-actions}
\setcounter{equation}{0}

Following the standard procedure that was originally used for locally
supersymmetric vector multiplet actions in Minkowski space
\cite{deWit:1984wbb,deWit:1984rvr}, on chooses two functions
$\mathcal{F}^\pm(X_\pm)$ depending on several fields $X_\pm{\!}^I$
which are the lowest components of a set of vector multiplets labeled
by an index $I$. According to \eqref{eq:vect-embedd-in chiral} the
chiral and the anti-chiral multiplets both involve all the vector
multiplet fields, albeit in mutually different ways. The functions
$\mathcal{F}^\pm(X_\pm)$ may also depend on some independent
(anti-)chiral multiplets which can serve as a background. For
simplicity we suppress such background fields for now and restrict
ourselves to vector multiplets only. The inclusion of background
fields dependence, crucial for the construction of higher-derivative
actions, will be presented later in this section.

In order to have local supersymmetry, the functions $\mathcal{F}^\pm$
must be homogeneous functions of degree two, so that
\begin{equation}
  \label{eq:homogenuity}
  \mathcal{F}^\pm(\lambda\, X_\pm{\!}^I) = \lambda^2\,
  \mathcal{F}^\pm(X_\pm{\!}^I)\,. 
\end{equation}
Using \eqref{eq:chiral-function-comp} and \eqref{eq:vect-embedd-in
  chiral} we determine the various components of the (anti-)chiral
multiplets defined by $\mathcal{F}^\pm(X_\pm{\!}^I)$ which have Weyl
weight $w=2$.  Subsequently one uses the invariant density formula
\eqref{eq:chiral-density} to obtain the Lagrangian,
\begin{equation}
  \label{eq:vector-Lagrangian}
  \mathcal{L}= \mathcal{L}_+ + \mathcal{L}_-\,, 
\end{equation}
which leads to the following bosonic terms, 
\begin{align}
  \label{eq:vector-lagrangian-pm}
  e^{-1} \mathcal{L}_\pm = &\, -\mathcal{F}^\pm{\!}_I \,\Box_c\,
  X_\mp{\!}^I - \tfrac18 \mathcal{F}^\pm{\!}_I\,\big(F_{ab}{\!}^{\pm
    I} - \tfrac14\,X_\pm{\!}^I T_{ab}{\!}^\pm\big)\,T^{ab\,\pm}
  \nonumber\\[1mm]
  &\, +\tfrac14\,\mathcal{F}^\pm{\!}_{IJ}\, \big(F_{ab}{\!}^{\mp\,I} -
  \tfrac14\,X_\mp{\!}^I\, T_{ab}{\!}^\mp\big) \,\big(F^{ab\,\mp\, J}
  - \tfrac14\,X_\mp{\!}^{J}\, T^{ab\,\mp}\big) \nonumber \\
  &\, +\tfrac18\,\mathcal{F}^\pm{\!}_{IJ}\,Y^{ij\,I}\, Y_{ij}{\!}^J
  -\tfrac1{32}\,\mathcal{F}^\pm\, (T_{ab}{\!}^\pm)^2 \,,
\end{align}
where $\mathcal{F}^\pm{\!}_{IJ\cdots}$ denotes derivatives of
$\mathcal{F}^\pm$ with respect to $X_\pm{\!}^I,
X_\pm{\!}^J,\ldots$. Here the bosonic terms of the superconformal
d'Alembertian $\Box_c$ acting on $X_\mp{\!}^I$ are equal to
\begin{equation}
  \label{eq:dallembert}
  \Box_c X_\mp{\!}^I = \big(\partial_\mu \mp A_\mu\big)^2
  X_\mp{\!}^I  + \big(\tfrac16\,R - D\big)\, X_\mp{\!}^I , 
\end{equation}
where we have used the bosonic part of $f_\mu{\!}^a$ defined in
\eqref{eq:dependent-gf}. Note that the dependence on the dilatational
gauge field $b_\mu$ has disappeared and $R$ denotes the standard Ricci
scalar. Upon dropping a total derivative the kinetic term for the
scalars can be written as
\begin{equation}
  \label{eq:scalar-kinetic}
  e^{-1} \mathcal{L}_\mathrm{kinetic}=  
  N_{IJ} \big[ (\partial_\mu +  A_\mu) X_+{\!}^I \,( \partial^\mu -
  A^\mu) X_-{\!}^J - \big(\tfrac16\,R - D\big)\, X_+{\!}^I \,X_-{\!}^J\,
  \big] \,,
\end{equation}
where 
\begin{equation}
  \label{eq:def-N}
  N_{IJ} = \mathcal{F}^+{\!}_{IJ} +  \mathcal{F}^-{\!}_{IJ} \,, 
\end{equation}
which is a symmetric tensor that can be written in a suggestive
way as the second derivative of $K(X_+,X_-)$, defined by
\begin{equation}
  \label{eq:p-kahler}
  N_{IJ} = \frac{\partial^2 K(X_+,X_-)}{\partial X_+{\!}^I\,\partial
    X_-{\!}^J}\,,
\end{equation}
where
\begin{equation}
  \label{eq:kahler-pot}
  K(X_+,X_-)= X_-{\!}^I\,\mathcal{F}^+{\!}_I(X_+) +
  X_+{\!}^I\,\mathcal{F}^-{\!}_I(X_-) \, , 
\end{equation}
These equations look rather similar to the ones that one encounters in
Minkowski space, except that $\mathcal{F}^+$ and $\mathcal{F}^-$ are
in principle two unrelated homogeneous functions of a number of
separate real variables. The quantity \eqref{eq:kahler-pot} has
already appeared in \cite{Cortes:2003zd}, where it was shown that it
defines a prepotential for the metric \eqref{eq:def-N} on a (affine)
special para-K\"{a}hler manifold.

Apart from the fact that we have two independent real functions
$\mathcal{F}^\pm(X_\pm)$ in the Euclidean theory, the structure
of these Lagrangians is very similar to the structure of the Minkowski
theory (see, e.g. \cite{deWit:1984wbb,deWit:1984rvr}). Therefore we
will not try to clarify various other terms in these
Lagrangians. We will also not consider the case of non-abelian vector
multiplet Lagrangians, which can be derived along the same lines. 

An interesting feature of these Lagrangians is that both the Euclidean
and the Minkowkian $4D$ gauge theories are subject to
electric-magnetic duality. Since there are subtle differences between
Euclidean and Minkowskian duality (for instance a selfdual tensor is
real in Euclidean space, while in Minowski space it is complex), it is
of interest to investigate electric-magnetic duality in the Euclidean
context. Hence let us start from a Lagrangian $\mathcal{L}(F)$
depending on $n$ abelian self-dual and anti-self-dual field strengths
$F_{\mu\nu}{\!}^{\pm I}$ (but not on their derivatives) and possibly on
other fields. The field equations are defined in terms of the tensors
\begin{equation}
  \label{eq:dual-G}
  G^{\mu\nu\pm}{\!}_I = \pm \frac{1}{e}\, \frac{\partial
    \mathcal{L}}{\partial F_{\mu\nu}{\!}^{\pm I}} \;.  
\end{equation}
For instance, the tensors $G^{\mu\nu\pm}{\!}_I$ for the model at
the beginning of this section read as follows, 
\begin{equation}
  \label{eq:example-G} 
  \pm G^{\mu\nu\pm}{\!}_I =\mathcal{F}^\mp{\!}_{IJ} \,
  F^{\mu\nu\,\pm J} -\tfrac14 N_{IJ} \,X_\pm{\!}^J \,
    T^{\mu\nu\,\pm} \,. 
\end{equation}
The corresponding Bianchi identities and equations of motion then take
a similar form, 
\begin{equation}
  \label{eq:bianch-eom}
 \partial^\mu\big( F_{\mu\nu}{\!}^{+I} - F_{\mu\nu}{\!}^{-I}\big)  = 0
 =  \partial^\mu\big( G_{\mu\nu}{\!}^+{\!}_I - G_{\mu\nu}{\!}^-{\!}_I
 \big) \,. 
\end{equation}
Obviously these equations are invariant under the electric-magnetic
duality transformations,
\begin{equation}
  \label{eq:emdual-rot}
  \begin{pmatrix} 
     F_{\mu\nu}{\!}^{\pm I}\\[2mm]
     G_{\mu\nu}{}^\pm{\!}_J 
    \end{pmatrix}
    \longrightarrow  
    \begin{pmatrix} 
     \tilde F_{\mu\nu}{\!}^{\pm I}\\[2mm]
    \tilde G_{\mu\nu}{}^\pm{\!}_J 
    \end{pmatrix}
    = 
  \begin{pmatrix}
   U^I{}_K&\ Z^{IL}\\[2mm]
    W_{JK}& V_J{}^L
   \end{pmatrix}
   \begin{pmatrix} 
     F_{\mu\nu}{\!}^{\pm K}\\[2mm]
     G_{\mu\nu}{}^\pm {\!}_L
    \end{pmatrix}\,,
\end{equation}
where $\tilde F_{\mu\nu}{\!}^{I\pm}$, and $\tilde G_{\mu\nu J}{}^\pm$
denote the transformed field strengths (and not the Hodge dual), and
the $n\times n$ submatrices $U^I{}_K$, $Z^{IL}$, $ W_{JK}$ and
$V_J{}^L$ are real. The relevant question is whether the rotated
tensors $\tilde G_{\mu\nu}{\!}^\pm{\!}_J$ can again follow from a new
Lagrangian $\tilde{\mathcal{L}} (\tilde F)$ in analogy with \eqref{eq:dual-G}. In
that case there may be a different Lagrangian that leads to an
equivalent set of Bianchi identities and equations of motion.  As it
turns out, this poses the following restriction on the matrices
in \eqref{eq:emdual-rot},
\begin{align}
  \label{eq:sympl-matrix}
  &U^\mathrm{T} V - W^\mathrm{T} Z = VU^\mathrm{T} - W Z^\mathrm{T} =
  \oneone\,, \nonumber\\
  & U^\mathrm{T} W = W^\mathrm{T} U\,,\qquad Z^\mathrm{T} V=
  V^\mathrm{T} Z\,. 
\end{align}
which are equivalent to
\begin{equation}
  \label{eq:sympl-prop-O}
  \begin{pmatrix} U&  Z\nonumber\\[1mm]  W &V\end{pmatrix} 
\begin{pmatrix} 0&-\oneone\\[1mm] \oneone &0 \end{pmatrix}  
  \begin{pmatrix} U&  Z\nonumber\\[1mm]  W
    &V\end{pmatrix}^{\!\!\mathrm{T}}  =  
\begin{pmatrix} 0&-\oneone\\[1mm] \oneone &0 \end{pmatrix}\\,
\end{equation}
Hence the electric-magnetic dualities form a group of equivalence
transformations that connect different Lagrangians which describe the
same physics. As is clear from this group is equal to
$\mathrm{Sp}(2n;\mathbb{R})$ which is the same group that is relevant
for the Minkowski theory \cite{Gaillard:1981rj}.

The above results follow from the observation that the two Lagrangians
$\tilde{\mathcal{L}}$ and $\mathcal{L}$ are related by
\begin{equation}
  \label{eq:F-versus-Ftilde}
  \tilde{\mathcal{L}}(\tilde F) - \tfrac14 \tilde{F}_{\mu\nu}{\!}^{+I}\, \tilde
  G^{\mu\nu}{\!}_I {\!}^+ - \tfrac14 \tilde {F}_{\mu\nu}{\!}^{-I}\, \tilde
  G^{\mu\nu}{\!}_I {\!}^- = \mathcal{L}(F) - \tfrac14  F_{\mu\nu}{\!}^{I+}\, 
  G^{\mu\nu}{\!}_I {\!}^+ - \tfrac14  F_{\mu\nu}{\!}^{I-}\, 
  G^{\mu\nu}{\!}_I {\!}^-\,,
\end{equation}
up to terms that are independent of
$F_{\mu\nu}{\!}^{\pm\,I}$. However, by carrying out the
electric-magnetic duality transformation one only redefines the field
strengths and not the other fields. As a consequence supersymmetry
will no longer be manifest. To avoid this unwelcome feature, one can
also make a similar transformation on the scalar fields. This
transformation takes the following form,
\begin{equation}
  \label{eq:emdual-rot-X}
  \begin{pmatrix} 
     X_\pm{\!}^I \\[2mm]
      \mp\mathcal{F}^\pm{\!}_J
    \end{pmatrix}
    \longrightarrow  
    \begin{pmatrix} 
     \tilde  X_\pm{\!}^I \\[2mm]
    \mp\tilde{\mathcal{F}}^\pm{\!}_J 
    \end{pmatrix}
    = 
  \begin{pmatrix}
   U^I{}_K&\ Z^{IL}\\[2mm]
    W_{JK}& V_J{}^L
   \end{pmatrix}
   \begin{pmatrix} 
      X_\pm{\!}^K \\[2mm]
      \mp\mathcal{F}^\pm{\!}_L
    \end{pmatrix}\,,
\end{equation}
where $\tilde X_\pm{\!}^I$ and $\tilde{\mathcal{F}}^\pm{\!}_J$ denote
the new scalar fields and the first derivatives of a new function
$\tilde{\mathcal{F}}^\pm(\tilde X_\pm)$. This function will only exist
provided \eqref{eq:sympl-matrix} is satisfied. This is all in line
with what is known from the Minkowski theory
\cite{deWit:1984wbb,Cecotti:1988qn,deWit:2001pz}.

We close this section by discussing the possibility of including an
additional (anti-)chiral superfield of Weyl weight $w$ into the
functions $\mathcal{F}^\pm$. This new field will not correspond to a
vector multiplet and can be considered as a background. In that case
the homogeneity requirement for $\mathcal{F}^\pm$ must involve the
background field,
\begin{equation}
  \mathcal{F}^\pm(\lambda X_\pm{\!}^I,\lambda^w \widehat{A}_\pm) = 
  \lambda^2\,\mathcal{F}^\pm(X_\pm{\!}^I,\widehat{A}_\pm) \, .
\end{equation}
We can then construct the corresponding invariant
action following the same procedure outlined at the beginning of this
section. The resulting bosonic action is \eqref{eq:vector-Lagrangian},
supplemented by terms that depend on the derivatives of the
prepotentials with respect to the (anti-)chiral superfield,
\begin{equation}
  \mathcal{L}_A = \mathcal{L}_{A+} + \mathcal{L}_{A-} \, ,
\end{equation}
with
\begin{align}
  \label{eq:chiral-bckgd-Lagrangian-pm}
  \mathcal{L}_{A\pm} =&\, -\tfrac12\,
  \mathcal{F}^\pm{\!}_{\widehat{A}}\,\widehat{C}_\pm -
  \tfrac14\,\mathcal{F}^\pm{\!}_{\widehat{A}I}\,\widehat{B}_\pm{\!}^{ij}
  \,Y_{ij}{\!}^I 
  +\tfrac12\,\mathcal{F}^\pm{\!}_{\widehat{A}I}\,
  \widehat{F}^{\,ab\,\mp}\big(F_{ab}{\!}^{\mp I} 
  - \tfrac14\,X_\mp{\!}^I
  T_{ab}{\!}^\mp\big) \nonumber \\[1mm]
  &\,
  +\tfrac18\,\mathcal{F}^\pm{\!}_{\widehat{A}\widehat{A}}\,
  \widehat{B}_\pm{\!}^{ij}\,\widehat{B}_\pm{\!}^{kl}
  \,\varepsilon_{ik}\,\varepsilon_{jl} +
  \tfrac14\,\mathcal{F}^\pm{\!}_{\widehat{A}\widehat{A}}\,\widehat{F}_{ab}{\!}^\mp
  \,\widehat{F}^{\,ab\,\mp} \, .
\end{align}  
The possibility of including such a dependence on an (anti-)chiral
superfield allows for the inclusion of higher-derivative terms in the
action of Euclidean $4D$, $N=2$ superconformal gravity coupled to
vector multiplets, by choosing the square of the Weyl multiplet as the
(anti-)chiral background.  In this case, the first term in
\eqref{eq:chiral-bckgd-Lagrangian-pm} contains the square of the
modified Lorentz curvature according to
\eqref{eq:weyl-squared-components}.

\section{Summary and conclusions}
\label{sec:summary-conclusions}
\setcounter{equation}{0}

In this paper we have derived off-shell supergravity in four Euclidean
dimensions with eight supersymmetries, together with an extended
superconformal multiplet calculus that can be used to construct a
large variety of invariant actions. Its general structure closely
resembles that of Minkowski $N=2$ supergravity. The two theories have
an equal number of supersymmetries and therefore they have a
corrreponding variety of supermultiplets. Nevertheless there are a
number of crucial differences. The Euclidean theory has symplectic
Majorana spinors and an $\mathrm{SU}(2)\times \mathrm{SO}(1,1)$
R-symmetry group, whereas the Minkowski theory has Majorana fermions
and an $\mathrm{SU}(2)\times \mathrm{U}(1)$ R-symmetry group. These
differences are then also reflected in different reality conditions on
the various fields. For instance the chiral and anti-chiral
supermultiplets are real and they are not related by complex
conjugation. The latter fact implies that the action of the theory is
built on two, not necessarily related, real functions for the chiral
and the anti-chiral multiplet sector. 

The non-linear transformations of Euclidean supergravity were
initially derived by time-like reduction from $5D$ Minkowski
supergravity, the results of which were then subsequently extended to a
complete superconformal multiplet calculus. As a by-product we have obtained a
dictionary between the five- and four-dimensional fields, but in this
paper the five-dimensional origin only plays a secondary role.

Just as for $N=2$ Minkowski supergravity the multiplet calculus is
built on superconformal multiplets. This implies that the resulting
actions are initially superconformally invariant. However, by
including the actions for an appropriate set of compensating
multiplets the combined action will become gauge equivalent to an
off-shell Poincar\'e supergravity. As a result there exist different
formulations of Poincar\'e supergravity corresponding to inequivalent
compensating multiplet configurations. In \cite{deWit:1982na} three
choices were pointed out. One always needs one compensating vector
multiplet and a second one which is either a hypermultiplet, a tensor
multiplet or a so-called non-linear multiplet. Each of these
multiplets contains $8\oplus 8$ bosonic and fermionic degrees of
freedom. Since the superconformal multiplet contains $24\oplus 24$
bosonic and fermionic degrees of freedom, off-shell Poincar\'e
supergravity will be based on $40\oplus 40$ degrees of freedom. Note
that this list is not necessarily exhaustive.

As it turns out the same situation exists for Euclidean
supergravity. There are also three off-shell field representations
with the first compensator equal to a vector multiplet and the second
one a hypermultiplet, a tensor multiplet or a non-linear
multiplet. The latter exists also in $5D$ Minkowski theory
\cite{Zucker:2003qv} and upon a time-like reduction can thus be
realized in $4D$ Euclidean supergravity as well. We have refrained from
giving its explicit transformation rules. Instead, in appendix
\ref{app:NL-supermult}, we have presented the $5D$ transformation
rules of this multiplet in the conventions of section
\ref{sec:euclid-superg-from-dim-red}.

The theory presented in this paper can be used to define Euclidean
supersymmetric theories on curved spaces upon freezing out the
supergravity fields, which is of interest to localization problems in
field theory. A related area of interest concerns the quantum entropy
of black hole solutions in supergravity obtained via localization of
the quantum entropy function, where the supergravity fields are not
frozen completely.  This question can now be considered without having
to rely on analytic continuation of the Minkowski theory. As a first
step in this program, one should derive the BPS black hole solutions
of the Euclidean theory that we have just presented. We intend to
return to this issue in a forthcoming paper.

We should point out that the analytically continued theories that
appeared in \cite{Klare:2013dka,Dabholkar:2010uh} are related to the
Euclidean supergravity theory of this paper modulo some
straightforward field redefinitions. Evidently, our results are also
in line with the theory of Euclidean vector multiplets presented in
\cite{Cortes:2003zd}, since both make use of a time-like reduction of
a $5D$ Minkowski theory.

\begin{appendix}
%
\section{Off-shell dimensional reduction; the Weyl multiplet}
\label{sec:off-shell-dim-red-Weyl}
\setcounter{equation}{0}
Starting from the off-shell superconformal transformations for $5D$
supermultiplets presented in section
\ref{sec:euclid-superg-from-dim-red}, we perform a reduction of the
time coordinate and obtain the $4D$ Euclidean superconformal
transformations for the corresponding supermultiplets. The Weyl
multiplet contains the gauge fields associated with the superconformal
transformations as well as additional supercovariant fields, which act
as a background for all other supermultiplets.  Therefore this
multiplet must be considered first. Here a subtle complication is that
the Weyl multiplet becomes reducible upon the reduction. In $D=5$ it
comprises $32\oplus 32$ bosonic and fermionic degrees of freedom, which, in
the reduction to $D=4$ dimensions, decompose into the Weyl multiplet
comprising $24\oplus 24$ degrees of freedom and a vector multiplet
comprising $8\oplus 8$ degrees of freedom.

In section \ref{sec:euclid-superg-from-dim-red} we also described the
Kaluza-Klein decomposition of the metric and the dilatational gauge
field that ensure that the $4D$ fields transform covariantly under the
$4D$ diffeomorphisms. Since these decompositions involve gauge choices
on the vielbein and the dilatational gauge field, compensating Lorentz
and special conformal transformations must be included when deriving
the $4D$ Q-supersymmetry transformations to ensure that these gauge
conditions are preserved. Here the parameter of the compensating Lorentz
transformation is most relevant. It is equal to
\begin{equation}
  \label{eq:comp-Lor}
  \varepsilon^{a5} = -\varepsilon^{5a} = \mathrm{i} \phi\,
  \bar\epsilon_i\gamma^a\psi^i \; \Longleftrightarrow \;
  \varepsilon^{a0} = -\varepsilon^{0a} =  \phi\,
  \bar\epsilon_i\gamma^a\psi^i \,,
\end{equation}
where we assumed the standard Kaluza-Klein decomposition on the
gravitino fields, 
\begin{equation}
  \label{eq:gravitino-KK}
  \psi_M{}^i =  \begin{pmatrix}\psi_\mu{}^i+ B_\mu \psi^i\\[4mm]
    -\mathrm{i}\psi^i \end{pmatrix}\;, 
\end{equation}
which ensures that $\psi_\mu{}^i$ on the right-hand side transforms as
a $4D$ vector. Owing to the factor $\mathrm{i}$ in this decomposition
both $\psi_\mu{\!}^i$ and $\psi^i$ are symplectic Majorana
spinors. Upon including this extra term, one can write down the Q- and
S-supersymmetry transformations on the $4D$ fields defined above. As a
result of this, the $4D$ and $5D$ supersymmetry transformation will be
different. For instance, the supersymmetry transformations of the $4D$
fields $e_\mu{}^a$, $\phi$ and $B_\mu$ read,
\begin{align}
  \label{eq:susy-e-B-phi}
  \delta e_\mu{}^a =&\,  \bar\epsilon_i\gamma^a\psi_\mu{}^i
  \,, \nonumber\\[.2ex]
  \delta\phi =&\, \mathrm{i} \phi^2\,\bar\epsilon_i\gamma^5\psi^i\,,
  \nonumber \\[.2ex]
  \delta B_\mu=&\,  -\phi^2 \,\bar\epsilon_i\gamma_\mu\psi^i -\mathrm{i}
  \phi \,\bar\epsilon_i\gamma^5\psi_\mu{}^i \,,
\end{align}
where the first term in $\delta B_\mu$ originates from the
compensating transformation \eqref{eq:comp-Lor}. Consequently the
supercovariant field strength of $B_\mu$ contains a term that is not
contained in the supercovariant five-dimensional curvature
$R(P)_{MN}{}^A$.  Therefore the $5D$ spin-connection components are
not supercovariant with respect to $4D$ supersymmetry, as is reflected
in the second formula below,
\begin{align}
  \label{eq:spin-connection}
  \omega_M{}^{ab} =&\, \begin{pmatrix} \omega_\mu{}^{ab} \\[4mm] 
    0 \end{pmatrix} - \tfrac12  \phi^{-2} \hat F(B)^{ab} \, 
  \begin{pmatrix} B_\mu \\[4mm] -\mathrm{i} \end{pmatrix} \;, \nonumber \\[.8ex]
  \omega_M{}^{a5} =&\, -\tfrac12 \mathrm{i}\begin{pmatrix} \phi^{-1}
    \hat F(B)_\mu{}^a - \phi\,\bar\psi_{\mu i}\gamma^a\psi^i \\[4mm]
    0 \end{pmatrix} -\mathrm{i} \phi^{-2} D^{a}\phi  \,    
  \begin{pmatrix} B_\mu \\[4mm] -\mathrm{i}\end{pmatrix} \;.
\end{align}
Here we introduced a supercovariant field strength and derivative
(with respect to $4D$ supersymmetry), 
\begin{align}
  \label{eq:supercov-FB-Dphi}
  \hat F(B)_{\mu\nu} =&\, 2\,\partial_{[\mu} B_{\nu]} + \phi^2\,
  \bar\psi_{[\mu i}\gamma_{\nu]} \psi^i +\tfrac12 \mathrm{i}
  \phi\,\bar\psi_{\mu i} \gamma_5 \psi_{\nu}{}^i \,,\nonumber\\
  D_\mu\phi =&\, (\partial_\mu -b_\mu) \phi -\tfrac12\mathrm{i} \phi^2
  \,\bar\psi_{\mu i}\gamma_5 \psi^i \,.
\end{align}
We should mention that the dilatational gauge field (as
well as the composite gauge fields, such as $\omega_\mu{}^{ab}$, that
depend on it) will not necessarily acquire the form that is familiar
from $4D$. This may require to include an  additional compensating
conformal boost transformation. 

Subsequently one writes corresponding Kaluza-Klein decompositions for
some of the other fields of the Weyl multiplet, which do not require
special gauge choices,
\begin{equation}
  \label{eq:1Weyl-KK}
    {V}_{M i}{}^j=
  \begin{pmatrix}{V}_{\mu i}{}^j+ B_\mu
    {V}_{i}{}^j\\[4mm] 
    -\mathrm{i} {V}_{i}{}^j \end{pmatrix}\;,\qquad
  \phi_M{}^i =  \begin{pmatrix}\phi_\mu{}^i+ B_\mu \phi^i\\[4mm]
    -\mathrm{i}\phi^i \end{pmatrix}\;\qquad 
   T_{AB} = \begin{pmatrix} T_{ab} \\[4mm]  T_{a5}\equiv \tfrac16
     \mathrm{i} A_a \end{pmatrix}  \,. 
\end{equation}

Hence we are now ready to consider the Q- and S-supersymmetry
transformations of the spinor fields originating from the $5D$
gravitino fields. Up to possible higher-order spinor terms, one
derives the following results from \eqref{eq:Weyl-susy-var},
\begin{align}
  \label{eq:susy-W-gravitino}
  \delta(\phi^2\,\psi^i) =&\, -\tfrac12
   \big[-\hat F(B)_{ab} + \gamma_5 \phi (
  3\,T_{ab} +\tfrac14  \phi^{-1} \hat F(B)_{ab}\gamma_5
  )\big]   \gamma^{ab}\epsilon^i \nonumber\\
  &\,
  +\mathrm{i}\big[ \Slash{D} \phi \,\gamma^5 - 
  \Slash{A}\phi\big] \epsilon^i - \phi^2 
  {V}^i{}_j \,\epsilon^j \nonumber\\[.2ex]
  &\,
    +\gamma_5 \phi \big[ \eta^i
    -\tfrac13\mathrm{i} \Slash{A}\gamma_5\epsilon^i  -\tfrac1{8} 
  \gamma_5\phi^{-1}(\hat F(B)_{ab}-4\phi
   T_{ab}\gamma_5)\gamma^{ab} \epsilon^i\big]  \,, \nonumber\\[.2ex] 
  \delta\psi_\mu{}^i =&\, 2\,\big(\partial_\mu
  -\tfrac14\omega_\mu{}^{ab}\gamma_{ab}+\tfrac12 b_\mu
  +\tfrac12 e_\mu{}^a \,A_a \gamma_5 \big)\epsilon^i
  +  {V}_{\mu j}{}^i \epsilon^j \nonumber\\
  &\, + \tfrac12 \mathrm{i}\big[3\, T_{ab} +\tfrac14 
  \phi^{-1} \hat F(B)_{ab}\gamma_5  \big] \gamma_{ab} \gamma_\mu
  \epsilon^i  \nonumber\\
  &\,
  - \mathrm{i} \gamma_\mu \big[\eta^i
   -\tfrac13\mathrm{i}\Slash{A}\gamma_5 \epsilon^i  -\tfrac1{8}
   \gamma_5\phi^{-1}(\hat F(B)_{ab}-4\phi
   T_{ab}\gamma_5)\gamma^{ab} \epsilon^i \big]\,.  
\end{align}
Clearly, the fields $e_\mu{}^a$ and $\psi_\mu{}^i$ must belong to the
Weyl multiplet, whereas $\phi$, $B_\mu$ and $\phi^2\psi^i$ correspond
to the Kaluza-Klein vector multiplet, as the transformations shown in
\eqref{eq:susy-e-B-phi} and \eqref{eq:susy-W-gravitino} have many
features in common with the expected $4D$ transformations of these
supermultiplets. Note that we have multiplied $\psi^i$ with a factor
$\phi^2$ to give it the expected Weyl weight ${w=\tfrac32}$. At this
stage we have only identified one of the two $w=1$ scalars that must reside
in a $4D$ vector multiplet.  The field $A_a$ seems to play
the role of an R-symmetry connection because it appears to
covariantize the derivatives on $\phi$ and $\epsilon^i$ in
\eqref{eq:susy-W-gravitino}.  Furhtermore, a particular
linear combination of the $5D$ tensor components $T_{ab}$ and the
(dual) supercovariant field strength $\hat F(B)_{ab}$ appears in the
transformations \eqref{eq:susy-W-gravitino} in precisely the same form
as the $4D$ auxiliary tensor $T_{ab}$, so that the latter is not just
proportional to the original $5D$ tensor field. The same combination
will also appear in other transformation rules, as we shall see in,
for instance, appendix \ref{sec:shell-dimens-reduct-matter}. Finally,
S-supersymmetry transformations are accompanied by extra contributions
characterized by a field-dependent parameter proportional to
$\epsilon^i$.  

However, the result \eqref{eq:susy-W-gravitino} is not yet complete as
we have suppressed variation contributions quadratic in the spinor
fields. First of all we did not include the non-covariant term in
\eqref{eq:spin-connection} and we ignored the compensating Lorentz
transformation \eqref{eq:comp-Lor}. Secondly we ignored the variation
of the field $B_\mu$ in the decomposition of the $4D$ gravitino
\eqref{eq:gravitino-KK}, and thirdly the multiplication of $\psi^i$ with
$\phi^2$ will also generate a variation quadratic in $\psi^i$ . Since
these terms will play an important role we summarize them below,
\begin{align}
  \label{eq:susy-W-gravitino-nonlin}
  \delta(\phi^2\psi^i)\big\vert_\mathrm{non-linear} =&\, \tfrac12
  \mathrm{i} \phi^3\,
  \bar\epsilon_j\gamma^a\psi^j\, \gamma_a\gamma_5\psi^i   +
  2\mathrm{i}\, \phi^3 \,\bar\epsilon_j\gamma^5\psi^j\, \psi^i\,,
  \nonumber\\[.2ex]  
  \delta\psi_\mu{}^i\big\vert_\mathrm{non-linear} =&\, -\tfrac12
  \mathrm{i} \phi \,\bar\psi_{\mu j} \gamma^a\psi^j \, \gamma_a\gamma^5
  \epsilon^i 
  + \tfrac12 \mathrm{i} \phi\, \bar\epsilon_j\gamma^a\psi^j\,
  \gamma_a\gamma_5\psi_\mu{\!}^i \nonumber\\
  &\,+\big(\phi^2
  \,\bar\epsilon_j\gamma_\mu\psi^j +\mathrm{i} \phi
  \,\bar\epsilon_j\gamma^5\psi_\mu{}^j \big) \psi^i \,.
\end{align}

The systematic pattern already noticed in \cite{Banerjee:2011ts} for
the space-like reduction is that  the $5D$ supersymmetry
transformations can uniformally  be decomposed in terms of the $4D$
supersymmetry transformations, and S-supersymmetry and
$\mathrm{SU}(2)$ R-symmetry transformations with field-dependent
parameters. Since the derivation is
identical to what was carried out in \cite{Banerjee:2011ts}, we just
present the universal formula for $5D$ Q-supersymmetry transformations
of fields $\Phi$ that transform covariantly in the $4D$ setting,
\begin{equation}
  \label{eq:D5-D4-decomp}
  \delta_\mathrm{Q}(\epsilon)\big|^\mathrm{reduced}_{5D} \Phi =
  \delta_\mathrm{Q}(\epsilon)\big|_{4D} \Phi + 
  \delta_\mathrm{S}(\tilde\eta)\big|_{4D} \Phi +
  \delta_{\mathrm{SU}(2)}(\tilde\Lambda)\big|_{4D}\Phi  +
  \delta^\prime(\tilde\Lambda^0) \Phi\,.  
\end{equation}
Here the first term on the right-hand side defines the $4D$
supersymmetry transformation, while $\tilde\eta$ and $\tilde\Lambda$
denote the (universal) field-dependent parameters of accompanying
S-supersymmetry and $\mathrm{SU}(2)$ R-symmetry transformations. The
last variation denoted by $\delta^\prime(\tilde\Lambda^0)$ is a linear
transformation on the fields $\Phi$ that signals the emergence of an
extra component in the $4D$ Euclidean R-symmetry group. Note that
$\tilde\eta$, $\tilde\Lambda$ and $\tilde\Lambda^0$ are all linearly
proportional to the supersymmetry parameter $\epsilon^i$. The explicit
form of these field-dependent parameters is as follows, 
\begin{align}
  \label{eq:Q-susy decom-par}
  \tilde{\eta}^i =&\,-\tfrac13\mathrm{i} \Slash{A}\gamma_5\epsilon^i
  -\tfrac1{8}\gamma_5\phi^{-1}\big(\hat F(B)_{ab}-4\phi
  T_{ab}\gamma_5\big)\gamma^{ab} \epsilon^i \nonumber\\
  &-\tfrac14\mathrm{i}\,\phi^2\big(\bar{\psi}_j\gamma^5\psi^i\gamma_5
    -\bar{\psi}_j\psi^i + \bar{\psi}_j\gamma^a\psi^i\gamma_a
    +\tfrac12\bar{\psi}_k\gamma^5\gamma^a\psi^k\gamma_5
    \gamma_a\delta_j{}^i\big)\epsilon^j\, , \nonumber\\
  \tilde{\Lambda}_j{}^i =&\,- \mathrm{i}\,\phi\big(\bar{\epsilon}_j\gamma^5\psi^i
    -\tfrac12\delta_j{}^i\bar{\epsilon}_k\gamma^5\psi^k\big)\, ,
  \nonumber\\
  \tilde{\Lambda}^0 =&\,\mathrm{i}\,\phi\,\bar{\epsilon}_k\psi^k \,. 
\end{align}
After verifying that the decomposition is universally realized these
extra symmetries with field-dependent coefficients can be dropped
provided they define local symmetries of the $4D$ theory.

Evaluating the terms of higher order in the fermions is subtle; here we
can only partly rely on the results of \cite{Banerjee:2011ts} because
the phases of the spinor bilinears cannot always be converted directly
from $5D$ (as noted below equation \eqref{eq:bilinear}). It leads
to the following redefinitions of the various bosonic fields,
\begin{align}
  \label{eq:field-redef}
   \hat A_\mu =&\, A_a \,e_\mu{}^a -\tfrac12\mathrm{i}\, \phi\,
  \bar\psi_j\psi_\mu{}^j-\tfrac1{4} \phi^2\,
  \bar\psi_j\gamma^5\gamma_\mu\psi^j \,, \nonumber \\[.2ex]
  \hat{T}_{ab}=&\, 24\,T_{ab} - \phi^{-1}\,\varepsilon_{abcd}\,
  \hat F(B)^{cd} 
   +\mathrm{i} \phi^{2}\, \bar\psi_i\gamma_{ab} \psi^i
   \,,\nonumber \\[.2ex] 
   \hat V_j{}^i=&\, \phi^2\, V_j{}^i + \tfrac32\mathrm{i} \phi^3\, 
   \bar\psi_j\,\gamma^5\psi^i \,,\nonumber\\ 
   \hat V_\mu{}_j{}^i =&\, {V}_\mu{}_j{}^i
  +\mathrm{i}\phi \big( \bar\psi_{\mu j} \gamma^5 \psi^i
   - \tfrac12 \delta_j{}^i\,
  \bar\psi_{\mu k} \gamma^5 \psi^k  \big)
  +\tfrac12  \phi^2\,\bar\psi_{j} \gamma_\mu \psi^i \,.
\end{align}
Note that in the last two equations possible contributions
proportional to $\bar\psi_k\gamma^5\psi^k$ and
$\bar\psi_k\gamma_\mu\psi^k$ do not appear as they vanish owing to the 
symplectic Majorana condition. 

The modifications given in \eqref{eq:field-redef} lead to important
changes in the supersymmetry transformations. For instance, the
S-supersymmetry transformations are given by
\begin{align}
  \label{eq:S-ATV}
   \delta\hat A_\mu=&\,\tfrac12\mathrm{i}\bar\psi_{\mu j}\gamma^5 \eta^j\,,
   \nonumber \\[.2ex] 
   \delta\hat T_{ab} =&\, 0\,, \nonumber \\[.2ex]
   \delta \hat V_j{}^i=&\, 0\,,  \nonumber\\[.2ex]
   \delta\hat V_{\mu j}{}^i=&\,  -2\mathrm{i}\big(\bar\psi_{\mu j}\,\eta^i
   -\tfrac1{2}\delta_j{}^i\,\bar\psi_{\mu k}\,\eta^k \big) \,. 
\end{align}
In particular, note that the factor in the variation of $\hat V_{\mu
  i}{}^j$ has now changed as compared to the corresponding $5D$
S-variation given in \eqref{eq:Weyl-susy-var}. Furthermore, $\hat
A_\mu$ is not supercovariant because its Q-supersymmetry variation
contains a term proportional to the derivative of the supersymmetry
parameter. This suggest that $\hat A_\mu$ will be related to a gauge
field associated with an extra $4D$ R-symmetry, which will indeed
be consistent with the fact that $\hat A_\mu$ transforms into the
gravitino fields under S-supersymmetry.

Let us now present the supersymmetry transformations for the redefined
fields, suppressing the field-dependent S-supersymmetry and
$\mathrm{SU}(2)$ transformations indicated in
\eqref{eq:D5-D4-decomp}. For the vierbein and gravitini, we find
\begin{align}
  \label{eq:susy-weyl1}
  \delta e_\mu{}^a =&\,  \bar\epsilon_i\gamma^a\psi_\mu{}^i
  \,, \nonumber\\[.2ex]
    \delta\psi_\mu{}^i =&\,2\,\big(\partial_\mu
  -\tfrac14\omega_\mu{}^{ab}\gamma_{ab}+\tfrac12 b_\mu
  +\tfrac12  \hat A_\mu \gamma_5 \big)\epsilon^i
  +  {\hat V}_{\mu j}{}^i\, \epsilon^j 
  + \tfrac1{16} \mathrm{i}\,\hat T_{ab} \gamma^{ab} \gamma_\mu
  \epsilon^i    -\mathrm{i} \gamma_\mu \eta^i \,.
\end{align}
For the scalar $\phi$, the spinor $\hat\psi^i\equiv \phi^2 \psi^i$ and
the Kaluza-Klein photon field $B_\mu$ we have the following Q- and
S-supersymmetry transformations,
\begin{align}
  \label{eq:susy-KK}
    \delta\phi =&\, \mathrm{i}\,\bar\epsilon_i\gamma^5\hat\psi^i\,,
  \nonumber \\[.2ex]
  \delta B_\mu=&\, -\bar\epsilon_i\gamma_\mu\hat \psi^i - \mathrm{i}
  \phi \,\bar\epsilon_i\gamma^5\psi_\mu{}^i \,,\nonumber
  \\[.2ex]  
  \delta\hat\psi^i  =&\, \tfrac12 \big[\hat
  F(B)_{ab}-\tfrac18 \phi\, \hat T_{ab}\gamma_5 \big]
  \gamma^{ab}\epsilon^i  \nonumber\\
  &\, -\mathrm{i} \gamma^5 \gamma^\mu \big[\mathcal{D}_\mu\phi
  -\tfrac12\mathrm{i}(\bar\psi_{\mu j} 
  \gamma^5 \hat\psi^j -\bar\psi_{\mu j}\hat \psi^j \gamma^5) - \hat
    A_\mu \phi\,\gamma^5 \big] \epsilon^i 
    +  {\hat V}_j{}^i \,\epsilon^j +\phi \gamma^5 \eta^i\,,
\end{align}
where the derivative $\mathcal{D}_\mu$ is covariant with respect
to $4D$ local Lorentz, dilatation and $\mathrm{SU}(2)$ transformations.

At this point we briefly make a few observations. First of all, we
have suppressed the chiral transformations proportional to the
field-dependent parameter $\tilde\Lambda^0$, 
\begin{equation}
  \label{eq:field-dep-so11}
  \delta\psi_\mu{}^i =  - \tfrac12 \tilde\Lambda^0 \,\gamma^5
  \psi_\mu{}^i\,, \quad  \delta\hat\psi^i  =  -\tfrac12
  \tilde\Lambda^0 \, \gamma^5\hat\psi^i \,.
\end{equation}
Clearly we were not allowed to drop these terms in view of the fact
they do not correspond to local transformation of the $4D$ theory at
this stage. Furthermore, the variations of $\hat\psi^i$ proportional
to $\bar\psi_{\mu j} \hat \psi^j$ are not part of a supercovariant
derivative of the field $\phi$. And finally the field $\hat A_\mu$ is
not a gauge field associated with the chiral transformations (although
it appears in a suggestive way). However, it is not a proper matter
field either as it does not transform supercovariantly. As it turns
out these issues have a common origin.

Before resolving these questions it is better to first proceed and
take a closer look at a composite fermionic gauge field
$\hat\phi_\mu{}^i$ that serves as a $4D$ connection for
S-supersymmetry. It is the solution of the equation (in the ensuing
analysis we will not exhibit terms quadratic in the spinor fields)
\begin{equation}
  \label{eq:intermediate-4d-connection}
  \gamma^\mu \Big[\big(\mathcal{D}_{[\mu} + \tfrac12 \hat A_{[\mu} \gamma^5
  \big)\psi_{\nu]}{\!}^i  
  -\tfrac12 \mathrm{i} \,\gamma_{[\mu} \,\hat\phi_{\nu]}{}^i   
  +\tfrac1{32} \mathrm{i}\,\hat T_{ab} \gamma^{ab} \,\gamma_{[\mu}
  \,\psi_{\nu]}{}^i \Big]  =0\,,
\end{equation} 
and transforms under S- and Q-supersymmetry as 
\begin{align}
  \label{eq:delta-hat-phi-mu}
  \delta\hat \phi_\mu{}^i =&\, 2\,\big(\mathcal{D}_\mu -\tfrac12 \hat
  A_\mu \gamma_5 \big)\eta^i + \tfrac{1}{48}\mathrm{i}\gamma_\mu
  \hat{T}_{ab}\gamma^{ab}\eta^i \nonumber\\
  &\, + 2\mathrm{i} \,\hat f_\mu{}^a\gamma_a\epsilon^i
  +\tfrac1{16}\bigl(\gamma^\nu \gamma^{ab}\gamma_\mu -
  \tfrac13\gamma_\mu\gamma^{ab}\gamma^\nu\bigr)D_\nu\hat
  T_{ab}\epsilon^i \nonumber\\
  &\, -\tfrac14\mathrm{i}\bigl(\gamma^{ab}\gamma_\mu -
  \tfrac13\gamma_\mu\gamma^{ab}\bigr)R(\hat V)_{abj}{}^i\epsilon^j
  - \tfrac12\mathrm{i}\bigl(\gamma^{ab}\gamma_\mu +
  \tfrac13\gamma_\mu\gamma^{ab}\bigr)R(\hat A)_{ab}\gamma^5\epsilon^i \,,
\end{align}
where $\hat f_\mu{}^a$ reads
\begin{equation}
  \label{eq:def-hat-f}
  \hat f_\mu{}^a =\tfrac12  R(\omega,e)_\mu{}^a - \tfrac1{12} R(\omega,e)
  \,e_\mu{}^a -\tfrac12 \widetilde R(\hat A)_\mu{}^a
  -\tfrac1{128}  (\hat T- \widetilde{\hat T})_{\mu b} \,(\hat T+\widetilde{\hat T})^{ba} \,, 
\end{equation}
where $R(\omega,e)_\mu{}^a = R(\omega)_{\mu\nu}{\!}^{ab}\,e_b{}^\nu$
is the generalized (non-symmetric) Ricci tensor. Its anti-symmetric
part is equal to $R(\omega,e)_{[\mu\nu]} =R(b)_{\mu\nu}= \partial_\mu b_\nu
-\partial_\nu b_\mu$.  This follows from the identity
$R(\omega)_{[ab,c]}{}^d = -R(b)_{[ab}\,\delta_{c]}{}^d$, which
reflects the fact that the spin connection $\omega_\mu{\!}^{ab}$ depends on
the dilatational gauge field $b_\mu$. As a result
the generalized Riemann tensor $R(\omega)_{\mu\nu}{\!}^{ab}$ is not
symmetric under pair-exchange,
\begin{equation}
    \label{eq:pair-exchange}
    R(\omega)_{ab,cd}-R(\omega)_{cd,ab}  =
    -2\,\eta_{[a[c}\,R(\omega,e)_{d]b]} +2\,\eta_{[c[a}\,R(\omega,e)_{b]d]}\,.    
\end{equation}
Finally, $\widetilde R(\hat A)_{\mu\nu}$ denotes the dual of
$R(\hat A)_{\mu\nu}= \partial_\mu \hat A_\nu -\partial_\nu\hat A_\mu$.  

The $5D$ S-supersymmetry gauge field $\phi_M{}^i$ follows from the
fermionic conventional constraint given in
\eqref{eq:conv-constraints-5} and can be decomposed as follows under
the $4D$ reduction, 
\begin{align}
  \label{eq:S-susy-gauge fields}
  &\phi_\mu{}^i \vert_{5D}  - 
  \tfrac16\mathrm{i}\, \Slash{\hat{A}}\gamma_5\psi_\mu{}^i +
  \tfrac1{96}\hat{T}_{ab}\gamma^{ab}\psi_\mu{}^i
  -\tfrac1{12}\phi^{-1}\hat{F}_{ab}\gamma^{ab}\gamma_5\psi_\mu{}^i \nonumber\\
   &= \tfrac12\hat \phi_\mu{}^i
   +\tfrac13\phi^{-1}\gamma_5 D_\mu\hat{\psi}^i +
  \tfrac1{12}\phi^{-1}\gamma_\mu \gamma_5\Slash{D}\hat{\psi}^i
  - \tfrac23\phi^{-1}\hat{A}_\mu\hat{\psi}^i
  +\tfrac16\phi^{-1}\gamma_\mu\Slash{\hat{A}}\hat{\psi}^i \nonumber\\
  &\,\quad +\tfrac1{3}\mathrm{i}\phi^{-2}\gamma^\nu\hat{F}_{\mu\nu}\hat{\psi}^i
  - \tfrac1{24}\mathrm{i}
  \phi^{-2}\gamma_\mu\hat{F}_{ab}\gamma^{ab}\hat{\psi}^i -
  \tfrac1{96}\mathrm{i}\phi^{-1}
  \hat{T}_{ab}\gamma^{ab}\gamma_\mu\gamma_5\hat{\psi}^i \nonumber\\
  &\,\quad-\tfrac23\phi^{-2}\gamma_5\big(\mathcal{D}_\mu\phi - \hat{A}_\mu\phi\gamma_5\big)\hat{\psi}^i -
  \tfrac16\phi^{-2}\gamma_\mu\gamma_5
  \big(\Slash{\mathcal{D}}\phi-\Slash{\hat{A}} \phi\gamma_5\big)\hat{\psi}^i \, .
\end{align}
The right-hand side of this equation contains only supercovariant $4D$
expressions, with the exception of the field $\hat \phi_\mu{}^i$ which
is a gauge field. For instance $D_\mu\hat\psi^i$ is the $4D$ fully
supercovariant derivative given by (at linear order in the spinor
fields)
\begin{align}
  \label{eq:cov-der-hat-psi}
  D_\mu\hat{\psi}^i =&\, \big(\mathcal{D}_\mu
  +\tfrac12\hat{A}_\mu\gamma^5\big)\hat{\psi}^i
  -\tfrac12\phi\,\gamma^5\hat\phi_\mu{}^i -\tfrac14 \big[\hat
  F(B)_{ab}-\tfrac18 \phi\, \hat T_{ab}\gamma_5 \big]
  \gamma^{ab}\psi_\mu{}^i  \nonumber \\
  &\, +\tfrac12\mathrm{i} \gamma^5 \gamma^\nu
  \big[\mathcal{D}_\nu\phi - \hat A_\nu \phi\,\gamma^5 \big]
  \psi_\mu{}^i - \tfrac12{\hat V}_j{}^i \,\psi_\mu{}^j\, ,
\end{align}
which also contains the S-supersymmetry gauge field
$\hat\phi_\mu{}^i$. The terms on the left-hand side of
\eqref{eq:S-susy-gauge fields} that depend explicitly on
$\psi_\mu{}^i$ seem to affect the covariance under
Q-supersymmetry. However, they are to be expected because, according
to \eqref{eq:D5-D4-decomp}, the $5D$ Q-supersymmetry differs from the
$4D$ one by a field dependent S-supersymmetry transformation
parametrized by $\tilde \eta^i$ given in \eqref{eq:Q-susy
  decom-par}. 

The correctness of this result can be verified by considering the Q-
and S-supersymmetry variations of the $4D$ $\mathrm{SU}(2)$ gauge
fields $\hat{V}_{\mu\,i}{}^j$. After taking into account the
Kaluza-Klein decomposition, one has to correct for the field-dependent
S-supersymmetry transformation indicated in \eqref{eq:D5-D4-decomp},
which precisely cancels against the terms in \eqref{eq:S-susy-gauge
  fields} that depend explicitly on $\psi_\mu{\!}^i$. Furthermore one
has to take into account the redefinitions in \eqref{eq:field-redef}
and the field-dependent $\mathrm{SU}(2)$ transformation in
\eqref{eq:D5-D4-decomp}. Their combined effect will only lead to terms
such as
\begin{equation}
  \label{eq:delta-gravitino}
  \mathrm{i}(\delta \bar \psi_{\mu i} -2\,\mathcal{D}_\mu \epsilon_i)
  \,\gamma^5 \phi^{-1} \hat\psi{}^j\,,\quad
  -2\mathrm{i}\,\bar\epsilon_i \,[\gamma^5 D_\mu (\phi^{-1} \hat\psi^j)
  +\tfrac12 \hat \phi_\mu{}^j] \,, \quad
  \phi^{-1} \bar{\hat\psi}_i \,\delta(\phi^{-1} \hat\psi^j) \,,  
\end{equation}
where the derivative $D_\mu$ is supercovariant. Combining this with the
result of the Kaluza-Klein decomposition and with
\eqref{eq:S-susy-gauge fields}, one obtains
\begin{equation}
  \label{eq:delta-V-hat-su2}
      \delta\hat{V}_{\mu\,i}{}^j =
      2\mathrm{i}\,\bar{\epsilon}_i\hat{\phi}_\mu{}^j  
      - 2\, \bar{\epsilon}_i\gamma_\mu\hat{\chi}^j - 2\mathrm{i}\,\bar\eta_i
      \psi_\mu{}^j- \tfrac12\delta_i{}^j\bigl(2\mathrm{i} \,
      \bar{\epsilon}_k\hat{\phi}_\mu{}^k - 2\,
      \bar{\epsilon}_k\gamma_\mu\hat{\chi}^k - 2\mathrm{i}\,\bar\eta_k
      \psi_\mu{}^k \bigr) \, , 
\end{equation}
where $\hat{\chi}^i$ is a supercovariant spinor field equal to 
\begin{align}
  \label{eq:hat-chi}
  \hat{\chi}^i =&\, 8\,\chi^i\big\vert_{5D} -
  \tfrac1{4}\mathrm{i}\phi^{-1}\gamma^5\Slash{D}\hat{\psi}^i 
  - \tfrac12\phi^{-2}\hat{V}_k{}^i\hat{\psi}^k \nonumber\\ 
  &\,+ \tfrac1{8}\phi^{-2}\bigl[\hat{F}_{ab} 
  - \tfrac14\phi\hat{T}_{ab}\gamma^5\bigr]\gamma^{ab}\hat{\psi}^i -\tfrac12
  \mathrm{i}\phi^{-1}\Slash{\hat{A}}\hat{\psi}^i \, . 
\end{align}

Let us subsequently turn to the Q- and S-supersymmetry transformations
of the field $\hat\chi^i$, which contains the remaining independent
fermion field $\chi^i\vert_\mathrm{5D}$ of the $5D$ Weyl multiplet
according to the equation above. When writing its variation in terms
of the $4D$ quantities, we naturally obtain terms that depend
exclusively on the $4D$ Weyl multiplet components and others that will
involve both the Weyl multiplet and the Kaluza-Klein vector
multiplet. The latter terms should then cancel by the variations of
the additional terms in \eqref{eq:hat-chi}, because $\hat\chi^i$ must
vary exclusively into the components of the $4D$ Weyl multiplet. Here
one should again compensate for the composite S-supersymmetry
variation parametrized in terms of $\tilde\eta^i$. This leads to the
following expression,
\begin{align}
  \label{eq:QS-transfo-chi5}
  \delta\hat\chi^i =&\,8\,\delta\chi^i\big\vert_{5D} - \tfrac3{2}T_{AB}\,\gamma^{AB}
  \tilde\eta^i   -\tfrac1{4}\mathrm{i} \, \delta\big[\phi^{-1} \gamma^5
  \Slash{D} \hat\psi^i\big] \nonumber\\ 
  &\,-\tfrac18\,  \delta \big[ 
  4\, \phi^{-2} \hat V_j{}^i \,\hat\psi^j
  - \phi^{-2} [ \hat F(B)^{ab} -\tfrac1{4} \phi\,
  \hat T^{ab} \gamma^5]\gamma_{ab}   \,\hat\psi^i  +4\mathrm{i}\,
  \phi^{-1} \Slash{\hat A}\, \hat\psi^i \big] \,,
\end{align}
where we use the definition \eqref{eq:cov-der-hat-psi} for the
supercovariant derivative of $\hat\psi^i$ based on the S-supersymmetry
gauge field $\hat\phi_\mu{\!}^i$. Eventually we will make another,
more suitable, choice for this composite gauge field, but for the
moment we adopt this definition.

Restricting ourselves to terms linearly proportional to fermion fields, the
variation of $\hat\chi^i$ takes the following form, 
\begin{align}
  \label{eq:delta-hat-chi}
  \delta\hat\chi^i=&\, \tfrac1{24}\hat T_{ab} \gamma^{ab} \eta^i  +
  \tfrac1{6}R(V)_{abj}{}^i \,\gamma^{ab} \epsilon^j + \tfrac1{24}
  \mathrm{i} \,\gamma^{ab} \, \Slash{\mathcal{D}}\hat T_{ab}\epsilon^i
  -\tfrac13 R(A)_{ab} \,\gamma^{ab} \gamma^5 \epsilon^i  + \hat D\, \epsilon^i \,,
\end{align}
where $\hat D$ is defined as (up to terms quadratic in spinor fields)
\begin{align}
  \label{eq:def-hat-D}
  \hat{D} =&\, 4\,D\big\vert_{5D} - \tfrac1{4} \phi^{-2} \,\hat
  V_j{}^k\, \hat V_k{}^j + \tfrac14 \phi^{-1}\,\bigl[(\mathcal{D}_a)^2
  + \tfrac16 R(\omega,e)\bigr] \phi
  - \tfrac1{12} (\hat A_a)^2  \nonumber \\
  &\, - \tfrac1{12}\phi^{-2}\hat{F}^{ab}\hat{F}_{ab} +\tfrac1{192}
  \phi^{-1}\epsilon_{abcd}\,\hat{T}^{ab}\hat{F}^{cd}
  +\tfrac1{384}\hat{T}^{ab}\,\hat{T}_{ab} \, ,
\end{align}
where we have made use of~\eqref{eq:def-hat-f}.  Note that all 
bosonic terms in \eqref{eq:delta-hat-chi} have been included. 

We conclude this part of the analysis by giving the Q- and
S-supersymmetry transformations for the remaining fields, where we
give also some further details about terms quadratic in the spinor
fields,
\begin{align}
  \delta b_\mu =&\, \tfrac12 \mathrm{i}\, \bar\epsilon_i\,\hat
  \phi_\mu{}^i -\tfrac12 \epsilon_i\gamma_\mu\hat\chi^i
  +\tfrac12 \mathrm{i}\,  \bar\eta_i \,\psi_\mu{\!}^i  \,, \nonumber\\
  \delta\hat{T}_{ab} =&\,
  -8\mathrm{i}\,\bar{\epsilon}_i\,R(Q)_{ab}{}^i +
  4\mathrm{i}\,\bar{\epsilon}_i\gamma_{ab}\hat{\chi}^i +\tfrac12
  \varepsilon_{abcd}\, \hat T^{cd} \, \tilde\Lambda^0 \,,\nonumber \\
  \delta\hat{A}_\mu =&\;
  \bar{\epsilon}_i\gamma_\mu\gamma^5\hat{\chi}^i
  -\tfrac12\mathrm{i}\,\bar{\epsilon}_i\gamma^5\hat{\phi}_\mu{}^i -
  \tfrac12\mathrm{i} \,\bar\eta_i \gamma^5 \psi_\mu{}^i + \partial_\mu
  \tilde \Lambda^0   \,, \nonumber\\
  \delta \hat V_j{}^i=&\, 2\,\bar\epsilon_j (\Slash{D}\hat{\psi}^i
  -\mathrm{i} \gamma^5 \phi \,\hat \chi^i) - \delta_j{}^i\,
  \bar\epsilon_k (\Slash{D}\hat{\psi}^k -\mathrm{i}
  \gamma^5 \phi \,\hat \chi^k)  \,,  \nonumber\\
  \delta\hat{D} =&\, \bar{\epsilon}_i \,\Slash{D}\hat{\chi}^i +
  \cdots\,.
\end{align}
In the derivation of the first result for $\delta b_\mu$ we note that
the same phenomenon takes place as when deriving the transformation
rules for $\hat V_{\mu\,i}{}^j$ in \eqref{eq:delta-V-hat-su2}. Namely
the S-supersymmetry transformation with field-dependent parameter
$\tilde\eta^i$ in \eqref{eq:D5-D4-decomp} cancels against the terms
in \eqref{eq:S-susy-gauge fields} that depend explicitly on
$\psi_\mu{\!}^i$. After that we use the definition of $\hat\chi^i$ in
\eqref{eq:hat-chi}, and the remaining terms are absorbed into the $4D$
conformal boost transformation. Since $b_\mu$ is the only field that
transforms under conformal boosts, this will only affect the explicit
form of the supersymmetry algebra. The transformation rules of $\hat
T_{ab}$, $\hat A_\mu$ and $\hat V_j{}^i$ do not involve further
subtleties, except that $\hat A_\mu$ does not seem to transform
supercovariantly. The transformation rule of $\hat D$, however, cannot
be reliably calculated at this stage, because we have not yet
determined the contributions quadratic in the spinor fields in its
definition \eqref{eq:def-hat-D}.  In view of the fact that the
original $5D$ theory as well as its reduced $4D$ version is consistent,
there is no doubt that the present calculation can be completed to all
orders.

We have thus shown in considerable detail how the $5D$ Weyl multiplet
reduces to the $4D$ Euclidean Weyl multiplet and a Kaluza-Klein vector
supermultiplet. However, the latter multiplet involves only seven
bosonic and eight fermionic degrees of freedom, so that one bosonic
field seems to be missing in the Kaluza-Klein vector multiplet. A
similar counting for the Weyl multiplet reveals that the Weyl
multiplet has twenty-five bosonic and twenty-four fermionic degrees of
freedom (in off-shell counting one always corrects for the number of gauge
invariances, so that, for instance, each gravitino contributes only
eight fermionic degrees of freedom).

The reason for the mismatch is well known; under dimensional reduction
one obtains the lower-dimensional theory in a partially gauge-fixed
form. The R-symmetry is extended to
$\mathrm{SU}(2)\times\mathrm{SO}(1,1)$, where the non-compact
$\mathrm{SO}(1,1)$ factor acts by a chiral transformations on the
fermions (it will also act on some of the bosonic fields). At this
point the $\mathrm{SO}(1,1)$ group is, however, not realized as a
local invariance. Although the vector field $\hat A_\mu$ seems to play
the role of an $\mathrm{SO}(1,1)$ gauge field, it is not transforming
under a corresponding gauge symmetry and represents four bosonic
dergrees of freedom. This is the underlying reason why the
Kaluza-Klein supermultiplet is not yet realized as an irreducible
multiplet. 

Full irreducibility can be obtained by introducing a compensating scalar
field $\varphi$ and writing
\begin{equation}
  \label{eq:def-A-gauge}
  \hat A_\mu= A_\mu - \partial_\mu \varphi \,,
\end{equation}
where $A_\mu$ and $\varphi$ transform under {\it local}
$\mathrm{SO}(1,1)$ gauge transformations as
\begin{equation}
  \label{eq:so-11-gauge transf}
  A_\mu \to A_\mu + \partial_\mu\Lambda^0 \,, \qquad
  \varphi \to \varphi + \Lambda^0\,,
\end{equation}
so that $\hat A_\mu$ remains invariant. Under supersymmetry we assume
that $\varphi$ changes according to 
\begin{equation}
  \label{eq:transf-varphi}
  \delta\varphi = - \tilde\Lambda^0 = -\mathrm{i}
  \phi^{-1}\bar\epsilon_i\,\hat\psi^i\,. 
\end{equation}
Subsequently one uniformly redefines all fields and parameters with a
suitable $\varphi$-dependent $\mathrm{SO}(1,1)$ transformation, which
will remove all explicit terms in the transformation rules
proportional to $\tilde\Lambda^0$ and furthermore resolve various
questions raised previously (see e.g. the discussion following
\eqref{eq:field-dep-so11}). Upon imposing the gauge condition
$\varphi=0$, all the $\tilde\Lambda^0$-terms will re-emerge in the
form of compensating gauge transformations.

We now summarize all the $\varphi$-dependent field redefinitions. The
R-covariant spinors, transforming under local
$\mathrm{SU}(2)\times\mathrm{SO}(1,1)$ R-symmetry transformations, are
as follows, 
\begin{equation}
  \label{eq:compensating-chiral-tr}
  \begin{array}{rcl}
  \epsilon^i\vert^\mathrm{Rcov} &\!\!\!\!=\!\!\!& 
  \exp[-\tfrac12\varphi\,\gamma^5]\, 
  \epsilon^i\,,\\[2mm] 
  \eta^i\vert^\mathrm{Rcov} &\!\!\!\!=\!\!\!&
  \exp[\tfrac12\varphi\,\gamma^5]\,\eta^i \,,\\[2mm] 
  \chi^i\vert^\mathrm{Rcov} &\!\!\!\!=\!\!\!&
  \exp[-\tfrac12\varphi\,\gamma^5]\, \hat\chi^i \,,\\
   \end{array}
  \qquad
  \begin{array}{rcl}
  \psi_\mu{}^i\vert^\mathrm{Rcov} &\!\!\!\!=\!\!\!&
  \exp[-\tfrac12\varphi\,\gamma^5]\,\psi_\mu{}^i\,,\\[2mm] 
    \phi_\mu{\!}^i\vert^\mathrm{Rcov} &\!\!\!\!=\!\!\!&
  \exp[\tfrac12\varphi\,\gamma^5]\, \hat\phi_\mu{\!}^i \,,\\[2mm]
  \psi^i\vert^\mathrm{Rcov} &\!\!\!\!=\!\!\!&
  \exp[-\tfrac12\varphi\,\gamma^5]\, \hat\psi^i \,.\\
  \end{array}
\end{equation}
Also some of the bosons will have to be redefined so that they
transform covariantly under $\mathrm{SO}(1,1)$. First of all the
tensor fields $\hat T_{ab}$, when decomposed into the self-dual and
anti-self-dual (real) components, take the form 
\begin{equation}
  \label{eq:cov-T}
  T_{ab}{\!}^{\pm\mathrm{Rcov}} = \exp[\pm\varphi]\, \hat
    T_{ab}{\!}^\pm\,.  
\end{equation}
Furthermore the scalars $\phi$ and $\varphi$ are combined into the
fields $\phi \,\exp[\mp\varphi]$ which transform into R-covariant
spinors according to 
\begin{equation}
  \label{eq:cov-phi-varphi}
  \delta\big ( \phi  \,e^{\mp\varphi} \big) = \pm \mathrm{i} \big[
  \bar\epsilon_i (1\pm\gamma^5) \psi^i\big]^\mathrm{Rcov} \,,
\end{equation}
where the fermions on the right-hand side are R-covariant. 
After these last redefinitions the Weyl multiplet has become
irreducible. It now includes the $\mathrm{SO}(1,1)$ gauge field $A_\mu$,
so that it comprises $24\oplus 24$ off-shell bosonic and fermionic degrees of
freedom. The compensator $\varphi$ belongs to the Kaluza-Klein vector
multiplet, which is defined in a background consisting of the Weyl
multiplet and comprises $8\oplus 8$ degrees of freedom.

In the next subsection we describe how to bring the theory in a form
that is closely related to the $4D$ Minkowski theory, first by modifying the
expressions for the dependent fields of the Weyl multiplet,
$\phi_\mu{\!}^i$, and $f_\mu{\!}^a$, which implies that we modify the
conventional constraints. In a second subsection
we will then briefly disucuss the vector, tensor and hyper multiplets.

\subsection{S-invariant conventional constraints }
\label{sec:mod-conventional-and chiral}
At this stage we will make some field redefinitions to bring the
results in closer contact with the Minkowski version of $N=2$
supergravity. First of all we will redefine the S-supersymmetry gauge field
according to 
\begin{equation}
    \label{eq:S-convetional constraint}
    \phi_\mu{}^i = \phi_\mu{}^i\vert_\mathrm{old} - \tfrac12\mathrm{i}\gamma_\mu\chi^i \,.
\end{equation}
This will correspond to a different conventional constraint (the
previous one was given by \eqref{eq:intermediate-4d-connection}) that
is S-supersymmetric. At the same time we make use of the R-covariant
fields defined previously. As a result the transformation rules will
acquire a simpler form.

Because of the redefinition \eqref{eq:S-convetional constraint} the explicit
expressions for the the dependent gauge fields $\phi_\mu{\!}^i$ and
$f_\mu{\!}^a$ become 
\begin{align}
  \label{eq:def-phi-f}
  \phi_\mu{\!}^i =&\,-\mathrm{i} \big(\tfrac12 \gamma^{\rho\sigma}\gamma_\mu
  -\tfrac16 
  \gamma_\mu\gamma^{\rho\sigma} \big) \,\big(\mathcal{D}_\rho
  \psi_\sigma{\!}^i + \tfrac1{32}\mathrm{i}\, T_{ab} \gamma^{ab} \gamma_\rho\psi_\sigma{\!}^i
  +\tfrac14 \gamma_{\rho\sigma} \chi^i \big) \,,  \nonumber\\[1mm]
    f_\mu{\!}^a =&\,  \tfrac12  R(\omega,e)_\mu{}^a -
    \tfrac14\bigl({D} + \tfrac13 R(\omega,e)\bigr) 
  \,e_\mu{}^a -\tfrac12 \tilde R(A)_\mu{}^a
  -\tfrac1{32}  T_{\mu b}{\!\!}^- \,T^{+ba} \,,
\end{align}
where, in the last expression, we have only shown the bosonic terms. 
Note that we have been using R-covariant spinors and tensor fields in
the above results. Furthermore we have suppressed the carets on 
the various fields and therefore use the proper $4D$ covariant fields
wherever possible. The derivative $D_\mu$ is fully
supercovariant whereas the derivative $\mathcal{D}_\mu$ is covariant
with respect to local Lorentz, dilatations and the full R-symmetry
$\mathrm{SO}(1,1)\times \mathrm{SU}(2)$. All the field strengths are
also supercovariant. The presence of the fully
supersymmetric covariant quantities has not been verified in every
possible detail, but these covariantizations are implied by the
supersymmetry algebra. 

In view of the fact that we are now dealing with a Euclidean theory,
we prefer to change the definition of the Dirac conjugate fields
accordingly, so that  $\bar \chi_i \equiv \chi_i^\dagger$.  This
requires to replace all the spinors $\bar \chi_i$ in previous
equations by $\bar\chi_i\,\gamma^5$. The symplectic Majorana
condition on the spinors then preserves its form, provided the
charge conjugation matrix is redefined according to
\begin{align}
  \label{eq:new-C}
  C\vert^\mathrm{new} = C\,\gamma^5\,.
\end{align}
The correponding symplectic Majorana condition for the spinors is
given in \eqref{eq:Majorana-4D}. As a result of the
redefinition~\eqref{eq:new-C} the hermitian gamma matrices $\gamma_a$
now satisfy
\begin{equation}
  \label{eq:gamma-C}
  C\,\gamma_a\,C^{-1} = -\gamma_a{\!}^\mathrm{T}\,, \qquad
  (a=1,2,3,4). 
\end{equation}
For spinor bilinears $\bar\psi_i \Gamma \varphi^j$, a useful formula
is given in \eqref{eq:euclidean-bilinear}. 

With these new conventions  we summarize the supersymmetry transformations in
terms of the R-covariant fields, 
\begin{align}
    \delta e_\mu{}^a =&\,  \bar\epsilon_i\,\gamma^5\gamma^a\psi_\mu{}^i
  \,, \nonumber\\[.1ex]
    \delta\psi_\mu{}^i =&\,2\,\mathcal{D}_\mu \epsilon^i  
    + \tfrac1{16} \mathrm{i}  \,(T_{ab}{\!}^+ +T_{ab}{\!}^- ) \gamma^{ab} \gamma_\mu
    \epsilon^i    -\mathrm{i} \gamma_\mu \eta^i \,, \nonumber\\[1mm]
   \delta b_\mu =&\, \tfrac12 \mathrm{i}\, \bar\epsilon_i\,\gamma^5
   \phi_\mu{}^i -\tfrac34 \bar\epsilon_i\,\gamma^5\gamma_\mu \chi^i
   +\tfrac12 \mathrm{i}\,  \bar\eta_i \,\gamma^5\psi_\mu{\!}^i  +
   \Lambda_\mathrm{K}{\!}^a \,e_{\mu a} \,,
   \nonumber\\[1mm]
   \delta{A}_\mu =&\,
   -\tfrac12\mathrm{i}\,\bar{\epsilon}_i \,\phi_\mu{}^i -
   \tfrac34\,\bar{\epsilon}_i\,\gamma_\mu \chi^i 
  -\tfrac12 \mathrm{i}\,\bar\eta_i\,\psi_\mu{\!}^i \, ,
  \nonumber\\[1mm]
        \delta\mathcal{V}_{\mu}{}^i{}_j =&\, 
      2\mathrm{i}\,\bar{\epsilon}_j\,\gamma^5\phi_\mu{}^i
      - 3\, \bar{\epsilon}_j\,\gamma^5\gamma_\mu{\chi}^i - 2\mathrm{i}\,\bar\eta_j
      \,\gamma^5\psi_\mu{}^i\nonumber\\
      &\,- \tfrac12\delta^i{}_j\bigl(2\mathrm{i} \,
      \bar{\epsilon}_k\,\gamma^5{\phi}_\mu{}^k - 3\,
      \bar{\epsilon}_k\,\gamma^5\gamma_\mu{\chi}^k
      - 2\mathrm{i}\,\bar\eta_k\,\gamma^5  \psi_\mu{}^k \bigr) \, ,  \nonumber\\[1mm]
      \delta{T}_{ab}{\!}^\pm =&\, -8\mathrm{i}\,
    \bar\epsilon_i\,\gamma^5 R(Q)_{ab}{\!}^{i\,\pm} \,,  \nonumber \\[1mm]
    \delta\chi^i=&\,  \tfrac1{24}
  \mathrm{i} \,\gamma^{ab} \, \Slash{D}
  (T_{ab}^+ + T_{ab}^-) \gamma^{ab} \epsilon^i+
  \tfrac1{6}R(\mathcal{V})_{ab}{\!}^i{}_j \,\gamma^{ab} \epsilon^j
  -\tfrac13 R(A)_{ab} \,\gamma^{ab} \gamma^5 \epsilon^i \nonumber\\ 
  &\,  + D\,
  \epsilon^i +\tfrac1{24}(T_{ab}^+ + T_{ab}^-) \gamma^{ab} \eta^i \,, \nonumber\\[1mm]
    \delta{D} =&\, \bar{\epsilon}_i \,\gamma^5 \Slash{D}\chi^i \,,
\end{align}
where we changed notation for the $\mathrm{SU}(2)$ R-symmetry gauge
field to remain in line with the $4D$ Minkowski theory and used the
(anti-)self-dual components of the field strength $R(Q)_{ab}{\!}^{i}$,
\begin{align}
  \label{eq:new-v-r-Q}
  \mathcal{V}_\mu{}^i{}_j =&\, V_{\mu \,j}{}^i \,, \nonumber\\
   R(Q)_{\mu\nu}{\!}^i =&\, 2\,\mathcal{D}_{[\mu} \psi_{\nu]}{\!}^i   
  -\mathrm{i} \,\gamma_{[\mu} \,\phi_{\nu]}{}^i   
  +\tfrac1{16} \mathrm{i}\,(T_{ab}{\!}^+ +T_{ab}{\!}^- )\gamma^{ab} \,\gamma_{[\mu}
  \,\psi_{\nu]}{}^i \,.
\end{align}
We also present the transformation rule for the two dependent gauge
fields, $\phi_\mu{\!}^i$ and $f_\mu{\!}^a$, up to terms quadratic in
fermions:
\begin{align}
    \label{eq:delta-phi-f}
 \delta\phi_\mu{}^i =&\, 2\,\mathcal{D}_\mu \eta^i + 2\mathrm{i}
  \,f_\mu{}^a\gamma_a\epsilon^i 
  +\tfrac1{16}\Slash{D}  (T_{ab}{\!}^++T_{ab}{\!}^-) \gamma^{ab}\gamma_\mu 
 \,\epsilon^i \nonumber\\
  &\, -\tfrac14\mathrm{i}\gamma^{ab}\gamma_\mu
  R(\mathcal{V})_{ab}{\!}^i{}_j \,\epsilon^j
  - \tfrac12\mathrm{i}\gamma^{ab}\gamma_\mu \gamma^5 
  R(A)_{ab}\,\epsilon^i -\mathrm{i}\,\Lambda_\mathrm{K}{\!}^a\,\gamma_a 
  \psi_\mu{\!}^i  \, , \\[1mm]
  \delta f_\mu{\!}^a=&\,
  \tfrac14\mathrm{i}\,\bar\epsilon_i\,\gamma^5\psi_\mu{\!}^i D_b (T^{+ba} + T^{-ba})
  -   \tfrac34 e_\mu{\!}^a\, \bar\epsilon_i \gamma^5\Slash{D}\chi^i -\tfrac34
  \bar\epsilon_i \,\gamma^5\gamma^a \psi_\mu{\!}^i \,D \nonumber\\
   &\, +\bar\epsilon_i\,\gamma^5 \gamma_\mu D_b R(Q)^{ba\,i} 
   + \tfrac12\,\bar\eta_i \,\gamma^5\gamma^a\phi_\mu{\!}^i 
   + \mathcal{D}_\mu\Lambda_\mathrm{K}{\!}^a \, . \nonumber  
\end{align}

For completeness we also list the transformation rules for the
Kaluza-Klein multiplet,
\begin{align}
  \label{eq:susy-KK}
  \delta\big ( \phi \,e^{\mp\varphi} \big) =&\,  \mathrm{i} 
  \bar\epsilon_i (1\pm\gamma^5) \hat\psi^i  \,,  \nonumber \\[1mm]
  \delta B_\mu=&\, -\bar\epsilon_i\,\gamma^5\gamma_\mu \hat\psi^i - \mathrm{i}
  \,\bar\epsilon_i \, \phi\, \mathrm{e}^{\varphi\gamma^5}
  \psi_\mu{}^i \,,\nonumber
  \\[1mm]
  \delta\hat{\psi}^i =&\, \tfrac12 \big[\hat{F}(B)_{ab} -\tfrac18
  \phi\,\hat T_{ab}\gamma_5 \big]
  \gamma^{ab}\epsilon^i -\mathrm{i} \gamma^5 \Slash{D}\big(\phi\,
  \mathrm{e}^{\varphi\gamma^5}\big) \epsilon^i + \hat{V}_j{}^i
  \,\epsilon^j +\phi\,\mathrm{e}^{-\varphi\gamma^5}\gamma^5\eta^i\,,
  \nonumber\\[1mm]
  \delta \hat V_j{}^i=&\, 2\,\bar\epsilon_j \,\gamma^5 \Slash{D}\hat\psi^i 
  - \delta_j{}^i\,  \bar\epsilon_k\,\gamma^5 \Slash{D}\hat\psi^k  \,,
\end{align}
where the supercovariant field strength associated with the vector
field $B_\mu$
\begin{equation}
  \label{eq:KK-F}
   \hat{F}(B)_{\mu\nu} = 2\,\partial_{[\mu} B_{\nu]} + \bar\psi_{[\mu
     i}\,\gamma^5\gamma_{\nu]} \hat\psi^i +\tfrac12 \mathrm{i} 
  \bar\psi_{\mu i} \, \phi\, \mathrm{e}^{\varphi\gamma^5}  \psi_{\nu}{}^i \, .
\end{equation}
Observe  that for the Kaluza-Klein vector multiplet we still use the fields $\hat\psi^i$, $\hat
V_j{}^i$ and $\hat T_{ab}$ but the spinors are R-covariant fields.
%
\begin{table}[t]
\centering
\begin{tabular}{|c||cccc||cccc||cc|} 
\hline 
 &  \multicolumn{4}{c||}{vector multiplet}&\multicolumn{4}{c||}
 {tensor multiplet}& \multicolumn{2}{c|} {hypermultiplet} \\ \hline \hline 
 field & $\sigma$ & $W_M$ & $\Omega_i$  & $Y_{ij}$ &  $L^{ij}$&
 $E_A$&$\varphi^i$& $N$&   $\quad A_i{}^\alpha$&
 $\zeta^\alpha\quad$  \\[.4mm] \hline 
$w$  & $1$ & $0$ &$\tfrac{3}{2}$ & $2$& $3$ & $4$& $\tfrac72$& $4$ &
                                $\quad\tfrac{3}{2}$ & $2\quad$ \\[.4mm] 
\hline 

\end{tabular}
\vskip 2mm
\renewcommand{\baselinestretch}{1}
\parbox[c]{11cm}
{\caption{\footnotesize Weyl weights $w$ of the
    vector multiplet, the tensor multiplet and the hypermultiplet
    component fields in five 
    space-time dimensions. \label{tab:w-weights-matter-5D}}}   
\end{table}

\section{Off-shell dimensional reduction; matter multiplets}
\label{sec:shell-dimens-reduct-matter}
\setcounter{equation}{0}
We briefly consider the dimensional reduction for the vector and
tensor multiplets and for the hypermultiplet. We first present their
$5D$ transformation rules and then apply the reduction to the
transformations of the corresponding $4D$ Euclidean multiplets.  Using
the standard Kaluza-Klein ansatz outlined earlier, one obtains the Q-
and S-supersymmetry transformation rules upon including the
compensating Lorentz transformation \eqref{eq:comp-Lor}, and, at the
end, suppressing the composite S-supersymmetry, $\mathrm{SU}(2)$
R-symmetry and $\mathrm{SO}(1,1)$ transformations given in
\eqref{eq:D5-D4-decomp}. The dictionary that expresses the $4D$ fields
in terms of the $5D$ ones will then be presented to indicate how the
$4D$ results were derived.

\subsection{The vector supermultiplet}
\label{sec:vect-superm}
The $5D$ vector supermultiplet consists of a real scalar $\sigma$, a
gauge field $W_M$, a triplet of (auxiliary) fields $Y^{ij}$, and a
fermion field $\Omega^i$. Under Q- and S-supersymmetry these fields
transform as follows,
\begin{align}
  \label{eq:sc-vector-multiplet}
  \delta \sigma =&\,
  \mathrm{i}\,\bar{\epsilon}_i\,\Omega^i \,,
  \nonumber \\ 
  \delta\Omega^i =&\,
  - \ft12 (\hat{F}_{AB}- 4\,\sigma T_{AB}) \gamma^{AB} \epsilon^i
  -\mathrm{i} \Slash{D} \sigma\,\epsilon^i -2\varepsilon_{jk}\,
  Y^{ij} \epsilon^k 
  + \sigma\,\eta^i \,,   \nonumber\\ 
  \delta W_M =&\,
  \bar{\epsilon}_i\gamma_M\Omega^i - \mathrm{i}
  \sigma \,\bar\epsilon_i \psi_M{}^i  \,, \nonumber\\ 
  \delta Y^{ij}=&\,  
   \varepsilon^{k(i}\, \bar{\epsilon}_k\, \Slash{D} \Omega^{j)} 
  + 2{\mathrm{i}}\,\varepsilon^{k(i}\, \bar\epsilon_k (-\ft1{4}
  T_{AB}\gamma^{AB}\Omega^{j)}+ 4 \sigma \chi^{j)})
  -\ft1{2}{\mathrm{i}}  \varepsilon^{k(i}\, \bar{\eta}_k\,
  \Omega^{j)} \,.  
\end{align}
where $(Y^{ij})^\ast\equiv Y_{ij}= \varepsilon_{ik}\varepsilon_{jl}
Y^{kl}$, and the supercovariant field strength $\hat F_{MN}(W)$ is defined as, 
\begin{equation}
  \label{eq:W-field-strength}
 \hat F_{MN}(W) = 2\, \partial_{[M} W_{N]}  -
 \bar\Omega_i\gamma_{[M} \psi_{N]}{}^i +\ft12
 \mathrm{i}\sigma\,\bar\psi_{[M i} \psi_{N]}{}^i \,.
\end{equation}
The fields behave under local scale transformations according to the
weights shown in table~\ref{tab:w-weights-matter-5D}. 

The dimensional reduction proceeds in the same way as before, except
that we now have the advantage that we have already identified some of
the $4D$ fields belonging to the $4D$ Weyl multiplet. We decompose the
$5D$ gauge field $W_M$ into a four-dimensional gauge field $W_\mu$ and
a scalar $W= W_5$. The result for the $4D$ Euclidean supermultiplet
thus involves two real fields $X_+$ and $X_-$, one spinor $\Omega^i$,
the gauge field $W_\mu$ and an auxiliary field $Y^{ij}$. The overall
normalization of the multiplet is defined by the requirement that the
$4D$ gauge field $W_\mu$ is precisely the one that follows from the
$5D$ one upon adopting the Kaluza-Klein ansatz. The supersymmetry
transformations of the $4D$ multiplet components take the following
form,
\begin{align}
  \label{eq:vector-4D-app}
  \delta X_\pm=&\, \pm\mathrm{i}\,\bar\epsilon_{i\pm}\Omega^i{\!}_\pm \, , \nonumber\\[1mm]
  \delta W_\mu=&\, \bar\epsilon_{i+}  \big(\gamma_\mu \Omega^i{\!}_-
  - 2\mathrm{i}\,X_-\psi_\mu{\!}^i{\!}_+ \big) - \bar\epsilon_{i-}
  \big(\gamma_\mu \Omega^i{\!}_+  - 2\mathrm{i}\,X_+\psi_\mu{\!}^i {\!}_-\big)
  \nonumber\\[1mm] 
  \delta \Omega^i{\!}_\pm =&\, -2\mathrm{i}\,\Slash{D} X_\pm\,
  \epsilon^i{\!}_\mp  -\tfrac12[\hat F(W)_{ab}^\mp -\tfrac14 X_\mp\,
  T_{ab}^\mp]\gamma^{ab} \epsilon^i{\!}_\pm  - \varepsilon_{kj}\, 
  Y^{ik}\epsilon^j{\!}_\pm + 2\,X^\pm\eta^i{\!}_\pm \, ,  \nonumber \\[1mm]
  \delta Y^{ij} =&\, 2\,\varepsilon^{k(i}\bar{\epsilon}_k\gamma^5\Slash{D}\Omega^{j)} \, .
\end{align}
The supercovariant field strength $\hat F(W)_{\mu\nu}$ is defined as
\begin{equation}
\hat{F}(W)_{\mu\nu} = 2\,\partial_{[\mu} W_{\nu]} + \bar{\psi}_{i[\mu}\gamma_{\nu]}\Omega^i{\!}_+ - \bar{\psi}_{i[\mu}\gamma_{\nu]}\Omega^i{\!}_- + \mathrm{i}\,X_- \bar{\psi}_{\mu\,i}\psi_\nu{\!}^i{\!}_+ - \mathrm{i}\,X_+\bar{\psi}_{\mu\,i}\psi_\nu{\!}^i{\!}_- \, .
\end{equation}

This result is based on the following identification of the $4D$
fields expressed in terms of the $5D$ ones,
\begin{align}
  \label{eq:2identif-vector}
  X_\pm =&\,\tfrac12 \mathrm{e}^{\mp\varphi} (\sigma
  \pm \phi W) \, , \nonumber\\ 
  \Omega^i =&\, \exp[-\tfrac12\varphi\,\gamma^5]\, 
  (\Omega^i + W\,\hat{\psi}^i)\,,  \nonumber\\
    Y^{ij} =&\, 2\, Y^{ij} -   W\,\hat{V}_k{}^{i} \, \varepsilon^{jk} +
  \mathrm{i}\,\phi^{-1}\,(\bar\Omega_k\gamma^5-\tfrac{1}{2}\,\sigma
  \phi^{-1}\, \bar{\hat\psi}_k)\,\hat\psi^{(i} \, \varepsilon^{j)k} \,.
\end{align}
For the Kaluza-Klein supermultiplet the corresponding identification is as follows
(c.f. \eqref{eq:susy-KK}), 
\begin{align}
  \label{eq:identif-KKvector}
  X_\pm =&\, \mp\tfrac12 \mathrm{e}^{\mp\varphi} \phi  \, , \nonumber\\
  W_\mu =&\, B_\mu \,,\nonumber \\ 
  \Omega^i =&\, - \hat\psi^i  \,,  \nonumber\\
    Y^{ij} =&\, \varepsilon^{ik} \,\hat V_k{}^j   \,.
\end{align}
Note that we have been dealing with the abelian vector multiplet. The
non-abelian extension of the vector multiplet will be given in
subsection  \ref{sec:vector-mult}. 

\subsection{The tensor supermultiplet}
\label{sec:linear-multiplet}
The linear multiplet consists of a triplet of scalars $L^{ij}$, a
rank-three antisymmetric tensor gauge field $E_{MNP}$, a fermion
doublet $\varphi^i$, and a real (auxiliary) scalar $N$. The supercovariant
field strength associated with the tensor field is denoted by $\hat
E_A$. The superconformal transformation rules for these
fields are as follows,
\begin{align}
  \label{eq:5D-tensor}
  \delta L^{ij} =&\, -2\mathrm{i} \,\varepsilon^{k(i}\,\bar\epsilon_k
  \varphi^{j)}
  \,,  \nonumber\\
  \delta \varphi^{i}=&\, - \mathrm{i}\, \varepsilon_{jk}\Slash{D}
  L^{ij}\epsilon^k+ (N-\mathrm{i}\hat{\Slash{E}})\epsilon^i
  + 3\,\varepsilon_{jk} L^{ij} \eta^k  \, ,\nonumber\\
  \delta E_{MNP}= &\, \bar\epsilon_i\gamma_{MNP}\varphi^i -
  3\mathrm{i} \,\bar\epsilon_i\gamma_{[MN} \psi_{P]}{}^k\,
  \varepsilon_{jk} L^{ij}\,, \nonumber\\
  \delta \hat E_A =&\, - \mathrm{i}\, \bar\epsilon_{i}\gamma_{AB} D^B
  \varphi^{i} +\tfrac{1}{4} \bar\epsilon_{i}(3\gamma_A\gamma^{BC} +
  \gamma^{BC}\gamma_A) \varphi^{i}\,T_{BC} -2 \,
  \bar\eta_{i}\gamma_A\varphi^{i} \, ,\nonumber\\
  \delta N =&\, \bar\epsilon_i \Slash{D} \varphi^i +
  \ft{3}{2}\mathrm{i} \bar\epsilon_i \gamma^{AB}\varphi^i\,T_{AB} -8
  \mathrm{i} \, \varepsilon_{jk}\,\bar\epsilon_i\chi^k L^{ij}
  +\tfrac32 \mathrm{i} \,\bar\eta_i\varphi^i \,,
\end{align}
where  the supercovariant field strength $\hat E^A$ is  is given by,
\begin{equation}
  \label{eq:E-field-strength}
  \hat E^M= \tfrac16 \mathrm{i}\,e^{-1}
  \varepsilon^{MNPQR} 
  \Big[\partial_N E_{PQR} -\tfrac12 \bar\psi_{N  i}\gamma_{PQR} \varphi^i +
  \tfrac34\mathrm{i}\,\bar\psi_{N i}\gamma_{PQ}
  \psi_R{}^k\,\varepsilon_{jk}L^{ij} \Big] \,. 
\end{equation}
Its corresponding Bianchi identity reads $D_A\hat E^A =0$. 
The behaviour under local scale transformations is indicated by the
weights shown in table~\ref{tab:w-weights-matter-5D}. 

To have the standard $4D$ Weyl weight, we redefine the field $L^{ij}$
with a multiplicative factor $\phi^{-1}$. The supersymmetry
transformation of $\phi^{-1} L^{ij}$ then yields the  corresponding
spinor field 
\begin{equation}
  \label{eq:tensor -spinor}
   \varphi^i\vert_{4D}  = \phi^{-1} \varphi^i - 
   L^{ik} \,\gamma^5 \psi^l\,\varepsilon_{kl} \,,
\end{equation}
where we took into account the composite $\mathrm{SU}(2)$
transformation specified in \eqref{eq:D5-D4-decomp}. Subsequently one
introduces the phase factors, just as before, and collect the
variations of the new spinor. 

In this way one obtains the supersymmetry transformations for the $4D$
Euclidean supermultiplet,
\begin{align}
  \label{eq:tensor-4D-app}
  \delta L^{ij} =&\, 2\mathrm{i} \, \bar\epsilon_k \,\gamma^5
  \varphi^{(i} \,\varepsilon^{j)k} \,,\nonumber\\
  \delta E_{\mu\nu}  =&\, \mathrm{i}\,\bar\epsilon_i\,\gamma_{\mu\nu}
  \,\varphi^i + 2\,\varepsilon_{jk}L^{ij}\,\bar{\epsilon}_i\gamma_{[\mu}
  \psi_{\nu]}{}^k\,, \nonumber\\
  \delta\varphi_\pm^i =&\, -\mathrm{i}\,\varepsilon_{jk}\Slash{D}L^{ij}
  \,\epsilon^k_\mp - \mathrm{i}\,\Slash{\hat E}\epsilon^i_\mp + 
  G_\pm\epsilon_\pm^i + 2\, \varepsilon_{jk} L^{ij} \,
  \eta^k_\pm \,, \nonumber \\
  \delta G_\pm =&\, \mp2\,\bar{\epsilon}_{i\mp}\Slash{D}\varphi^i_\pm
  \pm 6\mathrm{i}\,\varepsilon_{jk}L^{ij}\,\bar{\epsilon}_{i\mp}\chi^k_\mp
  \mp\tfrac18\mathrm{i}\,T_{ab}^\pm\,\bar{\epsilon}_{i\mp}\gamma^{ab}\varphi^i_\mp
  \pm2\mathrm{i}\,\bar{\eta}_{i\pm}\varphi^i_\pm \, ,
\end{align}
where $\hat E^\mu$ denotes the supercovariant field strength
associated with the tensor field $E_{\mu\nu}$. Its definition is
given in \eqref{eq:tensor-field-strength}; $G_\pm$ are two real scalars. 
This result is based on the following identification of the $4D$
fields in terms of the $5D$ fields,
\begin{align}
  \label{eq:identif-tensor}
  L^{ij} =&\, \phi^{-1} L^{ij}\, , \nonumber \\
  E_{\mu\nu} =&\, \mathrm{i}\,E_{\mu\nu\hat5}  \,,\nonumber\\
  \varphi^i =&\, \exp[\tfrac12\varphi\gamma^5] \,\big(\phi^{-1} \varphi^i - 
   L^{ik} \,\gamma^5 \psi^l\,\varepsilon_{kl} \big)\,, \nonumber\\
  G_\pm=&\, e^{\pm\varphi}\big(\phi^{-1}N \mp \mathrm{i}\,\hat{E}_{\hat5} - 
  \tfrac12\mathrm{i}\,\bar{\psi}_i\gamma^5\varphi^i \mp \mathrm{i}\,\bar{\psi}_i\varphi^i
  \pm \tfrac12\,\phi^{-2}\varepsilon_{jk}L^{ij}\,\hat{V}_i{}^k\big) \, .
\end{align}

\subsection{Hypermultiplets}
Hypermultiplets are associated with target spaces of dimension $4r$
that are hyper-K\"ahler cones \cite{deWit:1999fp}. The 5$D$ supersymmetry
transformations are most conveniently written in terms of the sections
$A_i{}^\alpha(\phi)$, where $\alpha= 1,2,\ldots,2r$,
\begin{align} 
  \label{eq:hypertransf}
  \delta A_i{}^\alpha=&\, 2\mathrm{i}\,\bar\epsilon_i\zeta^\alpha\,,
  \nonumber\\ 
  \delta\zeta^\alpha =&\, - \mathrm{i}\Slash{D}
  A_i{}^\alpha\epsilon^i   + \tfrac3{2} A_i{}^\alpha\eta^i \,.
\end{align}
The $A_i{}^\alpha$ are the local sections of an
$\mathrm{Sp}(r)\times\mathrm{Sp}(1)$ bundle. We also note the
existence of a covariantly constant skew-symmetric tensor
$\Omega_{\alpha\beta}$ (and its complex conjugate
$\Omega^{\alpha\beta}$ satisfying
$\Omega_{\alpha\gamma}\,\Omega^{\beta\gamma}= \delta_\alpha{}^\beta$),
and the symplectic Majorana condition for the spinors reads as
$C^{-1}\bar\zeta_\alpha{}^\mathrm{T} = \Omega_{\alpha\beta}
\,\zeta^\beta$. Covariant derivatives contain the $\mathrm{Sp}(r)$
connection $\Gamma_A{}^\alpha{}_\beta$, associated with rotations of
the fermions. The sections $A_i{}^\alpha$ are pseudo-real, i.e. they
are subject to the constraint, $A_i{}^\alpha \varepsilon^{ij}
\Omega_{\alpha\beta} = A^j{}_\beta\equiv (A_j{}^\beta)^\ast$. The
information on the target-space metric is contained in the so-called
hyper-K\"ahler potential. For our purpose the geometry of the
hyper-K\"ahler cone is not relevant. Hence we assume that the cone is
flat, so that the target-space connections and curvatures will
vanish. The extension to non-trivial hyper-K\"ahler cone geometries is
straightforward.

The supersymmetry transformations for the $4D$ Euclidean
hypermultipets read as follows,
\begin{align}
  \label{eq:hyper-4D}
  \delta A_i{}^\alpha = &\, 2\mathrm{i}\,  \bar\epsilon_i \,\gamma^5
  \zeta^\alpha \,,\nonumber\\
  \delta \zeta^\alpha  =&\, -\mathrm{i} \Slash{D} A_i{}^\alpha  +A_i{}^\alpha \eta^i \,. 
\end{align}
This result is based on the following identification of the $4D$
fields in terms of the $5D$ fields, 
\begin{align}
  \label{eq:identif-hyper}
  A_i{}^\alpha =&\, \phi^{-1/2} A_i{}^\alpha \,, \nonumber\\
  \zeta^\alpha  =&\, \exp[\tfrac12\varphi\gamma^5]
  \,\big( \phi^{-1/2} \zeta^\alpha -\tfrac12 \,
  \phi^{-3/2}   A_i{}^\alpha \,\gamma^5 \hat{\psi}^i \big)\,, 
\end{align}
In the deriviation of the transformation rules \eqref{eq:hyper-4D} we
again made use of the composite $\mathrm{SU}(2)$ transformation
specified in \eqref{eq:D5-D4-decomp}. Hypermultiplets charged under 
non-abelian vector multiplets are presented in subsection \ref{sec:hypers}.

\subsection{The non-linear multiplet} 
\label{app:NL-supermult}
Here we present the so-called non-linear multiplet in
$5D$ Minkowski space as presented in \cite{Zucker:2003qv}.  This
multiplet should correspond upon dimensional reduction to the
non-linear multiplets in $4D$ Minkowski \cite{deWit:1980tn} .Here we
present the multiplet in the conventions of section
\ref{sec:euclid-superg-from-dim-red}. 

\begin{table}[t]
\centering
\begin{tabular}{|c||cccc|} 
\hline 
 &  \multicolumn{4}{c|}{non-linear multiplet}
 \\ \hline \hline 
 field & $\Phi^i{\!}_\alpha$ & $\lambda^i$ & $M$  & $V_A$ \\[.4mm] 
 \hline 
$w$  & $0$ & $\tfrac12$ &$1$ & $1$ \\[.4mm] 
\hline 
\end{tabular}
\vskip 2mm
\renewcommand{\baselinestretch}{1}
\parbox[c]{10cm}
{\caption{\footnotesize Weyl weights $w$ of the
    non-linear multiplet  component fields in five 
    space-time dimensions. \label{tab:w-weights-NL-5D}}}   
\end{table}

The $5D$ multiplet is based on an $\mathrm{SU}(2)$ matrix of scalars
$\Phi^i{\!}_\alpha$ with $\alpha=1,2$, a symplectic Majorana spinor
$\lambda^i$, a scalar $M$ and a vector field $V_A$, in a
superconformal background. The weights of the fields under dilatations
are shown in table \ref{tab:w-weights-NL-5D}. The Q- and S
supersymmetry transformations and the conformal boost transformations
are as follows,
\begin{align}
  \label{eq:non-linear-mult}
  \delta\Phi^i{\!}_\alpha =&\; \mathrm{i}\,(2\,\bar\epsilon_j \,\lambda^i -
  \delta^i{\!}_j\,\bar\epsilon_k \,\lambda^k)\,
  \Phi^j{\!}_\alpha\,,\nonumber\\
  \delta\lambda^i =&\;  \mathrm{i}\,
  \Phi^i{\!}_\alpha\,\Slash{D}\Phi^\alpha{\!}_j \,\epsilon^j  +
  \tfrac12\,M\,\epsilon^i - \tfrac12\mathrm{i}\,\slash{V}  \epsilon^i  
  + 2\mathrm{i}\,(\bar\epsilon_j \lambda^i)\, \lambda^j 
  -\tfrac32\,\eta^i \,, \nonumber\\
  \delta M =&\; 2\,\bar{\epsilon}_i\,\Slash{D}\lambda^i 
  - \bar{\epsilon}_i\,\slash{V}\,\lambda^i + \mathrm{i}\,
  \bar{\epsilon}_i\,\lambda^i\,M 
  +  2\,\Phi^i{\!}_\alpha\,D_A\Phi^\alpha{\!}_j\,
  \bar{\epsilon}_i\,\gamma^A\,\lambda^j   
  + 4\mathrm{i}\,\bar{\epsilon}_i\,\chi^i 
  + \tfrac52\mathrm{i}\,\bar{\epsilon}_i\,
  \gamma^{AB}\,\lambda^i\,T_{AB}\nonumber\\ 
  \delta V_A =&\; -2\mathrm{i}\,\bar{\epsilon}_i\,\gamma_{AB}\,D^B\lambda^i 
  + \mathrm{i}\,\bar{\epsilon}_i\,\gamma_A\slash{V}\,\lambda^i 
  + \bar{\epsilon}_i\,\gamma_A\,\lambda^i\,M  
  - 2\mathrm{i}\,\Phi^i{\!}_\alpha\,
  D_B\Phi^\alpha{\!}_j\,\bar{\epsilon}_i\,\gamma_A\gamma^B\,\lambda^j
  \nonumber \\ 
  &\;-4\,\bar{\epsilon}_i\,\gamma_A\,\chi^i +
  6\,\bar{\epsilon}_i\,\gamma^B\,\lambda^i\,T_{AB} 
  +2\,\bar{\epsilon}_i\,\gamma_A\gamma^{BC}\,\lambda^i\,T_{BC} +
  \bar{\lambda}_i\,\gamma_A\,\eta^i + 6\,\Lambda_{\mathrm{K}\, A}\, .  
\end{align}
There is also a constraint on the divergence of the vector field,
\begin{equation}
  \label{constraint-nonlin-mult}
  D_AV^A -\tfrac12\,V_AV^A - 2\,D - \tfrac12\,M^2 +
  D_A\Phi^i{\!}_\alpha\,D^A\Phi^\alpha{\!}_i + \mathrm{fermions} = 0   \, , 
\end{equation}
which ensures that the multiplet comprises $8\oplus 8$ bosonic and
fermionic degrees of freedom and the transformation rules will close
off-shell. 

\end{appendix} 

\section*{Acknowledgement}
We thank Daniel Butter, Thomas Mohaupt, Paul Richmond and Alberto
Zaffaroni for useful comments and clarifications. This work was
partially supported by the ERC Advanced Grant no. 246974, {\it
  ``Supersymmetry: a window to non-perturbative physics''}. The work
of VR is also supported in part by INFN and by the ERC Starting Grant
637844-HBQFTNCER.
%
 
\providecommand{\href}[2]{#2}

\end{document}